\documentclass[aps,prd,nofootinbib,preprint,floatfix,superscriptaddress]{revtex4}
\usepackage[utf8]{inputenc}
\usepackage{amssymb}
\usepackage{amsmath}
\usepackage{amsfonts}
\usepackage{graphicx}
\usepackage{xspace}
\usepackage[normalem]{ulem} 
\usepackage{lscape}
\usepackage{ wasysym }
\usepackage{float}
\usepackage{mathrsfs}
\usepackage{slashed}
\usepackage{multirow,rotating}
\usepackage{xcolor}
\usepackage{hyperref}
\hypersetup{
  colorlinks=true,
  allcolors=darkpurple,
  pdfborder={0 0 0},
  linktocpage=true
}

\definecolor{darkpurple}{rgb}{0.5,0,0.5}
\definecolor{cambridgeblue}{rgb}{0.64, 0.76, 0.68}
\definecolor{darkraspberry}{rgb}{0.53, 0.15, 0.34}

\def\gsim{\raise0.3ex\hbox{$\;>$\kern-0.75em\raise-1.1ex\hbox{$\sim\;$}}}
\def\lsim{\raise0.3ex\hbox{$\;<$\kern-0.75em\raise-1.1ex\hbox{$\sim\;$}}}

\newcommand{\ba}[1]{\begin{eqnarray} \label{(#1)}}
\newcommand{\ea}{\end{eqnarray}}

\newcommand{\AddrAHEP}{\footnotesize AHEP Group, Instituto de F\'{\i}sica Corpuscular --
    CSIC/Universitat de Val{\`e}ncia, Apartado 22085,
  E--46071 Val{\`e}ncia, Spain}
  
\newcommand{\AddrAT}{\footnotesize Departament de F{\'i}sica Te{\`o}rica, Universitat de Val{\`e}ncia \\ 
    Instituto de F\'{\i}sica Corpuscular --
    C.S.I.C./Universitat de Val{\`e}ncia \\
Dr.~Moliner 50, E--46100 Burjassot, Spain}

  \newcommand{\AddrUFSM}{\footnotesize Departamento de F\' isica, Facultad de Ciencias, Universidad de La Serena, 
Avenida Cisternas 1200, La Serena, Chile 
}

\newcommand{\AddrSAPHIR}{\footnotesize Millennium Institute for Subatomic Physics at High Energy Frontier (SAPHIR), Fernández Concha 700, Santiago, Chile
}

\def\gsim{\raise0.3ex\hbox{$\;>$\kern-0.75em\raise-1.1ex\hbox{$\sim\;$}}}
\def\lsim{\raise0.3ex\hbox{$\;<$\kern-0.75em\raise-1.1ex\hbox{$\sim\;$}}}

\makeatletter
\g@addto@macro\bfseries{\boldmath}
\makeatother

\begin{document}

\preprint{FTUV-21-0528.5219}
\preprint{IFIC/21-18}

\title{Heavy neutral leptons in effective field theory \\
and the high-luminosity LHC}

\author{Giovanna Cottin}
\email{giovanna.cottin@uai.cl}

\affiliation{\footnotesize Departamento de Ciencias, Facultad de Artes Liberales, 
Universidad Adolfo Ib\'a\~{n}ez,
Diagonal Las Torres 2640, Santiago, Chile}
\affiliation{\AddrSAPHIR}
\author{Juan Carlos Helo} \email{jchelo@userena.cl}\affiliation{\AddrUFSM}\affiliation{\AddrSAPHIR}
\author{Martin Hirsch} \email{mahirsch@ific.uv.es}\affiliation{\AddrAHEP}
\author{Arsenii Titov}\email{arsenii.titov@ific.uv.es}\affiliation{\AddrAT}
\author{Zeren Simon Wang}\email{wzs@mx.nthu.edu.tw}
\affiliation{\footnotesize Department of Physics, National Tsing Hua University, 
Hsinchu 300, Taiwan}
\affiliation{\footnotesize Center for Theory and Computation, National Tsing Hua University,	Hsinchu 300, Taiwan}


\begin{abstract}

Heavy neutral leptons (HNLs) with masses around the electroweak scale
are expected to be rather long-lived particles, as a result of the
observed smallness of the active neutrino masses.  In this work, we
study long-lived HNLs in $N_R$SMEFT, a Standard Model (SM) extension
with singlet fermions to which we add non-renormalizable operators up
to dimension-6.  Operators which contain two HNLs can lead to a
sizable enhancement of the production cross sections, compared to the
minimal case where HNLs are produced only via their mixing with the SM
neutrinos.  We calculate the expected sensitivities for the ATLAS
detector and the future far-detector experiments: AL3X, ANUBIS,
CODEX-b, FASER, MATHUSLA, and MoEDAL-MAPP in this setup.  The
sensitive ranges of the HNL mass and of the active-heavy mixing
angle are much larger than those in the minimal case.  We study both,
Dirac and Majorana, HNLs and discuss how the two cases actually differ
phenomenologically, for HNL masses above roughly 100 GeV.

\end{abstract}



\maketitle

\tableofcontents
%



\section{Introduction}
\label{sect:intro}

So far, searches for physics beyond the standard model (BSM) at the
Large Hadron Collider (LHC) have focused on promptly decaying heavy
new states. However, no concrete discovery of any such field has been
announced yet, despite the roughly 140 fb$^{-1}$ of statistics
accumulated at both CMS and ATLAS. Perhaps mostly for this reason,
attention in recent years~---~from both experimental and theoretical
fronts~---~has shifted towards the study of fields that are light and
very weakly coupled to SM fermions, such that they have a relatively
long lifetime~\cite{Alimena:2019zri,Lee:2018pag,Curtin:2018mvb}.

Long-lived particles (LLPs), once produced at a collider, can lead to
variety of exotic signatures as their decay products are displaced
from their production position~\cite{Alimena:2019zri}.  For instance,
charged particles nearly mass degenerate with a neutral, stable dark
matter candidate can lead to a disappearing track signature. An
example of such exotic signature is predicted in the Minimal
Supersymmetric Standard Model (MSSM) with a compressed electro-weakino
spectrum (for searches at the LHC, see
Refs.~\cite{Aaboud:2017mpt,Sirunyan:2018ldc,Sirunyan:2020pjd}).
Electrically neutral LLPs, on the other hand, can lead to signals with
displaced vertices. The literature abounds with candidate
models. For example, in R-parity-violating
supersymmetry~\cite{Dreiner:1997uz,Barbier:2004ez,Mohapatra:2015fua}
the lightest SUSY particle can be long-lived
\cite{Porod:2000hv,Hirsch:2003fe} because of the smallness of the
R-parity-violating coupling due to the small neutrino masses.

A particularly simple and popular class of neutral LLP models are the
so-called `portal' models. In this class of models, a new scalar,
pseudoscalar, fermion or vector field is introduced, connecting the SM
sector with a hidden (or dark) sector. In particular, the
fermionic-portal mediator is conventionally called a `heavy neutral
lepton' (HNL) in the context of LLPs. Event numbers in such models are
expected to be very small. Nevertheless, LLP searches benefit from
considerably lower backgrounds than prompt searches. In addition,
given that the LHC will deliver up to 3 ab$^{-1}$ of statistics over the next
decade and a half, large parts of untested parameter space of LLP
models could be probed.

In the minimal HNL scenario, there is a small mixing of unspecified
origin of the HNLs with the active neutrinos. This setup is motivated
by the neutrino masses (and mixings) observed in oscillation
experiments~\cite{deSalas:2020pgw} (see also
Refs.~\cite{Esteban:2020cvm,Capozzi:2017ipn}), and the HNLs could be
identified with the sterile neutrinos in a `seesaw
mechanism'~\cite{Minkowski:1977sc,Yanagida:1979as,GellMann:1980vs,Mohapatra:1979ia,
  Schechter:1980gr}. In the classical seesaw type-I, for
$\mathcal{O}(1)$ Yukawa couplings sterile neutrinos are expected to be
as massive as $10^{15}$~GeV, which is well beyond the reach of any
foreseeable experiment. However, HNLs may have masses much closer to
the electroweak scale. For example, in the
inverse~\cite{Mohapatra:1986bd} or
linear~\cite{Akhmedov:1995ip,Akhmedov:1995vm} seesaw models, sterile
neutrinos can be light while keeping the mass for the active neutrinos
small at the same time.  Such HNLs are automatically long-lived for
small mixing angles and thus possibly within the reach of present and
upcoming collider and fixed-target experiments.

HNLs were searched for previously in a number of experiments.  The best
present limits, for HNL masses below roughly 100 GeV, come from
CHARM~\cite{Bergsma:1985is}, PS191~\cite{Bernardi:1987ek},
JINR~\cite{Baranov:1992vq}, and DELPHI~\cite{Abreu:1996pa}.  CMS at
the LHC has also published exclusion bounds for masses between 20~GeV
and 1600~GeV~\cite{Sirunyan:2018xiv}. ATLAS~\cite{Aad:2019kiz}
excludes a long-lived HNL in a mass range between 4.5 and 10 GeV, for
mixings as low as $\sim 10^{-6}$. In addition, $B$-factories such as
Belle~\cite{Abashian:2000cg} and Belle
II~\cite{Abe:2010gxa,Kou:2018nap} have been shown to possess excellent
sensitivity for HNLs lighter than $B$-mesons, for the copious
production of mesons and the $\tau$
lepton~\cite{Liventsev:2013zz,Kobach:2014hea,Dib:2019tuj}.  In the
high-luminosity phase of the LHC, both LHCb and
the main detectors CMS and ATLAS can extend existing limits for (or
provide discovery of) HNLs, using displaced
searches~\cite{Cottin:2018kmq,Cottin:2018nms,Drewes:2019fou,Bondarenko:2019tss,Liu:2019ayx}.

In addition, in the past few years a number of far detectors
specifically designed to search for LLP decays have been proposed:
ANUBIS~\cite{Bauer:2019vqk}, AL3X~\cite{Gligorov:2018vkc},
CODEX-b~\cite{Gligorov:2017nwh}, FASER and
FASER2~\cite{Feng:2017uoz,Ariga:2018uku}, MoEDAL-MAPP1 and
MAPP2~\cite{Pinfold:2019nqj,Pinfold:2019zwp}, and
MATHUSLA~\cite{Chou:2016lxi,Curtin:2018mvb,Alpigiani:2020tva}.  Many
papers already studied a variety of theoretical scenarios including
the minimal HNLs for these future experiments (see
e.g. Refs.~\cite{Kling:2018wct,Helo:2018qej,Dercks:2018wum,
  Hirsch:2020klk,deVries:2020qns}).

Besides the minimal case, HNLs also appear in various models extending
the SM further.  This can lead to new HNL production or decay modes,
affecting the sensitivity reaches at different experiments by large
factors. For instance, in a model extending the SM with a new
$U(1)_{B-L}$ gauge group~\cite{Deppisch:2018eth,Amrith:2018yfb}, the
SM Higgs boson can mix with a heavy scalar and thus decay to a pair of
HNLs. Additional production modes via a
$Z'$~\cite{Deppisch:2019kvs,Chiang:2019ajm} can lead to pairs of
displaced HNLs. Leptoquark models~\cite{Dorsner:2016wpm} can also
yield displaced HNL signatures~\cite{Cottin:2021tfo}.

The best systematic way to study such ``non-minimal'' HNLs is to apply
effective field theory (EFT). In EFT, non-renormalizable operators
(NROs) with mass dimension $d >4$ respecting the gauge symmetries
of the SM are introduced on top of the standard renormalizable terms.
NROs are suppressed by a new physics (NP) scale $\Lambda$ that is
assumed to be higher than the electroweak scale. In this work, we
denote the EFT of the SM augmented with sterile right-handed (RH)
neutrinos, $N_R$, as
$N_R$SMEFT~\cite{delAguila:2008ir,Aparici:2009fh,Liao:2016qyd} (see
also Refs.~\cite{Bell:2005kz,Graesser:2007yj,Graesser:2007pc}).  A
complete classification of $d\leq7$ operators containing $N_R$ can be
found in Ref.~\cite{Liao:2016qyd}.~%
\footnote{For a basis of the $d=6$ operators containing $N_R$ and
  independent off shell, i.e. before applying equations of motion, 
  called also Green basis, see Ref.~\cite{Chala:2020vqp}.}  

At $d=5$, there are two operators involving $N_R$.  Both of them
violate lepton number.  Their phenomenology, and in particular, that
of RH neutrino magnetic moments, was originally investigated in
Ref.~\cite{Aparici:2009fh}.  The effects of the $d=5$ Higgs-$N_R$
operator leading to a decay of the Higgs boson to two HNLs (each of
which subsequently decays to a charged lepton and two jets via
active-heavy neutrino mixing) were studied in
Ref.~\cite{Caputo:2017pit}.

Cross sections for lepton number conserving operators with $d=6$ were
initially discussed in Ref.~\cite{delAguila:2008ir} in the context of
Tevatron and LHC. These operators have been further analysed in
Ref.~\cite{Cirigliano:2012ab} in the context of both low-energy and
collider experiments.  In the last few years the phenomenology of
$N_R$SMEFT was studied in different regimes.  In particular, in
Ref.~\cite{Alcaide:2019pnf}, constraints on the four-fermion operators
involving $N_R$ have been set from LHC searches for associated
production of a charged lepton with missing energy, monojet searches,
as well as pion and tau decays, assuming $N_R$ to be stable at
collider scales. A different regime, in which $N_R$'s decay promptly
to an active neutrino and a photon via a dipole operator~%
\footnote{For a study of different decay modes of a Majorana HNL
  triggered by effective interactions see
  Refs.~\cite{Duarte:2015iba,Duarte:2016miz}.  }, was studied in
Ref.~\cite{Butterworth:2019iff} for operators involving the Higgs
boson, and in Ref.~\cite{Biekotter:2020tbd} for the four-fermion
operators.  Further collider studies of $N_R$SMEFT
include Refs.~\cite{Duarte:2014zea,Duarte:2016caz,Han:2020pff,
  Barducci:2020ncz,Barducci:2020icf}.

On the low-energy front, Ref.~\cite{Bischer:2019ttk} has derived
constraints on the $d=6$ operators coming from charged lepton flavor
violation (CLFV), beta decays and coherent elastic neutrino-nucleus
scattering (CE$\nu$NS).  The constraints originating from the
CE$\nu$NS as well as from meson decays have been also obtained in
Refs.~\cite{Han:2020pff,Li:2020lba,Li:2020wxi} and those from CLFV in
Ref.~\cite{Bolton:2020xsm}.  Furthermore, the effects of
non-renormalizable interactions on neutrinoless double beta decay were
investigated in Ref.~\cite{Dekens:2020ttz}.  
Very recently, Ref.~\cite{Cirigliano:2021peb} has shown that, under certain conditions, a $d=6$ operator inducing RH leptonic charged current 
can account for the observed values of the muon and electron anomalous magnetic moments.
Finally, the one-loop renormalization of $d=6$ operators containing $N_R$ was performed in
Refs.~\cite{Chala:2020pbn,Datta:2020ocb,Datta:2021akg} (see also
Ref.~\cite{Bell:2005kz} for earlier partial results).

Most closely related to this work, Ref.~\cite{deVries:2020qns} has
studied long-lived HNLs lighter than about 5 GeV, in the framework of
$N_R$SMEFT, taking advantage of the copious production of bottom and
charm mesons at the LHC.  More concretely, in each of the various
theoretical scenarios studied in that work, two separate EFT operators
each with one single HNL are switched on together with the underlying
electroweak charged and neutral currents, mediating the production and
decay of the HNLs.  Sensitivity reaches have been worked out for the
fixed-target experiment SHiP~\cite{Anelli:2015pba,SHiP:2018yqc}, the
local ATLAS detector~\cite{Aad:2008zzm}, as well as the far
detectors.

In the present work, we focus on operators containing a pair of HNLs
and a pair of quarks. Such interactions can enhance the production of
the HNLs from parton collisions with respect to the minimal HNL
scenario. In these scenarios, the decay of HNLs is mediated largely dominantly by
the usual mixing angle\footnote{The NRO's with a pair of HNLs can still induce HNL decays via suppression from the standard active-heavy neutrino mixing. We find that in general these contributions are negligible compared to the electroweak decays via the active-heavy neutrino mixing. Therefore, for the rest of this paper, we will ignore this possibility and write always that the HNL-pair NRO's can only induce the production but not the decay of the HNLs.}, allowing to probe HNL masses well beyond the
$B$-meson threshold.  In this theoretical framework, we derive the
sensitivity reach of both ATLAS and possible far detectors for the NP
scale $\Lambda$, the active-heavy mixing angle, as well as the HNL
mass. As we will show, $\Lambda$ in excess of 10 TeV and mixing angles
{\it below even type-I seesaw expectations} can be probed in this
setup.

The rest of this paper is organized as follows.  In
section~\ref{sect:eft}, we introduce the model basics for HNLs
including both $N_R$SMEFT and discussing briefly some ultraviolet (UV)
complete models.  In section~\ref{sect:exp}, we describe in detail the
simulation procedure and the analysis we perform for both ATLAS and
far-detector experiments.  Our numerical results are presented in
section~\ref{sect:results}.  Finally, in section~\ref{sect:sum}, we
summarize the work and provide an outlook. We added two appendices.
In the first, we present the effects of a possible timing cut at ANUBIS on our analysis, while the second discusses the reinterpretation of a
CMS search for HNLs \cite{Sirunyan:2018xiv} within our theoretical
setup.

\section{Effective field theory and heavy neutral leptons\label{sect:eft}}

In this section, we present some basic theoretical aspects of our
setup. First, as a motivation, we briefly discuss different variants
of electroweak seesaw mechanisms. We then turn to $d=6$ NROs in
``$N_R$SMEFT'', i.e.\ the EFT of the SM extended by (three generations
of) HNLs. Next, we briefly discuss some UV completions which can
generate the $d=6$ operators in the effective theory. While we will
present our numerical results only in terms of EFT parameters,
understanding the UV models has been a necessity for our calculation,
because of some limitation in \texttt{MadGraph5} \cite{Alwall:2007st,Alwall:2011uj,Alwall:2014hca}, as we will explain in
section~\ref{sect:exp}.

\subsection{Heavy neutral leptons, neutrino mixing, and neutrino mass
\label{subsect:MNu}}

Adding HNLs to the SM is usually motivated as an explanation for the
observed (mostly) active neutrino masses.  The simplest (and most
well-known) model realization of this idea is the type-I seesaw
mechanism~\cite{Minkowski:1977sc,Yanagida:1979as,Mohapatra:1979ia,GellMann:1980vs,Schechter:1980gr}.
In this case, the sterile neutrinos are Majorana particles. However,
there are other realizations of the seesaw mechanism, such as the
inverse~\cite{Mohapatra:1986bd} or linear
seesaw~\cite{Akhmedov:1995ip,Akhmedov:1995vm}. In these models, the
heavy sterile neutrinos form quasi-Dirac
pairs~\cite{Wolfenstein:1981kw,Petcov:1982ya,Valle:1982yw}.  Since
there are some interesting differences in our numerical results for
Dirac and Majorana cases (see section~\ref{sect:results}), here we want
to briefly discuss how either of the two possibilities can be realized
in concrete models.

If we add to the SM particle content two sets of singlets, call them
$N_R$ and $S_L$, the most general mass matrix of the neutral fermions,
in the basis ($\nu_L,N_R^c, S_L$)~%
\footnote{Here $N_R^c \equiv C \overline{N_R}^T$ represents the charge
  conjugate of $N_R$, with $C$ being the charge conjugation matrix.
  The number of singlets is, of course, arbitrary. For concreteness,
  here we consider the ``symmetric'' situation and add one flavour of
  $N_R$ and $S_L$ per SM generation of fermions. All submatrices are
  then (3,3) matrices. In the numerical part of the work we will
  consider only one generation, for simplicity.}, can be written as:
\begin{eqnarray}\label{eq:fullmat}
{\cal M} & = &  \left(
  \begin{array}{ccc}
   0 & m_D^T & \epsilon^T  \\ 
   m_D & M_M^{'} & M_R \\
   \epsilon & M_R^T & \mu
  \end{array}
\right).
\end{eqnarray}  
This setup contains a number of interesting limits. First of all, if
$(M_R,M_M^{'}) \ll \mu$ the fields $S_L$ decouple, and we find an
ordinary seesaw type-I with three heavy and three light states.  Light
and heavy neutrino masses as well as active-heavy mixing, in leading
order in $m_D/M_M$, are given by:
\begin{eqnarray}\label{eq:typeI}
m_{\nu} &=& - m_D^T \cdot M_M^{-1} \cdot m_D + \cdots \,, \\ \label{eq:typeIa}
M_{N} & =& M_M  + \cdots \,, \\ \label{eq:typeIb}
V  &=&  m_D^T \cdot M_M^{-1} + \cdots,
\end{eqnarray}
where $M_M = M_M' - M_R\cdot\mu^{-1}\cdot M_R^T$.~%
\footnote{ This setup \cite{Mohapatra:1986aw,Barr:2003nn} is sometimes
  called ``double seesaw'' \cite{Grimus:2013tva} in the literature.}
In this simplest possible setup we can assume $M_M$ is diagonal,
$\hat{M}_M$, without any loss of generality. All six neutral states
are Majorana particles in seesaw type-I.  $m_{\nu}$ is diagonalized by
a (3,3) matrix $U$, which corresponds to the mixing matrix observed in
neutrino oscillations.  By $V$ we denote the (3,3) matrix connecting
the active and heavy states.  Using Eqs.~(\ref{eq:typeI}) and
(\ref{eq:typeIb}), the mixing between the active and heavy sectors can be trivially estimated to be order of $V^2 \sim
m_{\nu}/\hat M_M$.  Note, however, that this estimate assumes that
there are no cancellations among the different terms contributing to
$m_{\nu}$, and if this assumption is dropped, larger values of $V^2$
can be obtained.

In the limit $M_M'=\epsilon=0$ and $\mu \ll M_R$ we find the inverse
seesaw, while the case $M_M'=\mu=0$ and $\epsilon \ll M_R$ corresponds
to the linear seesaw.  Consider, for example, the case of the inverse
seesaw. Again, in leading order in $m_D/M_R$, light and heavy sectors
mass matrices and active-heavy mixing are given as:
\begin{eqnarray}\label{eq:ISS}
m_\nu & = & m_D^T\cdot(\hat M_R^{T})^{-1}\cdot\mu\cdot \hat M_R^{-1}\cdot m_D \,,
\\ \label{eq:ISSa}
M_{\pm} & = &  
\Big( {\hat M}_R + \left\{m_D \cdot m_D^T,\hat M_R^{-1}\right\} \Big) 
\pm \frac{1}{2}\mu \,,
\\ \label{eq:ISSb}
V  &=&  m_D^T \cdot {\hat M_R}^{-1}\,.
\end{eqnarray}
Here, $\{a,b\}$ is the commutator of $a,b$, and the calculation is
done in the basis where $M_R$ is diagonal, $\hat M_R$. In this basis
both, $m_D$ and $\mu$, can have off-diagonal elements in general.  The
essence of the inverse seesaw is that, since $\mu$ is small compared
to $M_R$, $V$ is actually expected to be {\em larger} than in the case
of the standard type-I seesaw, as now we have $V^2 \sim
m_{\nu}/\mu$. More important for us is that in the limit $\mu \to 0$,
the three pairs of $N_{R}$ and $S_{L}$ form each a Dirac particle. Of
course, strictly speaking, $\mu=0$ is not possible for an inverse
seesaw model with a consistent phenomenology, since light neutrino
masses vanish in this limit. However, as was shown
in Ref.~\cite{Anamiati:2016uxp}, as long as the mass splitting of the heavy
states (proportional to $\mu$) is smaller than their decay width, the
heavy states are quasi-Dirac neutrinos. This implies that (i)~they
have only lepton-number-conserving decays, and (ii)~their production
cross section equals the one for Dirac neutrinos. Note that, while the
formulas for neutrino mass and for the mass splitting of the heavy
states are different, also in the linear seesaw the heavy neutrinos
form quasi-Dirac neutrinos, so the phenomenology is qualitatively
similar as for the inverse seesaw.

In our model implementations (to be detailed below), the calculation
of production cross sections with \texttt{MadGraph5} considers both
possibilities: HNLs as Dirac or Majorana states. Parameter fits to
neutrino data \cite{deSalas:2020pgw} (see
also Refs.~\cite{Esteban:2020cvm,Capozzi:2017ipn}) could be easily done for
either case: for the type-I seesaw with the well-known Casas-Ibarra
parametrization \cite{Casas:2001sr} and for more general cases with the
results of Refs.~\cite{Cordero-Carrion:2018xre,Cordero-Carrion:2019qtu}.
However, since in our numerical scans we consider only one HNL within
the experimentally accessible window, we will not fit neutrino
oscillation data explicitly. Rather, we treat the elements of the
active-heavy mixing matrix $V$ as free parameters.

Concerning decays of HNLs to SM particles via mixing, these have been
calculated several times in the literature. We will therefore not
discuss the details here. In our numerical work we use the decay
formulas of Refs.~\cite{Atre:2009rg,Bondarenko:2018ptm}.

\subsection{Non-renormalizable operators at $d=6$ in $N_R$SMEFT}

Adding $n_N=3$ copies of HNLs to the SM allows new
terms in the Lagrangian already at $d=3$ level. Depending on whether
the HNLs are Majorana or Dirac we can write down: 
\begin{align}\label{eq:Mass}
  {\cal L}_M &= -\frac{1}{2} \overline{N_R} M_M N_R^c + \text{h.c.} \hskip10mm
  {\rm Majorana,}
\\
{\cal L}_D &= - \overline{N_R} M_R S_L + \text{h.c.} \hskip15mm {\rm Dirac.}
\end{align}  
We use $S_L$ here as the symbol for the left-handed (LH) partner of $N_R$,
motivated by the inverse seesaw discussed above, i.e.\ the pair
$(S_L,N_R)$ forms a Dirac particle.  We want to distinguish the 
LH and RH components of this particle, because in the following
we will write down only NROs involving $N_R$.  This way, the Dirac
case of the NROs resembles more closely the Majorana case.~%
\footnote{A complete EFT of the SM extended with a Dirac HNL 
would contain operators involving $N_R$ and/or $S_L$. 
We would like to emphasise that in this work we assume the HNL itself
to be either Dirac or Majorana in nature, whereas usually in the literature 
on the $N_R$SMEFT, the Dirac case corresponds to the 
situation where $N_R$ is the RH counterpart of the SM LH neutrino $\nu_L$, 
such that the pair $(\nu_L,N_R)$ forms a Dirac particle.}
At $d=4$ one finds the standard Yukawa terms of the different seesaw 
realizations.

NROs in $N_R$SMEFT were studied previously in the literature at
$d = 5$~\cite{delAguila:2008ir,Aparici:2009fh,Graesser:2007yj},
$d=6$~\cite{delAguila:2008ir,Bell:2005kz,Graesser:2007pc},
$d=7$~\cite{Bhattacharya:2015vja,Liao:2016qyd}, and very recently at $d \leq 9$~\cite{Li:2021tsq}.  A complete list of
operators up to $d=7$ can be found, e.g.\ in Ref.~\cite{Liao:2016qyd}. 
As already mentioned in section~\ref{sect:intro}, at the $d=5$
level for Majorana $N_R$ there are two new types of
operators, in addition to the Weinberg operator~\cite{Weinberg:1979sa}.

At $d=6$ level, one finds four-fermion operators and operators
containing two fermions plus either derivatives, Higgses or field
strength tensors, see Refs.~\cite{delAguila:2008ir,Liao:2016qyd}. The
latter are not interesting to us here, since they are of minor
importance for the production cross sections of $N_R$ at the LHC.  We
therefore do not show these operators in table~\ref{tab:Ops}, where we
list baryon and lepton number conserving $d=6$ four-fermion operators,
containing either one or two $N_R$.  We give the particle contents of
the operators, without specifying the family indices, which are in
general four index tensors. For each operator structure, we provide
the number of independent real parameters for $n_N = 1$ and $n_N=3$
generations of $N_R$ (and $n_f = 3$ generations of the SM fermions).
We have checked these numbers using the Sym2Int
package~\cite{Fonseca:2017lem,Fonseca:2019yya}.  Finally, $\epsilon$
stands for the totally antisymmetric tensor connecting $SU(2)_L$
indices.
\begin{table}[t]  
 \begin{tabular}[t]{|c|c|c|c|}
    \hline
    Name & Structure & $n_N = 1$ & $n_N = 3$ \\ 
    \hline
    \hline
    ${\cal O}_{dN}$ &
    $\left(\overline{d_R}\gamma^{\mu}d_R\right)\left(\overline{N_R}\gamma_{\mu}N_R\right)$ & 
    9 &
    81 \\
    ${\cal O}_{uN}$ &
    $\left(\overline{u_R}\gamma^{\mu}u_R\right)\left(\overline{N_R}\gamma_{\mu}N_R\right)$ &
    9 & 
    81 \\
    ${\cal O}_{QN}$ &
    $\left(\overline{Q}\gamma^{\mu}Q\right)\left(\overline{N_R}\gamma_{\mu}N_R\right)$ &
    9 & 
    81 \\   
 ${\cal O}_{eN}$ &
 $\left(\overline{e_R}\gamma^{\mu}e_R\right)\left(\overline{N_R}\gamma_{\mu}N_R\right)$ & 
  9 &
  81 \\
  ${\cal O}_{NN}$ &
 $\left(\overline{N_R}\gamma_{\mu}N_R\right)\left(\overline{N_R}\gamma_{\mu}N_R\right)$ &
  1 &
  36 \\
  ${\cal O}_{LN}$ &
 $\left(\overline{L}\gamma^{\mu}L\right)\left(\overline{N_R}\gamma_{\mu}N_R\right)$ &
  9 &
 81 \\
 \hline
\end{tabular}
 \hfill
 \begin{tabular}[t]{|c|c|c|c|}
    \hline
    Name & Structure (+ h.c.) & $n_N = 1$ & $n_N = 3$ \\ 
\hline
\hline
${\cal O}_{duNe}$ &
$\left(\overline{d_R}\gamma^{\mu}u_R\right)\left(\overline{N_R}\gamma_{\mu}e_R\right)$ & 
   54  &
 162 \\ 
 ${\cal O}_{LNQd}$ &
 $\left(\overline{L}N_R\right)\epsilon\left(\overline{Q}d_R\right)$ &
  54 & 
  162 \\
  ${\cal O}_{LdQN}$ & 
  $\left(\overline{L}d_R\right) \epsilon \left(\overline{Q}N_R\right)$ &
   54 & 
   162 \\
  ${\cal O}_{LNLe}$ &
  $\left(\overline{L}N_R\right) \epsilon \left(\overline{L}e_R\right)$ &
  54 & 
  162 \\
 ${\cal O}_{QuNL}$ &
 $\left(\overline{Q}u_R\right)\left(\overline{N_R}L\right)$ &
  54 & 
  162 \\  
\hline
\end{tabular}
\caption{\it Lepton-number-conserving $d=6$ four-fermion operators in
  $N_R$SMEFT, containing either two~(left) or one~(right) HNL(s),
  $N_R$.  For each operator structure, we provide the number of
  independent real parameters for $n_N = 1$ and $n_N= 3$ generations
  of $N_R$.  Note that operators ${\cal O}_{LNQd}$ and ${\cal
    O}_{LdQN}$ have the same particle content, but represent
  independent Lorentz structures.
\label{tab:Ops}}
\end{table}

We note that for Majorana neutrinos there is also a lepton-number-violating operator involving four $N_R$, ${\cal O}_{N^cN}
=\left(\overline{N_R^c}N_R\right)\left(\overline{N_R^c}N_R\right)$.~%
\footnote{For $n_N = 1$ this operator vanishes identically, whereas
  for $n_N = 3$ it contains 12 real parameters.}  Further, there are
two types of operators which violate both, baryon number and lepton number.  We do not use
them, and thus, do not list them here.

Not all of the operators listed in table \ref{tab:Ops} are equally
important for LHC phenomenology. From the six types of operators
involving pairs of $N_R$, only the first three will significantly
affect the production cross sections for $N_R$ at the LHC.~%
\footnote{Operators ${\cal O}_{eN}$ and ${\cal O}_{LN}$ would be the
  equivalent for a lepton collider.}  It is important to note that
operators with pairs of $N_R$ can not lead to decays of the lightest
$N_R$. This has important consequences for phenomenology, since large
production cross sections do not imply fast decay for the lightest
HNL, if one considers only these operators.

Operators in the right part of table \ref{tab:Ops} contain only one
$N_R$. They can lead to sizeable production cross sections for $N_R$
too, but also mediate decays of (also the lightest) $N_R$ to SM
fermions at the same time. Thus, for these operators, cross sections
and decay lengths are related, which leads to important upper limits
on the cross sections, if we wish to study HNLs as long-lived
particles.

Thus, in this work we focus on the $N_R$ pair operators with quarks. 
The corresponding $d=6$ Lagrangian reads~%
\footnote{The operators are independent, 
  so $\Lambda$ can be different for each operator.
  We use a common $\Lambda$ here, to simplify the discussion.}
\begin{equation}
 \mathcal{L}_{6} \supset \frac{1}{\Lambda^2} \left(
 c_{dN} \mathcal{O}_{dN} + c_{uN} \mathcal{O}_{uN} + c_{QN} \mathcal{O}_{QN}
 \right)
 = \frac{1}{\Lambda^2} \left(
 c_{dN}^{ijkl} \mathcal{O}_{dN}^{ijkl} + 
 c_{uN}^{ijkl} \mathcal{O}_{uN}^{ijkl} + 
 c_{QN}^{ijkl} \mathcal{O}_{QN}^{ijkl}
 \right),
\end{equation}
where a sum over the flavour indices $i,j = 1,2,3$ of quarks and $k,l
= 1,\dots,n_N$ of HNLs is implied.
In the numerical part of the work, for simplicity we consider only one generation of the HNLs and only the first generation of the quarks, and thus, simplify the notation to $c_{dN}^{11}$, $c_{uN}^{11}$, $c_{QN}^{11}$ and $\mathcal{O}_{dN}^{11}$, $\mathcal{O}_{uN}^{11}$, $\mathcal{O}_{QN}^{11}$, with $``11"$ referring to the first quark generation.

\subsection{Ultraviolet completions for $d=6$ operators}

The operators of the effective theory in table~\ref{tab:Ops} can be
generated in the UV by either a $Z'$ or via leptoquarks (LQs).  It is
important to notice that no single BSM particle can generate all
operators.  Since we are mostly interested in studying the EFT limit,
here we will only discuss the $Z'$ option and three particular scalar
LQs.  For a complete list of all possible LQ states for generating
$d=6$ four-fermion operators see Ref.~\cite{Bischer:2019ttk}.

Let us start with the LQ case, see table~\ref{tab:Conv}. These three
LQs have been chosen, since they can generate the three different
$N_R$ pair operators with quarks in the left part of
table~\ref{tab:Ops}.  Table~\ref{tab:Conv} gives the naming convention
for the LQs and their quantum numbers under the SM gauge group in the
order $SU(3)_C\times SU(2)_L \times U(1)_Y$, and indicates which $N_R$
pair operator is generated by each of these states in the EFT.
\begin{table}[t]  
\begin{center}
  \begin{tabular}{|c|ccc|c|c|}
\hline
LQ state  & $SU(3)_C$ & $SU(2)_L$ & $U(1)_Y$ & Coupling & Operator  \\
\hline
\hline
$S_d$ & $\mathbf{3}$ & $\mathbf{1}$ & $-1/3$ & $g_{dN}$ & ${\cal O}_{dN}$ \\ 
\hline
$S_u$ & $\mathbf{3}$ & $\mathbf{1}$ & $\phantom{-}2/3$ & $g_{uN}$ & ${\cal O}_{uN}$ \\ 
\hline
$S_Q$ & $\mathbf{3}$ & $\mathbf{2}$ & $\phantom{-}1/6$ & $g_{QN}$ & ${\cal O}_{QN}$ \\ 
\hline
\end{tabular}
\end{center}
\caption{\it Scalar LQs and the $N_R$ pair operators they can generate
  in $N_R$SMEFT.
\label{tab:Conv}}
\end{table}

The different LQs can have the following Yukawa interactions:
\begin{eqnarray}\label{eq:LQQ}
{\cal L}_{S_Q} &=&  g_{QN}\overline{Q} N_R S_{Q} 
                   + g_{dL} \overline{d_R}L \epsilon S_{Q} + \text{h.c.}\,,
\\ \label{eq:LQu}
{\cal L}_{S_u} &=& g_{uN}\overline{u_R} N_R^c S_{u}  + \text{h.c.}\,,
\\ \label{eq:LQd}
{\cal L}_{S_d} &=& g_{dN}\overline{d_R} N_R^c S_{d} 
                  + g_{ue} \overline{u_R}e_R^c S_{d}
                  + g_{QL} \overline{Q} \epsilon L^c S_{d}  + \text{h.c.}
\end{eqnarray}
Here, we have not written down couplings of the LQs to quark pairs,
since these would lead to baryon-number-violating processes, if they
were present at the same time as the terms in
Eqs.~(\ref{eq:LQQ})--(\ref{eq:LQd}).  We note that all the couplings
are matrices in flavour space.

The first terms in each line of these equations will generate the
$N_R$ pair operators, while the simultaneous presence of two (or more)
terms in each line will generate single $N_R$ operators in the EFT.
For example, ${\cal L}_{S_Q}$ generates ${\cal O}_{QN}$ with the
matching condition $c_{QN}/\Lambda^2 = g_{QN}^2/(2 m_{S_Q}^2)$, where
again we have suppressed flavour indices; the factor of 2 arises from
a Fierz identity.  It also generates ${\cal O}_{LdQN}$ with
$c_{LdQN}/\Lambda^2 = g_{QN}g_{dL}/m_{S_Q}^2$.  Since these two
operators depend on different combinations of couplings in the LQ
model, it is clear that we can treat them as independent
operators. Similar comments apply to $S_d$.  Interestingly, $S_u$ can
generate {\em only a $N_R$ pair operator} $\mathcal{O}_{uN}$.

The $N_R$ pair operators of interest can also be generated by
integrating out a heavy $Z'$ boson.  Here, we do not specify a
concrete type of $Z'$, but rather treat it as a massive vector boson
with the following interaction Lagrangian:
\begin{equation}
 \mathcal{L}_{Z'} = g'_N \overline{N_R} \gamma^\mu N_R Z'_\mu + 
 g'_d \overline{d_R} \gamma^\mu d_R Z'_\mu + 
 g'_u \overline{u_R} \gamma^\mu u_R Z'_\mu +
 g'_Q \overline{Q} \gamma^\mu Q Z'_\mu \,.
\end{equation}
Switching on pairs of couplings $(g'_N,g'_d)$, $(g'_N,g'_u)$ or
$(g'_N,g'_Q)$, one generates $\mathcal{O}_{dN}$, $\mathcal{O}_{uN}$
or $\mathcal{O}_{QN}$, respectively, with the matching condition
$c_{qN}/\Lambda^2 = g'_q g'_N/m_{Z'}^2$, where $q = d,u,Q$.

Let us note that in this scenario, the pair production of $N_R$ would
proceed via $s$-channel diagrams with a resonance enhancement for
$m_{Z'} = 2 m_N$. Thus, for the EFT description to be valid, $m_{Z'}$
should be significantly larger than $2m_N$, contrary to the LQ case,
where the $N_R$ production occurs via $t$-channel diagrams.


\section{Simulations\label{sect:exp}}

\subsection{Model implementations\label{sect:models}}

In order to estimate the sensitivity reach of the various experiments
considered in this work, we make use of Monte-Carlo (MC) techniques to
perform numerical simulations. The first step, common to all
experiments, is the implementation of the different model setups in
\texttt{MadGraph5}
\cite{Alwall:2007st,Alwall:2011uj,Alwall:2014hca}. We built our models
in \texttt{FeynRules} \cite{Christensen:2008py,Alloul:2013bka}, which
generates UFO files \cite{Degrande:2011ua} as output, which we then
couple to \texttt{MadGraph5}. \texttt{MadGraph5} outputs LHE event
files, which will be further processed in different ways for ATLAS and
the far detectors.  These steps will thus be explained separately in
the following subsections.

We note that the \texttt{FeynRules} model data base already contains
two independent HNL implementations~\cite{Degrande:2016aje,Coloma:2020lgy}. However, neither of these contains NROs. To
include the operators of table~\ref{tab:Ops}, we have therefore
written our own model files, which are roughly based on the original
HNL model implementation of Ref.~\cite{Degrande:2016aje}.

The implementation including NROs is valid, strictly speaking, only
for Dirac neutrinos since, as mentioned in its
manual~\cite{Alwall:2014hca}, \texttt{MadGraph5} can not handle
Majorana fermions in operators with more than two fermions.  This
problem can be circumvented by implementing some particular
renormalizable SM extension, which includes the necessary scalar (or
vector) fields, that will generate the operators under consideration
in the infrared.  As discussed in section~\ref{sect:eft}, the simplest
model to generate the operators in table~\ref{tab:Ops} is adding three
different scalar LQs to the HNL extended SM.

\begin{figure}[t]
\includegraphics[width=0.48\textwidth]{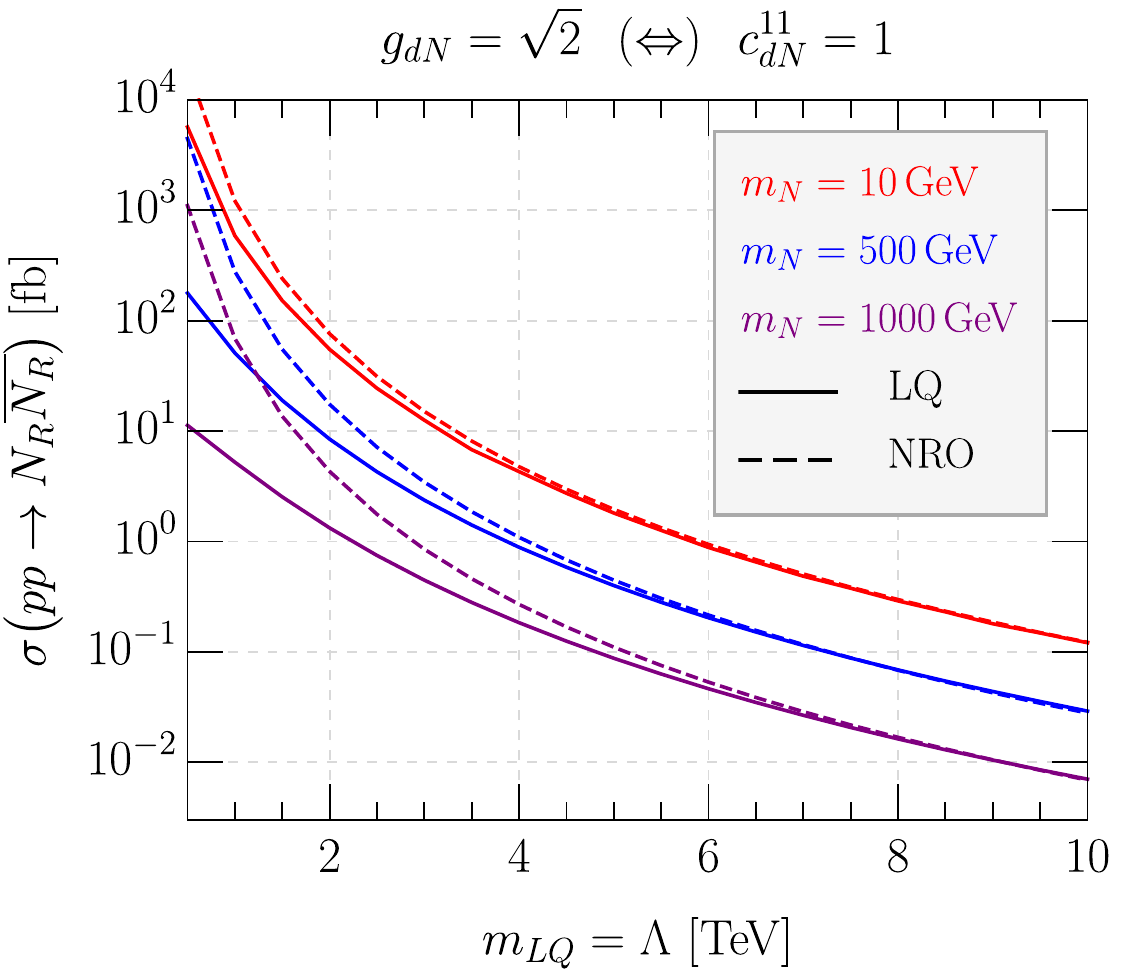}
\hspace{0.02\textwidth}
\includegraphics[width=0.48\textwidth]{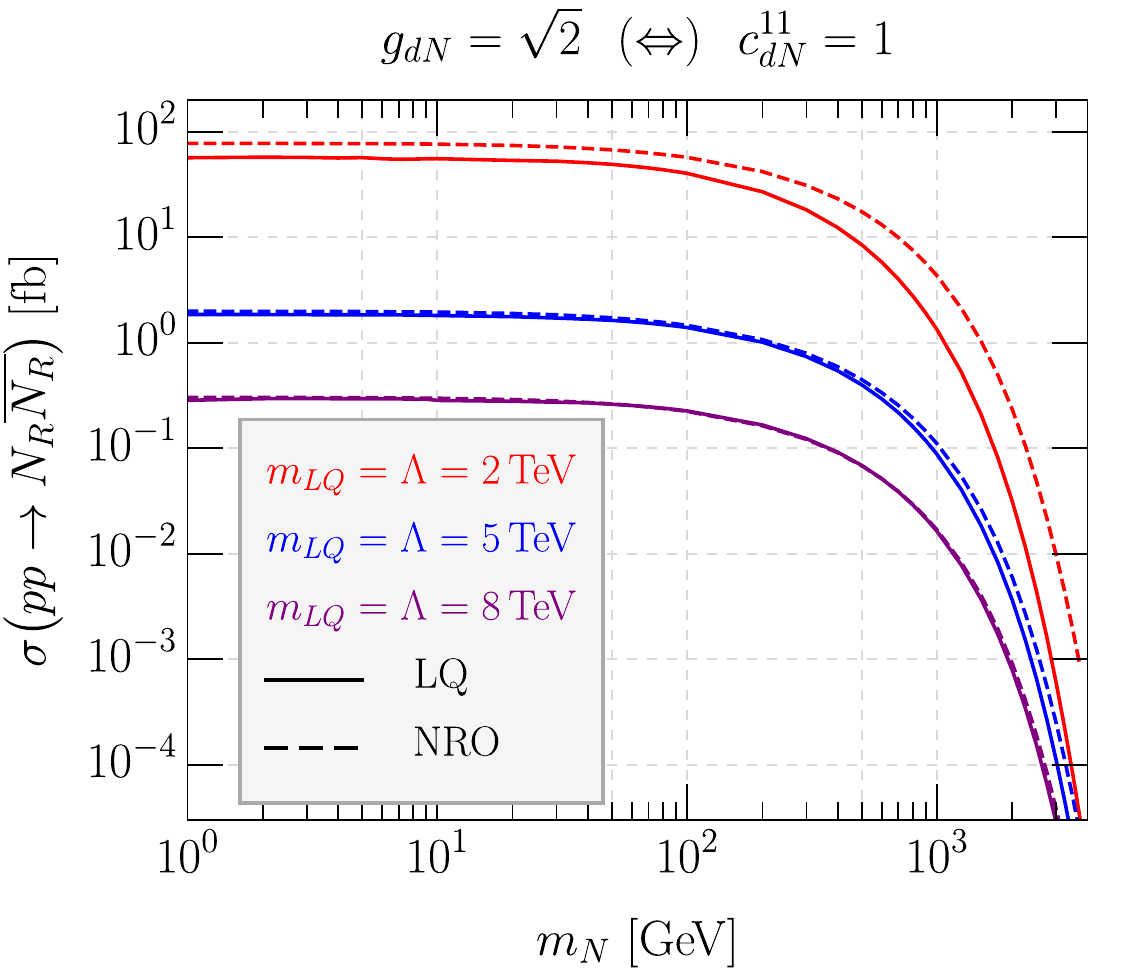}
\caption{Cross sections for pair production of HNLs $N_R$ via ${\cal
    O}_{dN}$ as function of $m_{LQ}=\Lambda$ (left) and $m_{N}$
  (right). The plots compare the calculation using 
  $c_{dN}^{11} = 1$ ($c_{dN}^{ij} = 0$ for $(i,j)\neq(1,1)$)
  with the cross sections of a LQ model
  (for $S_d$ with the first generation $g_{dN} = \sqrt{2}$). 
  The factor $\sqrt{2}$ is due
  to the matching of the two models via a Fierz transformation. The
  plots are valid for Dirac HNLs.
  \label{fig:XSec1}}
\end{figure}

We have written two of these model files, one for Dirac and one for
Majorana HNLs.~%
\footnote{For cross-checking the reliability of our
  \texttt{FeynRules} implementations, we have implemented the HNL+LQ
  models again in \texttt{SARAH} \cite{Staub:2012pb,Staub:2013tta}.
  \texttt{SARAH} can not handle NROs directly either, but we checked
  that cross section calculations, using the \texttt{SARAH} generated
  UFOs, agree with the \texttt{FeynRules} implementations of the
  HNL+LQ models.}
We have checked that for Dirac neutrinos the EFT model file containing NROs 
and the HNL+LQ implementation give the same cross sections in the appropriate
limits. This is shown in fig.~\ref{fig:XSec1}, where we compare
cross sections for HNL pair production calculated in EFT with those
calculated in a model with a scalar leptoquark, $S_d$. This example
calculation chooses only the first generation couplings non-zero.  We
have repeated this check for other generation indices. As can be seen,
the two calculations agree very well in the limit of large
$\Lambda=m_{LQ}$ and small HNL masses. The differences between the two
calculations can be sizeable, once $m_{N}$ approaches $m_{LQ}$, as
expected.~%
\footnote{We mention in passing, that we also implemented a
  $Z'$ model. However, $Z'$ exchange proceeds at the LHC in s-channel
  diagrams, while scalar LQs give t-channel diagrams. Thus, the EFT
  limit is only reached for values of $m_{Z'}$ larger than (roughly)
  10 TeV.} Since in the EFT total cross sections scale simply as
$\Lambda^{-4}$, one can calculate LHE event files for large values of
$\Lambda$ (or large values of $m_{LQ}$) and scale the total event
number to the desired value of $\Lambda$ to be simulated. Note that
this interchangeability is a necessary condition for us, in order to be
able to apply our Majorana HNL+LQ implementation to a simulation of Majorana HNLs in EFT, where, by setting the leptoquark masses to 10 TeV in our numerical simulation, the HNL+LQ model is effectively reduced to the EFT.

\begin{figure}[t]
\includegraphics[width=0.48\textwidth]{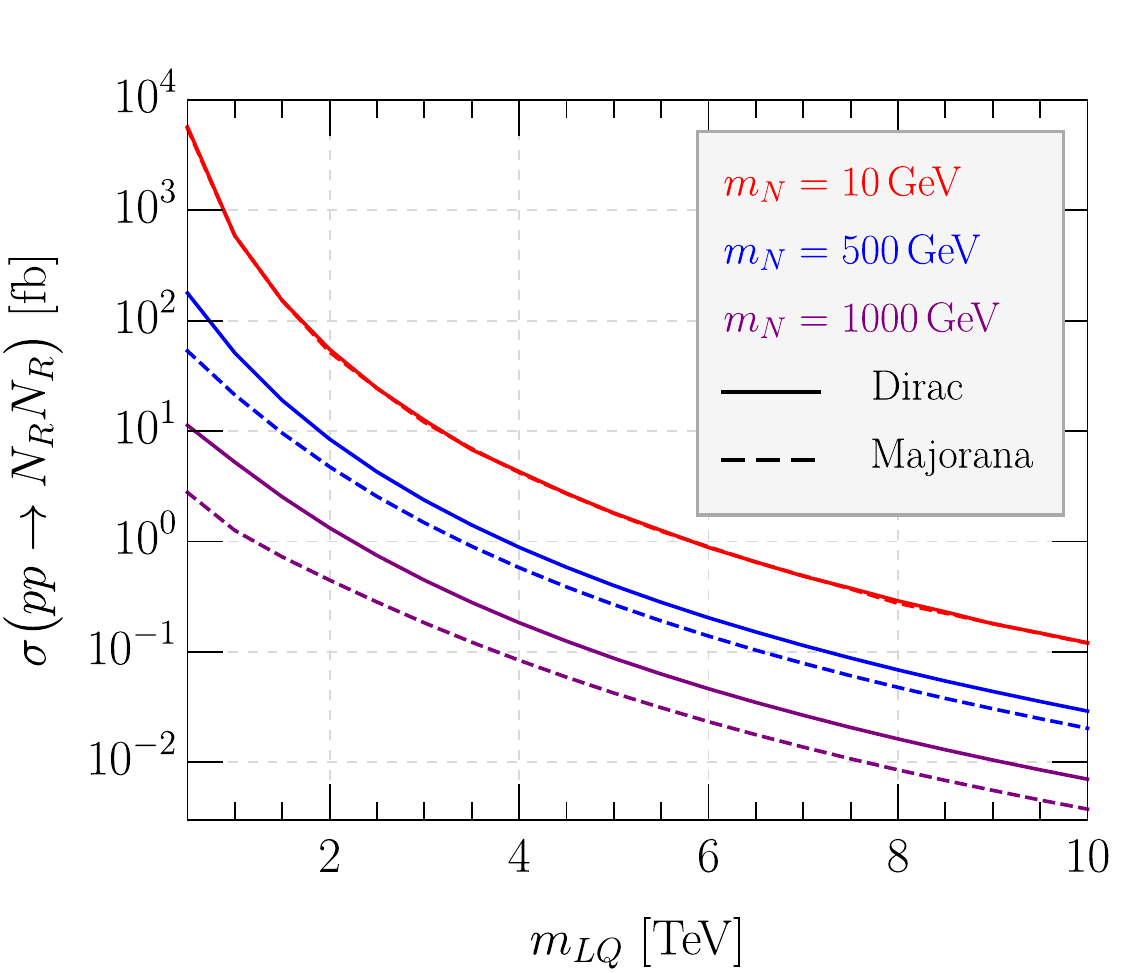}
\hspace{0.02\textwidth}
\includegraphics[width=0.48\textwidth]{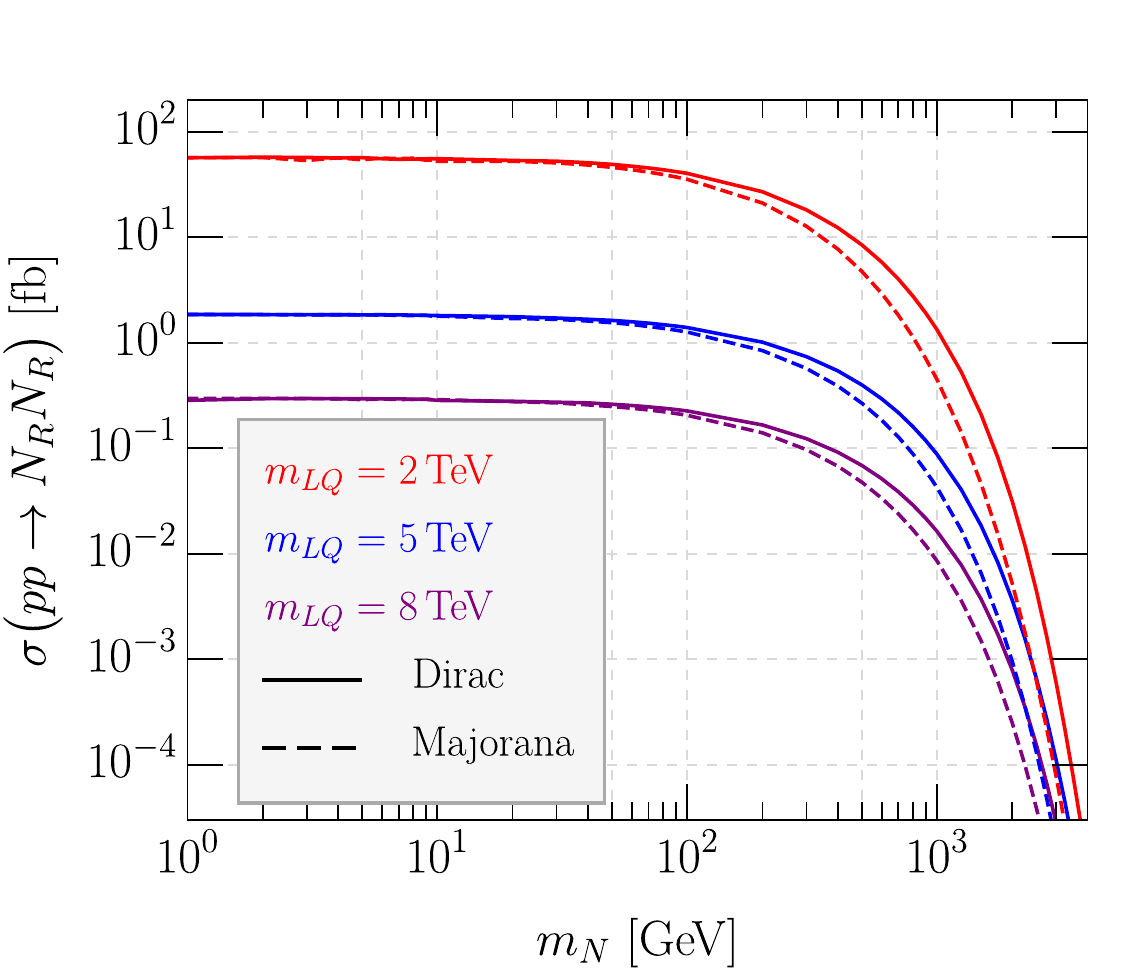}
\caption{Cross sections for pair production of HNLs $N_R$ as function
  of $m_{LQ}$ (left) and $m_{N}$ (right). The plots compare the
  calculation for HNLs which are Dirac particles to the one for
  Majorana HNLs. Only $g_{dN}=\sqrt{2}$ for first generation was
  chosen to be non-zero in this example.  Majorana $N_R$ cross
  sections are suppressed for $m_N \gtrsim 100$~GeV, see discussion.
    \label{fig:XSec2}}
\end{figure}

In fig.~\ref{fig:XSec2}, we compare cross sections for pair production
of HNLs for Dirac and Majorana cases. 
The calculation is done with the
HNL+LQ implementation and again with non-zero coupling to the first
generation quarks only.  For small HNL masses, the two calculations
coincide. However, for $m_{N}$ larger than about 100 GeV
numerically important differences show up. In order to understand this
behaviour, we have calculated analytically the cross section for
$\overline{d_R} d_R \to N_R N_R$ .  The results for Dirac and Majorana
HNLs are given by
\begin{eqnarray}\label{eq:sigD}
  \sigma_D\big(\overline{d_R} d_R \to \overline{N_R} N_R\big)
  &=& \frac{c_{dN}^2}{192 \pi \Lambda^4}
  s \sqrt{1-\frac{4 m_{N}^2}{s}}\left[1+ \frac{1}{3}\left(1-\frac{4 m_{N}^2}{s}\right)
    \right],
\\ \label{eq:sigM}  
  \sigma_M\big(\overline{d_R} d_R \to N_R N_R\big) &=& \frac{c_{dN}^2}{144 \pi \Lambda^4}
  s \left(1-\frac{4 m_{N}^2}{s}\right)^{3/2},
\end{eqnarray}  
respectively. From Eq.~(\ref{eq:sigM}) one can see directly the
suppression of the cross section at the 
larger values of $m_{N}$ in
the Majorana case. We traced the origin of this difference to the fact
that in the Majorana case the two final state $N_R$ are
interchangeable, leading to twice the number of diagrams than for the
Dirac case.
At larger $m_N$, the destructive interference effect becomes more important.
Equations~\eqref{eq:sigD} and \eqref{eq:sigM} apply to
partons. We have checked, however, that they agree with the numerical
result by \texttt{MadGraph}, once the PDFs are switched off in
\texttt{MadGraph}. The suppression of the cross sections at 
larger $m_N$ for Majorana neutrinos will lead to a reduced
sensitivity, when comparing the results for Majorana and Dirac cases,
see section~\ref{sect:results}.

\begin{figure}[t]
\includegraphics[width=0.48\textwidth]{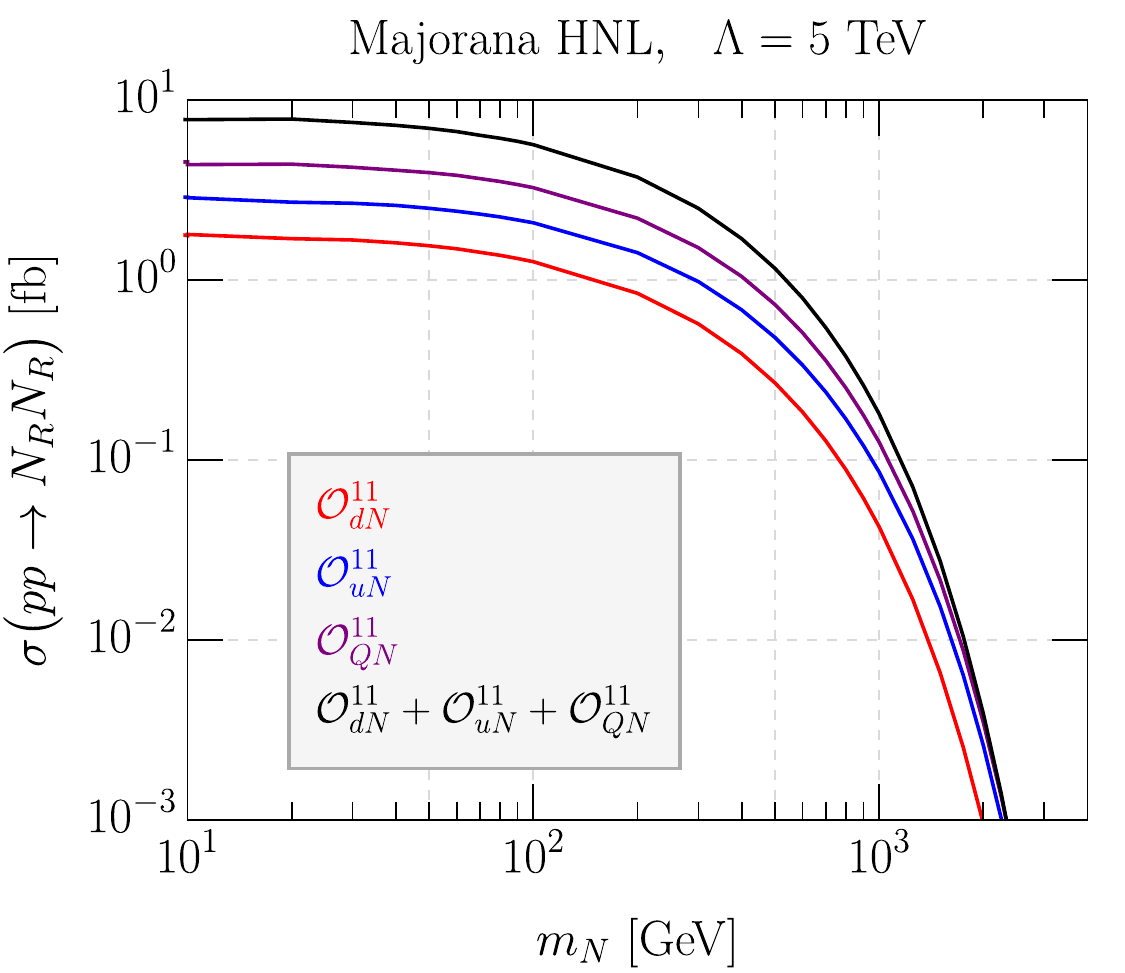}
\caption{Cross sections for pair production of HNLs $N_R$ as function
  of $m_{N}$. The plot compares different operators. For the single-operator calculation, only one coefficient has been chosen
  non-zero. For the three-operator calculation all $c_{qN}^{11}$ with
  $q=d,u,Q$, are set to one simultaneously.  This example is for a
  scale of $\Lambda=5$ TeV.
    \label{fig:XSec3}}
\end{figure}

We have also checked cross section calculations for different choices
of operators. Figure~\ref{fig:XSec3} shows some examples. Since the
LHC collides protons, one finds in general the cross section triggered
by $\mathcal{O}_{QN}$ to be larger than that induced by
$\mathcal{O}_{uN}$, that in turn is larger than the one for
$\mathcal{O}_{dN}$, as expected. In addition, we have calculated cross
sections with all three $N_R$ pair operators switched on at the same
time. Figure~\ref{fig:XSec3} shows an example with all couplings to
$N_R$ set to one. This choice produces the largest cross section
possible and thus represents the most optimistic assumption possible,
as far as sensitivity is concerned. Typically, this choice leads to
cross sections larger than the choice of only ${\cal O}_{dN}$, with
$c_{dN}^{11}=1$, by a factor of four or so.  These two cases therefore
span the ``smallest'' and the largest cross sections possible in our
setup.~%
\footnote{\label{foot:XS}Of course,
  this statement is for fixed $\Lambda$ and coefficients. One can
  always obtain smaller cross sections by choosing arbitrarily small
  coefficients. On the other hand, one can not choose arbitrarily
  large coefficients, since UV completions would then necessarily be
  non-perturbative.}


\subsection{ATLAS\label{subsect:atlas}}

For the ATLAS experiment, our analysis will focus on the HNLs decaying
to $ejj$.  To enforce the HNL decays into $ejj$ and to ensure
numerical stability for very small decay width computation, we process
the LHE file generated by \texttt{MadGraph5} for the HNL decay in
\texttt{MadSpin}~\cite{Artoisenet:2012st}.  The decayed LHE event
files are then read by \texttt{Pythia8}~\cite{Sjostrand:2014zea},
which performs showering and hadronization.  The displaced decay
positions of $N_R$ into $ejj$ can be extracted from the properly
simulated input LHE files.

We generate 100 thousand events at each grid point in a
two-dimensional scan in the $m_N$-$|V_{eN}|^2$ plane, for which we
consider 56 values of $m_N$ from 5 GeV up to 6 TeV, and 175 values of
$|V_{eN}|^2$ from $10^{-27}$ to $8\times 10^{-3}$, both in logarithmic
steps.

Event selection is then performed at the \texttt{Pythia}-event level,
with a customized detector simulation to quantify the detector
response to the physical objects of interest.  Our displaced search
design is inspired by the ATLAS inner tracker displaced-vertex (DV)
search in Ref.~\cite{Aad:2015rba}, for the ``DV + electron" channel.
In this experimental search channel, photon triggers are used for
final states involving displaced electrons.  As these triggers have no
cuts nor vetoes on tracks in the inner detector, displaced electrons
can be found within these data sets.  An electron candidate with
$p_{T}>120$ GeV is required to have triggered the event in this
channel~\cite{Aad:2015rba}.~%
\footnote{As the ATLAS ``DV + electron" analysis --- as well as our
  displaced signal --- does not possess a prompt isolated lepton,
  experimental triggers face the challenge to perform dedicated
  reconstruction techniques in order to enhance sensitivity to such
  LLP decay topologies.}  A similar search was proposed for displaced
HNLs from $Z'$ decays in Ref.~\cite{Chiang:2019ajm}.

Our event selection starts by identifying electrons with $|\eta|<2.47$
and $p_{T}>120$ GeV.  A flat efficiency of 70\% is assigned for
electron identification.  We then select events containing {\it{at
    least one}} reconstructed DV.  For this reconstruction, we
identify the $N_R$ displaced position at the truth-level and apply
parameterized detection efficiencies as a function of DV invariant
mass and number of tracks, as provided in the auxiliary material of
the ATLAS 13~TeV search in Ref.~\cite{Aaboud:2017iio}.~%
\footnote{For the purpose of our estimations in
  section~\ref{sect:results}, we assume these efficiencies remain the
  same at $14$ TeV.}  A reconstructed DV must lie within the
acceptance of the ATLAS inner tracker.  This means selecting displaced
vertices with transverse position, $r_{\text{DV}}$, to be within
$4~\text{mm} < r_\text{DV} < 300$ mm and longitudinal position,
$z_{\text{DV}}$, to be $|z_{\text{DV}}|< 300$ mm.

We require a displaced vertex to be made from at least four charged
particle tracks.  Each track coming from the DV is required to have
$p_{T}> 1$ GeV and to be displaced.  The latter is ensured by
requiring $|d_0|> 2$ mm, with $d_0= r\cdot\Delta\phi$ defined as
the approximate transverse impact parameter, with $r$ corresponding to
the transverse distance of the track from the interaction point (IP) and $\Delta\phi$ being the azimuthal angle between the track and direction of the long-lived $N_R$.

One of these displaced tracks must correspond to the electron that
passed the trigger. As our electrons are reconstructed without an
isolation requirement, we can match the truth-level index of the
electron track with one of the displaced tracks.

The invariant mass of the DV, $m_{\text{DV}}$, can then be
reconstructed from the above selected tracks, and it assumes all the
reconstructed tracks have the mass of a pion.  The additional
requirement of $m_{\text{DV}}\geq 5$ GeV is imposed in order to
eliminate SM background from $B$-mesons.  This last cut, together
with the requirement of a large track multiplicity from the DV,
defines a region where the signal is expected to be found free of
backgrounds~\cite{Aad:2015rba,Aaboud:2017iio}.~%
\footnote{Instrumental backgrounds to this search (which are very hard
  to simulate outside the experimental collaborations) can arise
  from a LLP track that interacts with the detector material and forms
  a fake DV. A material veto is included in the DV efficiencies we use
  \cite{Aaboud:2017iio}. Random track crossings can also form fake
  DVs~\cite{Aad:2015rba, Aaboud:2017iio}.}

\begin{figure}[t]
	\centering
	\includegraphics[width=0.48\textwidth]{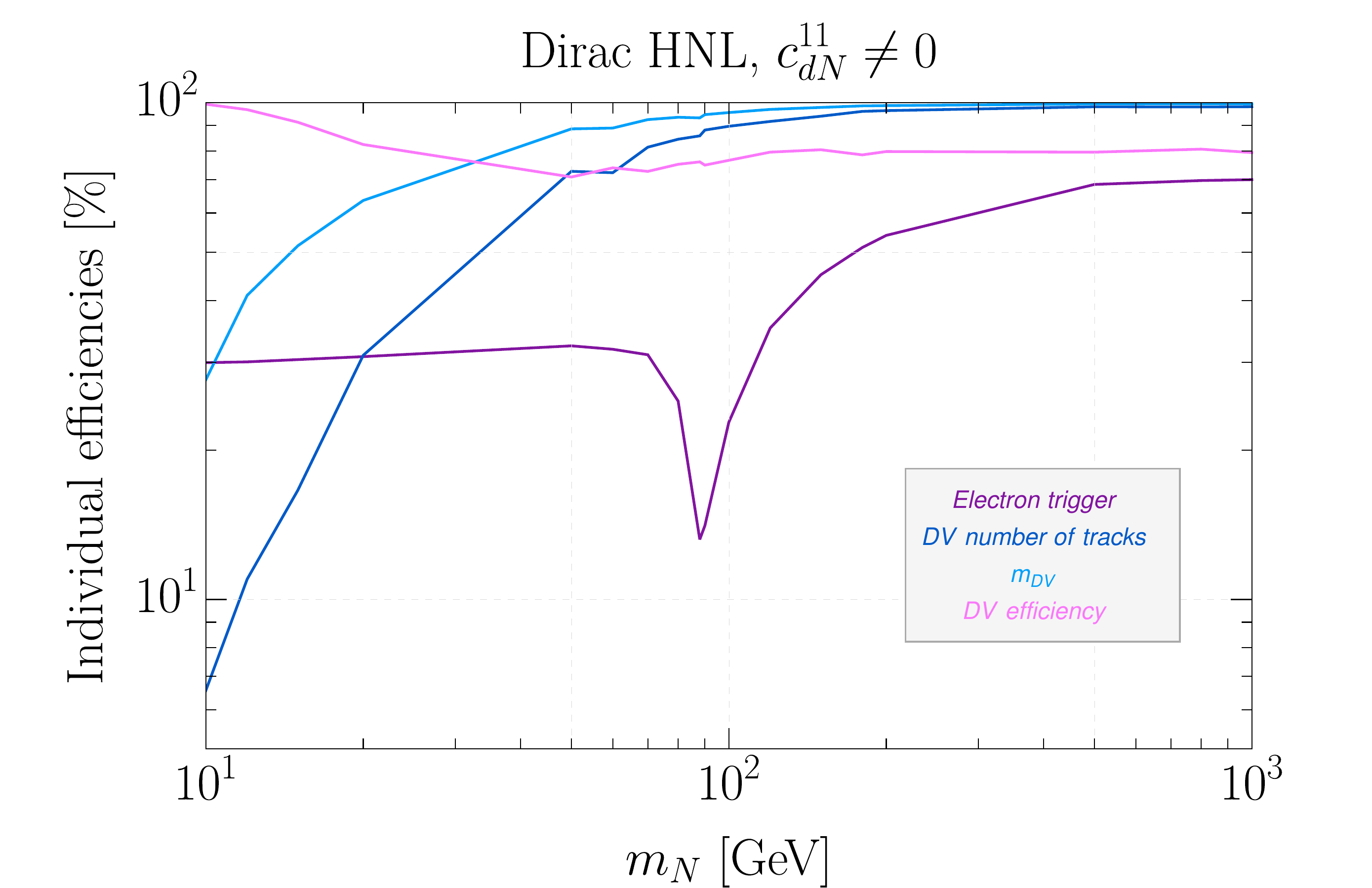}
	\hspace{0.02\textwidth}
	\includegraphics[width=0.48\textwidth]{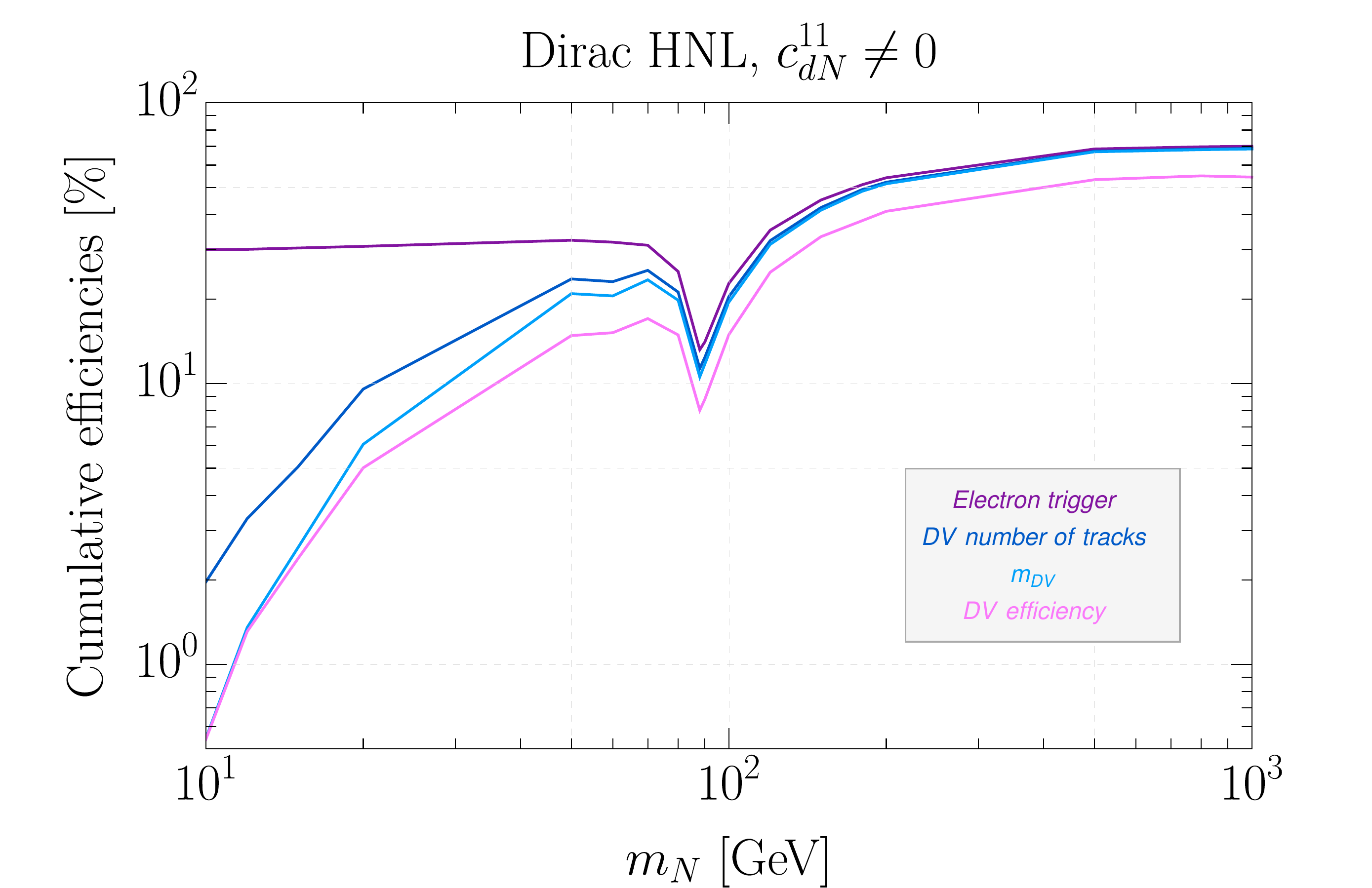}
	\caption{ATLAS selection efficiencies as a function of the
          mass of a Dirac HNL, excluding the requirement of the DV
          fiducial volume. Here as a sample scenario, only one single
          operator $\mathcal{O}_{dN}^{11}$ is switched on.  The left
          (right) panel is for individual (cumulative) cut
          efficiencies. In the right plot, each line is for the
          cumulative efficiency assuming the given cut as well as the
          previous cut(s) if any.
		\label{fig:efficiency}}
\end{figure}

In fig.~\ref{fig:efficiency}, we show in two plots the individual and
cumulative selection efficiencies of each cut for $m_N$ between 10 and
1000 GeV, for a Dirac HNL in a scenario with only the
$\mathcal{O}_{dN}^{11}$ operator switched on.  We have excluded the
selection effect of the fiducial volume for these efficiencies,
rendering the latter independent of the active-heavy mixing which has
no impact on the other cuts than the DV fiducial volume one.  We find a dip in the electron trigger efficiency at $m_N \sim m_W$, where the transition between two-body and three-body decays into an on-shell and off-shell $W$-boson, respectively, takes place. The individual efficiencies of requirements on the number of tracks
associated to the DV and on the DV invariant mass improve with
increasing HNL mass, as expected.  The parameterized DV efficiency cut
in general passes most of the events for the shown mass range.  As a
result of these individual selections, the cumulative efficiencies
tend to be enhanced for larger $m_N$, except for the drop around the $W$-boson mass.
Moreover, the $m_\text{DV}\geq 5$ GeV event requirement determines the lower mass reach.

The total number of signal events at the ATLAS is calculated with the following expression:
\begin{eqnarray}
N_{\text{sig.}}^{\text{ATLAS}}=\sigma \cdot  \mathcal{L} \cdot  \text{Br}\left( N_R\to ejj \right) \cdot 2 \cdot\epsilon\,,
\label{eq:Nsig_atlas}
\end{eqnarray}
where $\epsilon$ labels the efficiency of event selections including the ATLAS detector geometries, which depends on both the HNL mass and proper lifetime.
Also, the cross section $\sigma$ is a function of $m_N$ and the Wilson coefficients of the operators switched on, and Br$(N_R\to ejj)$ depends on $m_N$ only.
  This efficiency includes detector acceptance and corresponds to the efficiency for
reconstructing one displaced vertex in an event after all the
selections (whereas the efficiencies shown in fig.~\ref{fig:efficiency}
do not include the detector acceptance, i.e., equivalently, the
fiducial volume requirement). Note that the factor of $2$ arises from
the fact that we simply require either of the two $N_R$'s in each
event to decay to $ejj$ with a displacement while disregarding how the
other HNL decays.

Under the zero background assumption, 95\% C.L. exclusion limits can
be derived by requiring three signal events.
These limits are provided and discussed in section~\ref{sect:results}.
We note that constraints on mixing in the muon sector could also be obtained if the HNL decays to $\mu jj$.
In this case, as the lepton coming from the DV is a muon, muon triggers can be used to efficiently record events, and we would expect similar exclusion reach for the muon case~%
\footnote{This is the case, provided there are no explicit requirements that would veto displaced activity to the muon track identified in the inner tracker~\cite{Aad:2015rba,Aad:2020srt}.} (see for instance Ref.~\cite{Chiang:2019ajm}).
For mixing in the tau sector, we expect the sensitivity reach to be less powerful as a result of worse efficiencies and experimental difficulties in reconstructing (displaced) tau leptons.


\subsection{Far detectors\label{subsect:fd}}

Besides the default local experiments such as ATLAS and CMS, in recent
years a series of far-detector (FD) experiments have been proposed with a
distance of about 5--500 meters from the various IPs at the LHC: ANUBIS~%
\footnote{A brief discussion concerning a potential timing cut at
  ANUBIS is given in appendix~\ref{sect:appendixA}.}, AL3X, CODEX-b,
FASER and FASER2, MoEDAL-MAPP1 and MAPP2, and MATHUSLA.
In particular, MoEDAL-MAPP1 and FASER have been approved.  These
experiments are supposed to be shielded from the associated
IP by rock, lead, or other material, removing the SM
background events effectively.  The other background sources include
cosmic rays (especially for the on-the-ground experiment MATHUSLA),
which can be eliminated by directional cuts.  Consequently, in almost
all the cases, essentially zero background can be assumed, and
correspondingly the $95\%$ C.L. exclusion limits can be derived by
requiring three signal events.  It is also worth mentioning that these
experiments are proposed to be operated in different phases of the
(HL-)LHC and IPs, and so the projected integrated
luminosities vary.  For instance, MAPP1 is to receive 30
fb$^{-1}$ of integrated luminosity while FASER2 about 3 ab$^{-1}$.

The treatment of the sensitivity estimate at the far-detector
experiments is somewhat different from that for the ATLAS experiment.
In contrast to the ATLAS simulation, the decay of the HNLs is not
simulated in this case.  With \texttt{MadGraph5} we generate 10
million events for 81 mass values from 0.1 GeV to 6 TeV in logarithmic
steps.  The LHE event files are fed into \texttt{Pythia8} which simply
includes the effect of initial state radiation and final state
radiation.  For 175 values of the mixing angle squared, $|V_{eN}|^2$,
from $10^{-27}$ to $8\times 10^{-3}$ we then compute the decay
probability of each simulated HNL in the fiducial volume (f.v.) of
each far-detector experiment with the formulas below:
\begin{eqnarray}
	N_{\text{sig.}}^{\text{FD}} &=& 2 \cdot \sigma \cdot \mathcal{L} \cdot  \langle P[ N_R \text{ decay in f.v.} ] \rangle  \cdot \text{Br}(N_R \to \text{vis.})\,,\\
	 \langle P[ N_R \text{ decay in f.v.} ] \rangle  &=& \frac{1}{2k} \sum_{i=1}^{2k} P[ N_R^i \text{ decay in f.v.} ]\,,
\end{eqnarray}
where $ \langle P[ N_R \text{ decay in f.v.} ] \rangle$ denotes the
average decay probability of all the simulated HNLs (2 in each event)
inside the fiducial volume of a far detector, $k=10^7$ is the total number
of the simulated events, and $\text{Br}(N_R \to \text{vis.})$ labels the HNL decay branching ratio into visible
states, meaning all or some of the decay products are electrically
charged (here we only consider the tri-neutrino final state to be
invisible).
$P[ N_R^i \text{ decay in f.v.} ]$ is the individual
decay probability of the $i$th simulated $N_R$, and it is computed
with the help of the exponential decay distribution, kinematics (boost
factor and traveling direction of the HNL), and the proper decay
length $c\tau$ of the HNL.  These expressions have to be calculated
for each experiment with the corresponding cuts on the polar and
azimuthal angles, as each of them is to be placed at a different
location and thus has a different geometrical acceptance.  Essentially
$P[ N_R^i \text{ decay in f.v.} ]$ is calculated with the following
formula:
\begin{eqnarray}
  P[ N_R^i \text{ decay in f.v.} ] = e^{-L_1/\lambda^i}\cdot
                \left(1 - e^{-L_2/\lambda^i}\right),\label{eq:decayprobability}
\end{eqnarray}
where $L_1$ is the distance from the IP to the point where the
long-lived HNL reaches the detector, $L_2$ the distance the HNL
travels through the internal space of the detector given its direction if it does not decay before leaving the detector chamber, and $\lambda^i$ is
the decay length of the $i$th simulated HNL in the lab frame.
$P[ N_R^i \text{ decay in f.v.} ]=0$, of course, if $N_R^i$ would miss the detector volume.
We emphasize that here, Eq.~\eqref{eq:decayprobability} is given only in order to showcase the essence of the computation of decay probabilities of the long-lived HNLs in a far detector.
Of course, the far detectors considered in this work are proposed with various and even quite complicated geometries and relative positions from their respective IPs, for which one cannot directly apply Eq.~\eqref{eq:decayprobability}.
When we conduct the numerical simulation, we use more sophisticated and accurate formulas. We refer the reader to Ref.~\cite{deVries:2020qns} and the references therein, for a summary of these far-detector experiments and the explicit expressions for the computation of $P[ N_R^i \text{ decay in f.v.} ]$.



\section{Results\label{sect:results}}

\begin{figure}[t]
\centering
\includegraphics[width=0.49\textwidth]{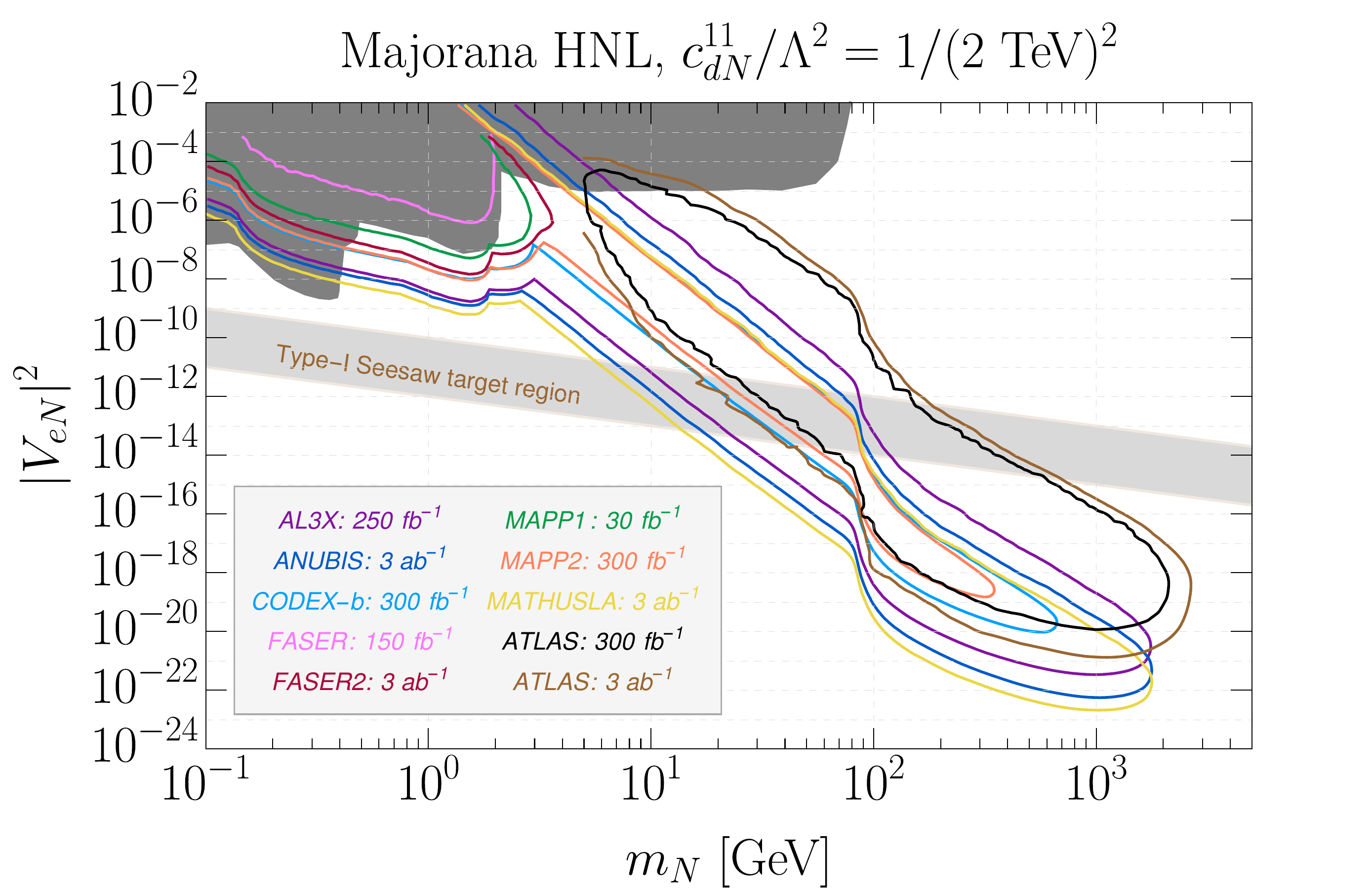}
\includegraphics[width=0.49\textwidth]{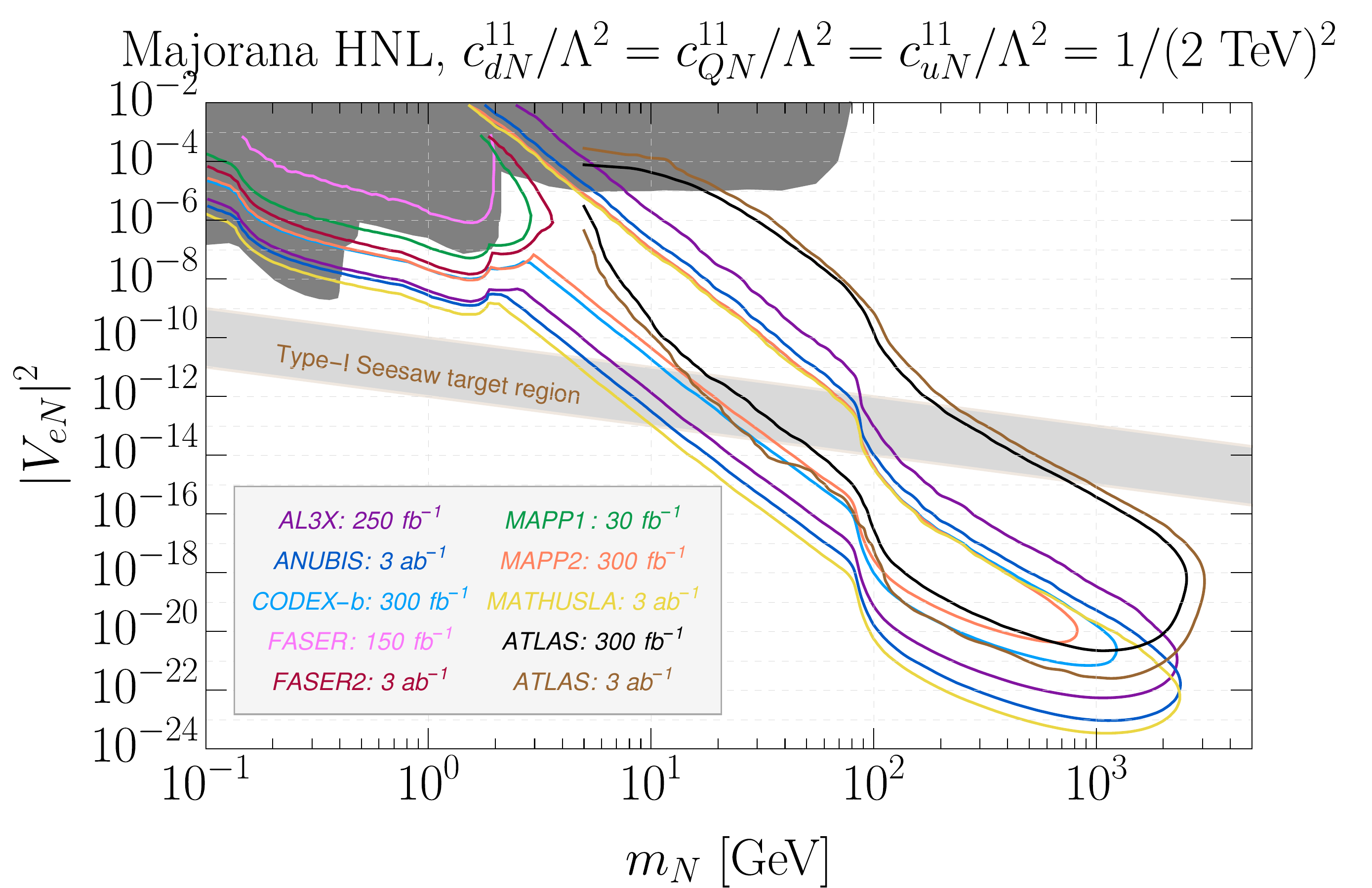}\\[0.4cm]
\includegraphics[width=0.49\textwidth]{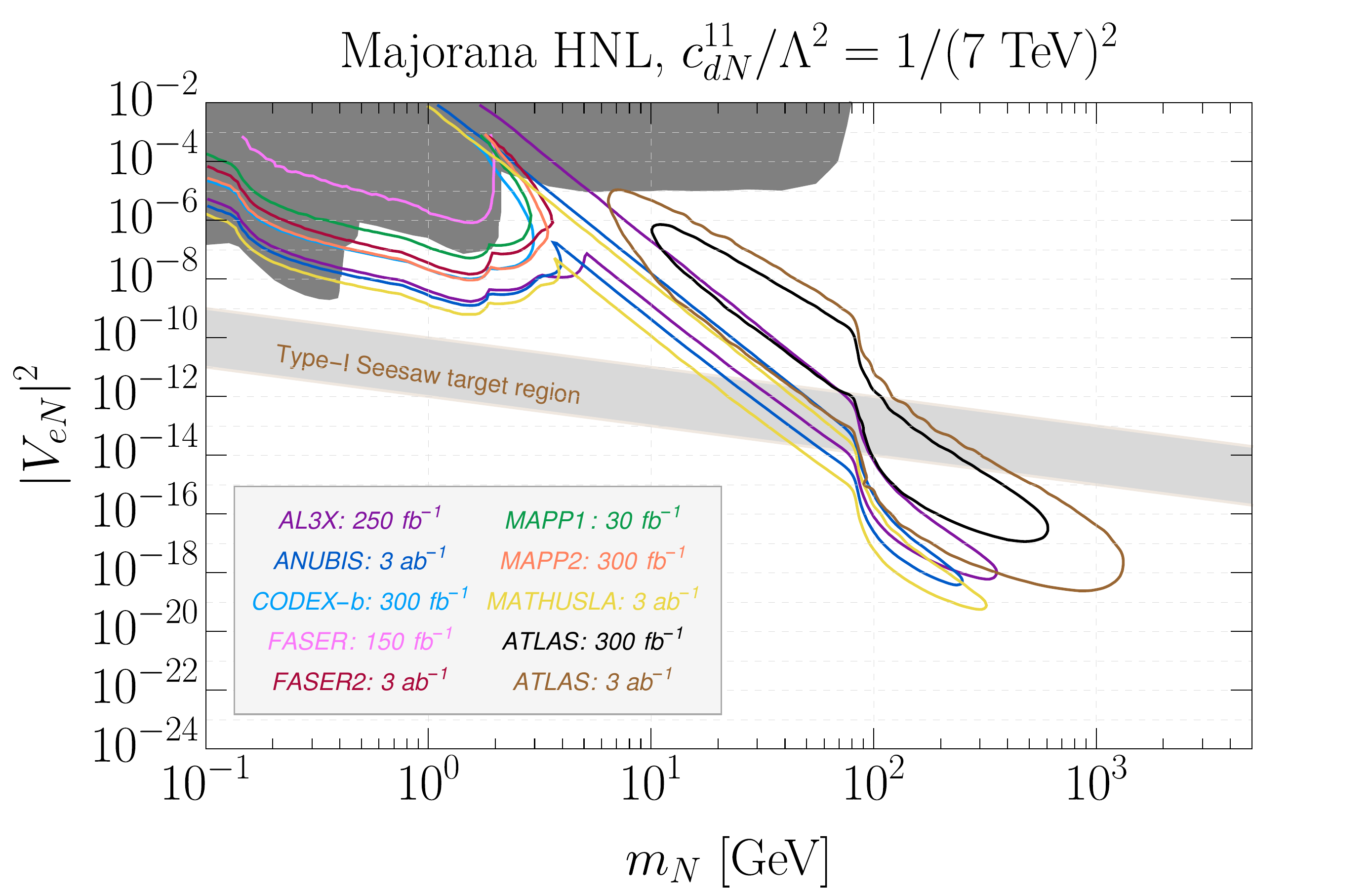}
\includegraphics[width=0.49\textwidth]{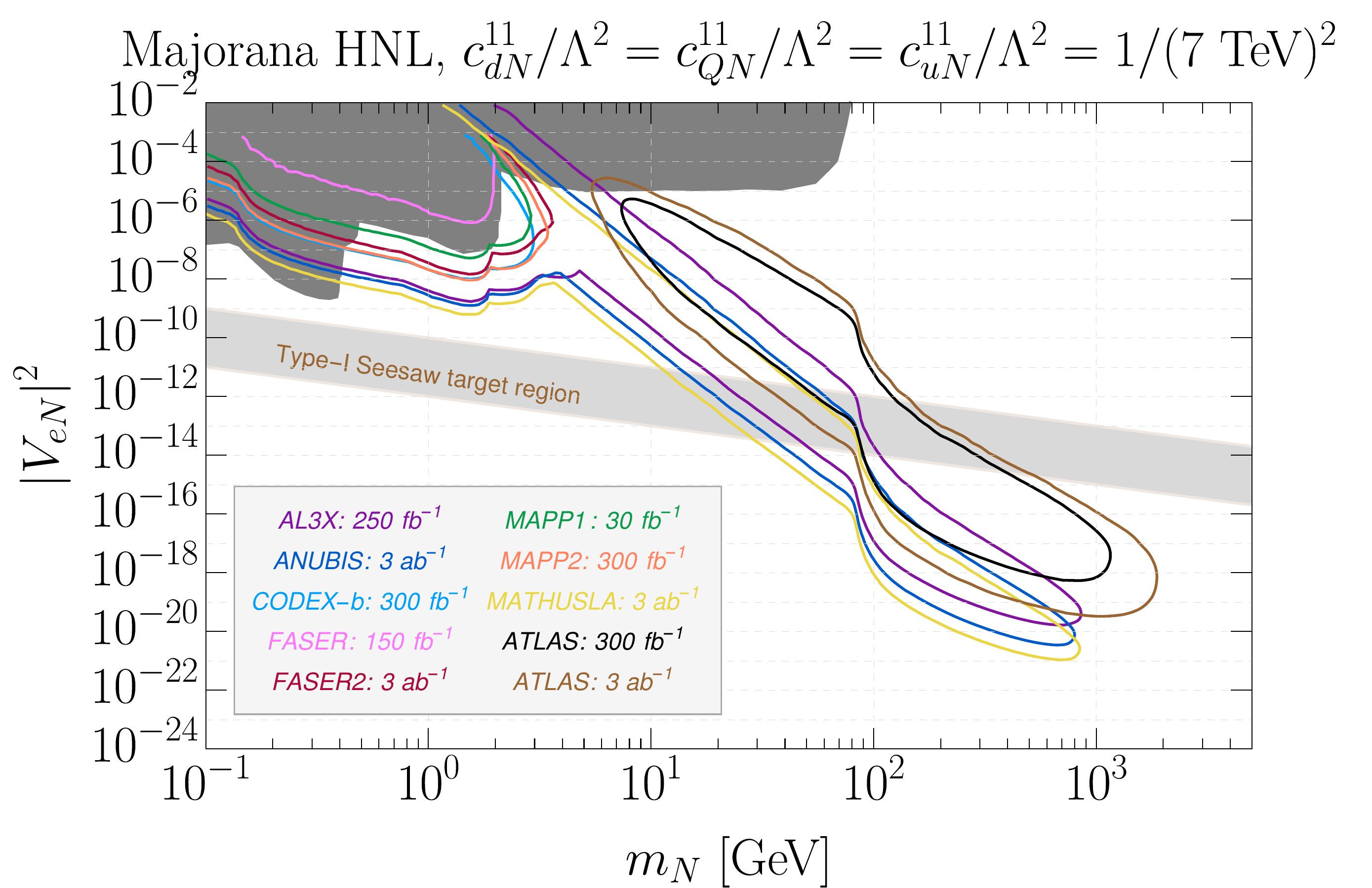}\\[0.4cm]
\includegraphics[width=0.49\textwidth]{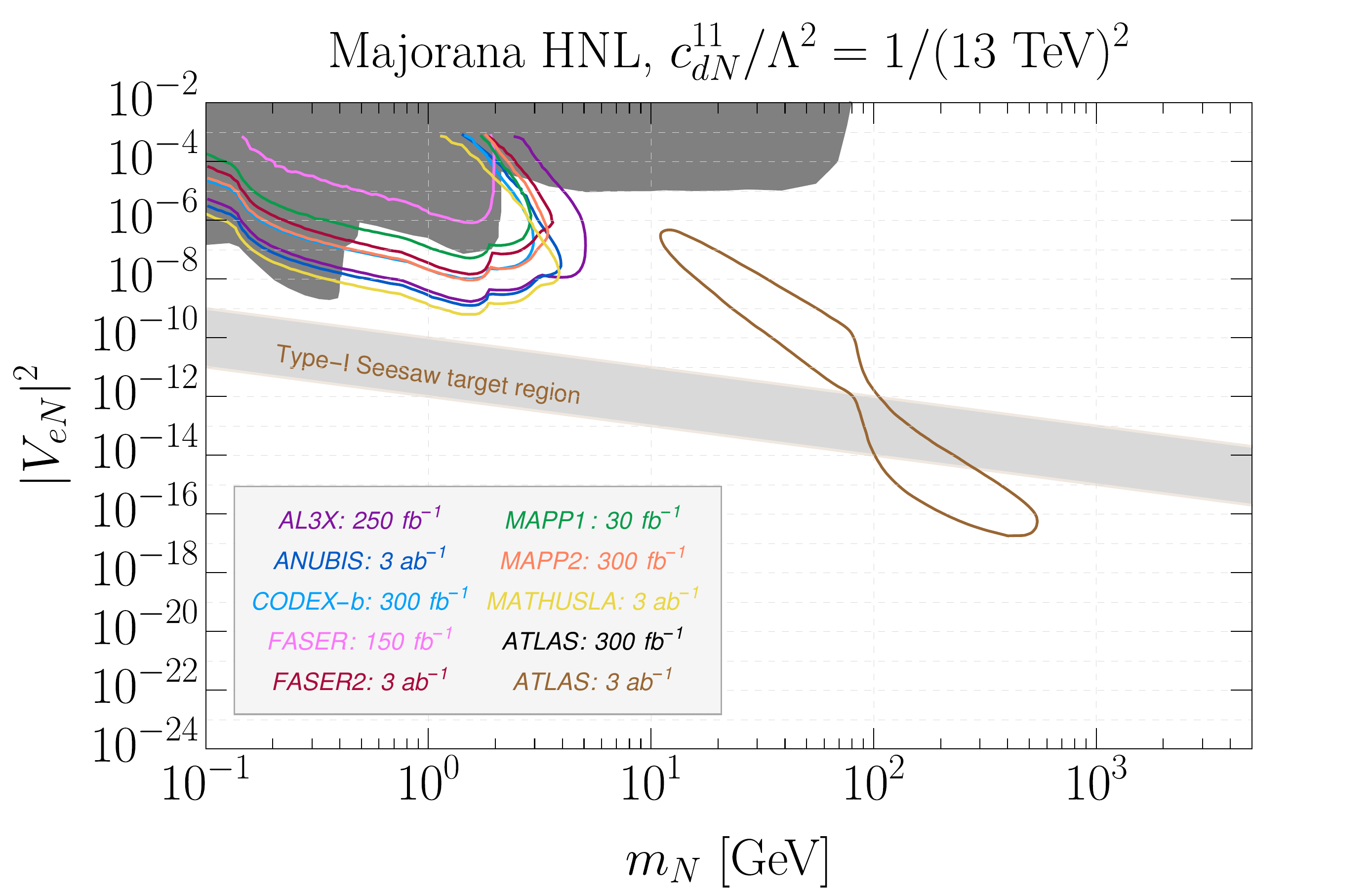}
\includegraphics[width=0.49\textwidth]{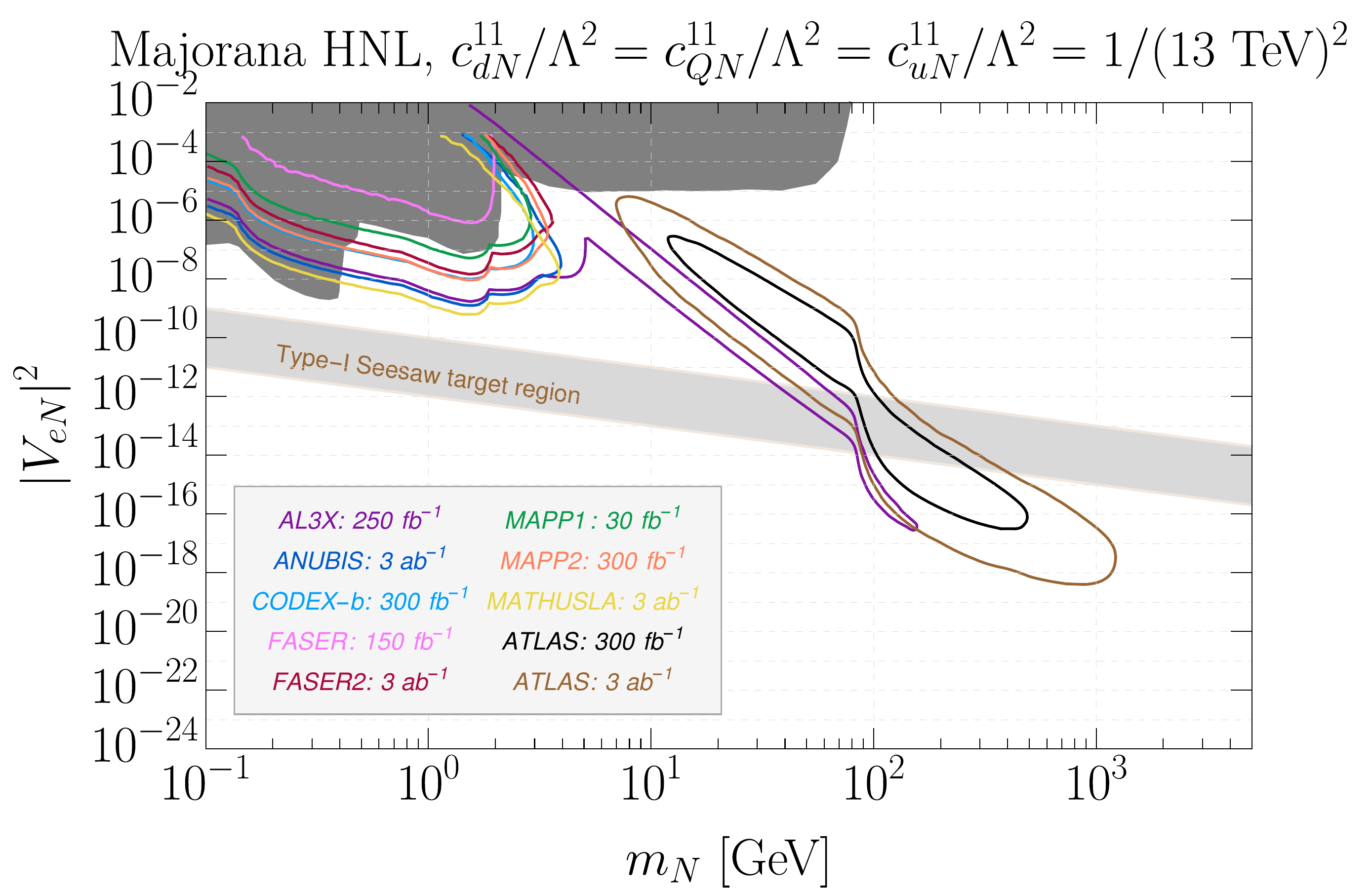}
\caption{Experimental sensitivity reach on $|V_{eN}|^2$ as a function
  of $m_N$ for Majorana HNLs.  The left panel contains results for one
  single operator ${\cal O}_{dN}$ only.  The plots on the right are
  for the case where all the three types of effective operators are
  switched on simultaneously.  The values of the respective operator
  coefficients are fixed to $c_{{\cal O}} / \Lambda^2 = 1/(2\text{
    TeV})^2$, $1/(7\text{ TeV})^2$ and $1/(13\text{ TeV})^2$.  The
  dark gray region has been excluded by various experiments including
  searches from CHARM \cite{Bergsma:1985is}, PS191
  \cite{Bernardi:1987ek}, JINR \cite{Baranov:1992vq}, and DELPHI
  \cite{Abreu:1996pa}, and the light gray band corresponds to the
  parameter space with the type-I seesaw relation assumed for active
  neutrino masses between $10^{-3}$ and
  $10^{-1}$~eV.  \label{Fig:VmN}}
\end{figure}

Based on the computation procedure as described in section
\ref{sect:exp}, we have estimated the 95\% exclusion limits under zero
background assumption (we will refer to them as ``experimental
sensitivities'') on the effective operators containing two HNLs and two
quarks ${\cal O}_{dN}$, ${\cal O}_{uN}$ and ${\cal O}_{QN}$, for both
ATLAS and future far detectors.  For this purpose, we have treated the
mixing with the active neutrinos $|V_{\alpha N}|^2$ with
$\alpha=e,\mu,\tau$, the HNL mass $m_N$, and the operator coefficients
$c_{{\cal O}}/\Lambda^2$ as independent parameters.  For simplicity,
we consider the case of $n_N=1$ generation of HNLs and focus on their
mixing with the active neutrinos in the electron sector only, i.e.,
$V_{\alpha N} = V_{eN}$.

In the theoretical framework considered in this paper, the HNL
production at the LHC may occur either via the effective operators
with a pair of HNLs or the mixing with the electron neutrinos
(mediated by $W$ and $Z$ bosons).  However, the decay of the HNLs can
only proceed via the standard active-heavy mixing, since the HNL-pair
EFT operators cannot induce the lightest HNL to decay.  The HNLs can
thus undergo either two-body or three-body (leptonic or semi-leptonic)
decays, depending on their mass.

In fig.~\ref{Fig:VmN}, we show the estimated experimental
sensitivities at the 14 TeV LHC to a long-lived Majorana HNL, in the
plane $|V_{eN}|^2$ versus $m_N$.
For our analysis we have considered different choices of operators.  In the left panel, we focus on one single operator, ${\cal O}_{dN}$, choosing only the coupling with down quarks, $c_{dN}^{11}$, to be non-zero.
In the right panel, we switch on all the three types of operators ${\cal O}_{dN}$, ${\cal O}_{uN}$ and ${\cal O}_{QN}$, taking the same value for all the couplings of the first quark-generation combination, $c_{dN}^{11} = c_{uN}^{11} = c_{QN}^{11}$.
These two choices lead to respectively the smallest and largest possible cross sections~\footnote{See the comment in
  footnote~\ref{foot:XS}.} for the HNL pair production at the LHC, and
thus cover the most conservative and the most optimistic scenarios.
For $m_N \lesssim 5 $ GeV, the HNLs are produced dominantly by $B$- and
$D$-mesons decays. These decays can proceed via mixing with active
neutrinos as well as through $d=6$ interactions. A detailed study of
the interplay between the minimal scenario and the effective operators
containing one $N_R$ has been performed in Ref.~\cite{deVries:2020qns}. Inclusion of the $N_R$ pair operators in the meson decay rates goes beyond the scope of the present work and is to be performed elsewhere.
Since we focus in our current numerical study on coefficients for quarks of the first generation only, we believe our figures will be affected by this approximation only in a minor way.
Therefore, the inclusion of effective operators does not affect the LHC sensitivities to $|V_{eN}|^2$ and in the region $m_N \lesssim 5$~GeV, we simply reproduce the results previously derived in the literature \cite{Kling:2018wct,Helo:2018qej,Dercks:2018wum,Hirsch:2020klk,deVries:2020qns}.

For $m_N \gtrsim 5$ GeV, the HNL production is dominated by the
effective operators.  In this case, the mixing is only important for
the HNL decays (but not for their production), and as a result, the
sensitivity to the mixing grows substantially because of the much
enhanced production rates of the HNLs.  When the mixing and mass lead
to a boosted decay length that falls well into the sensitive range of
a detector, a large acceptance rate is expected.  For larger $m_N$,
only a smaller mixing angle can roughly correspond to a fixed
acceptance rate.  As we can see in fig.~\ref{Fig:VmN}, for operator
coefficients $c_{{\cal O}}/\Lambda^2 = 1/(2 \text{ TeV})^{2}$, the
experiment MATHUSLA, for an integrated luminosity of 3 ab$^{-1}$, can
reach values of the mixing down to $|V_{eN}|^2 \sim 2\times10^{-23}$ for the single operator ${\cal O}_{dN}$ and $|V_{eN}|^2 \sim 3\times 10^{-24}$ for the most optimistic scenario in which all the three operators
${\cal O}_{dN}, {\cal O}_{uN}, {\cal O}_{QN}$ are simultaneously
switched on.  The MATHUSLA experimental sensitivity also extends
considerably for the HNL mass reaching up to $m_N \sim 1.8~(2.3)$ TeV
for the most conservative (optimistic) choice of effective operators.
Such a large mass reach is essentially allowed because the HNLs are produced from direct parton collisions of the center-of-mass energy 14 TeV.
Similar limits can be set by ANUBIS for the same integrated luminosity
and around one order of magnitude smaller by AL3X for an integrated
luminosity of 250~fb$^{-1}$.
ATLAS, on the other hand, can reach mixing limits down to $|V_{eN}|^2 \sim 10^{-21}~(2.6\times 10^{-22})$ and masses up to $m_N \sim 2.6~(3)$ TeV, for an integrated luminosity of 3~ab$^{-1}$, for the most conservative (optimistic) scenario.
Its lower mass reach is at about 5 GeV, which is due to the $m_{\text{DV}}>5$ GeV event selection we apply.
ATLAS has a sensitivity on the mixing $|V_{eN}|^2$ worse than MATHUSLA by almost two orders of magnitude, but it can reach larger masses of the
HNL.  This is mainly due to the distance at which MATHUSLA is planned
to be located ($\sim 100$~m from the interaction point) which makes it
sensitive to higher values of the lifetime, and in turn to smaller
mixing $|V_{eN}|^2$ and masses $m_N$.  As we can see, with the
inclusion of the effective operators, these experiments can reach
limits on the mixing $|V_{eN}|^2$ several orders of magnitude better
than the current experimental bounds, represented here with a dark
gray area at the top of each plot in fig.~\ref{Fig:VmN}.~%

Figure~\ref{Fig:VmN} also shows in each plot a light gray band labeled
``Type-I Seesaw target region'' corresponding to the values of the
mixing $|V_{eN}|^2$ and mass $m_N$ that lead to $m_\nu$ between
$10^{-3}$ and $10^{-1}$ eV.  For $c_{{\cal O}}/\Lambda^2 = 1/(2\text{
  TeV})^{2}$ this region can be accessed by MATHUSLA and ANUBIS for
HNL masses $5 \lesssim m_N \lesssim 100$~GeV, and by AL3X for masses
of $10 \lesssim m_N \lesssim 120$~GeV.  ATLAS, on the other hand, can
cover the gray band for HNL masses around $ 15 \lesssim m_N \lesssim
700$ GeV when we work with one single operator ${\cal O}_{dN}$ and $
14 \lesssim m_N \lesssim 1000$ GeV when we simultaneously switch on
the three types of operators.  For $c_{{\cal O}}/\Lambda^2 = 1/(7
\text{ TeV})^{2}$ this region can be accessed by MATHUSLA and ANUBIS
for HNL masses $16 \lesssim m_N \lesssim 90$ GeV, while AL3X can do
so for masses of $ \lesssim m_N \lesssim 120$ GeV, and ATLAS for $
30 \lesssim m_N \lesssim 500 $ GeV.  For $c_{{\cal O}}/\Lambda^2 =
1/(13\text{ TeV})^{2}$ only ATLAS with 3~ab$^{-1}$ integrated
luminosity can be sensitive to this region for
$80 \lesssim m_N \lesssim 300$ GeV when we work with one single
operator ${\cal O}_{dN}$, in virtue of its larger geometrical
acceptance.  When we simultaneously switch on all the three types of
operators AL3X can cover the band for $35 \lesssim m_N
\lesssim 90$~GeV, while ATLAS can do so for $50 \lesssim m_N \lesssim
400$ GeV.  Thus, the inclusion of the effective operators has a great
impact on the experimental sensitivity reach to the values of 
mixing $|V_{eN}|^2$ and mass $m_N$ required by the type-I seesaw
mechanism to explain the neutrino masses.  The future far detectors
and ATLAS are complementary to each other in this regard.

\begin{figure}[t]
	\centering
	\includegraphics[width=0.49\textwidth]{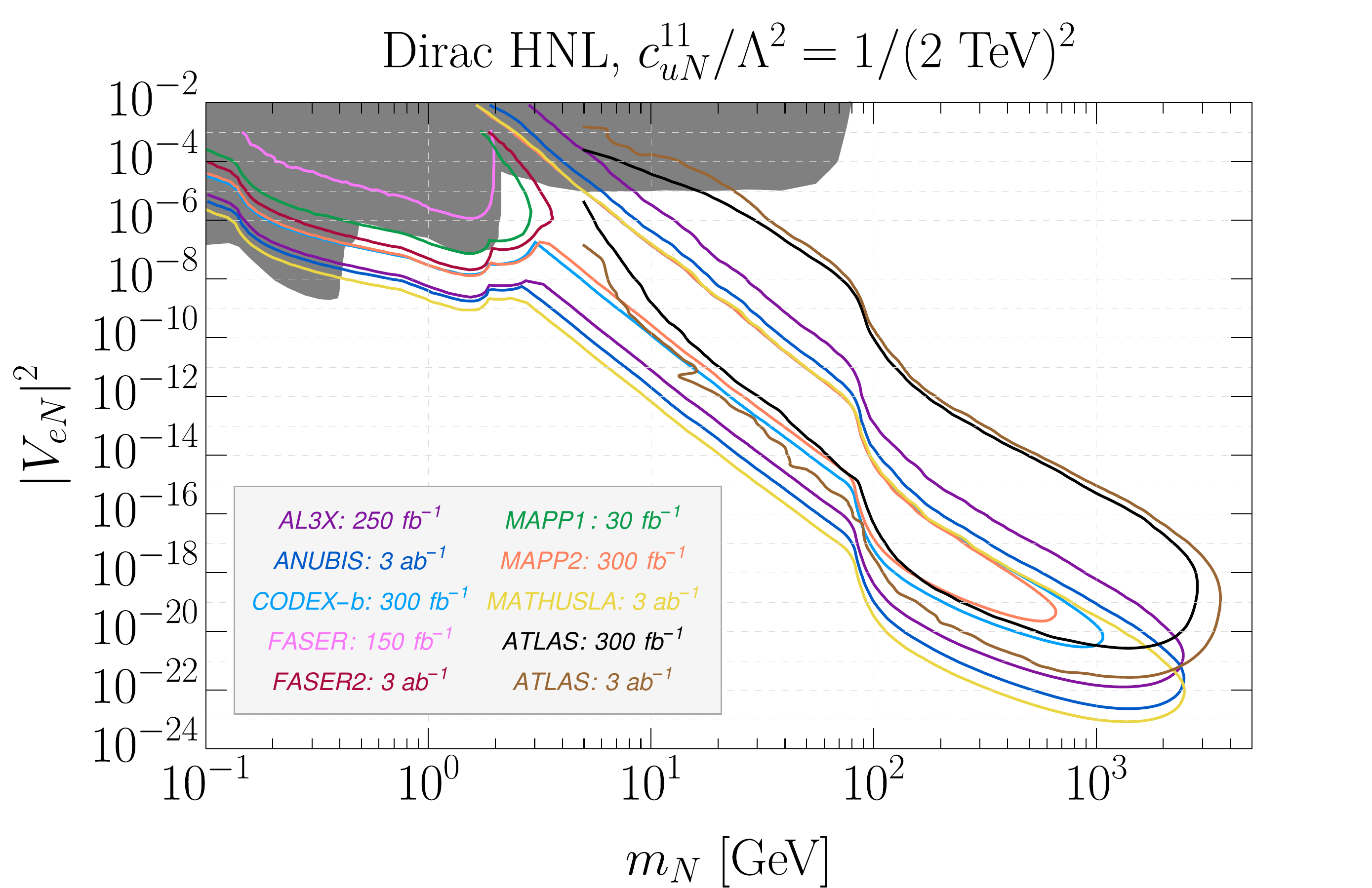}
	\includegraphics[width=0.49\textwidth]{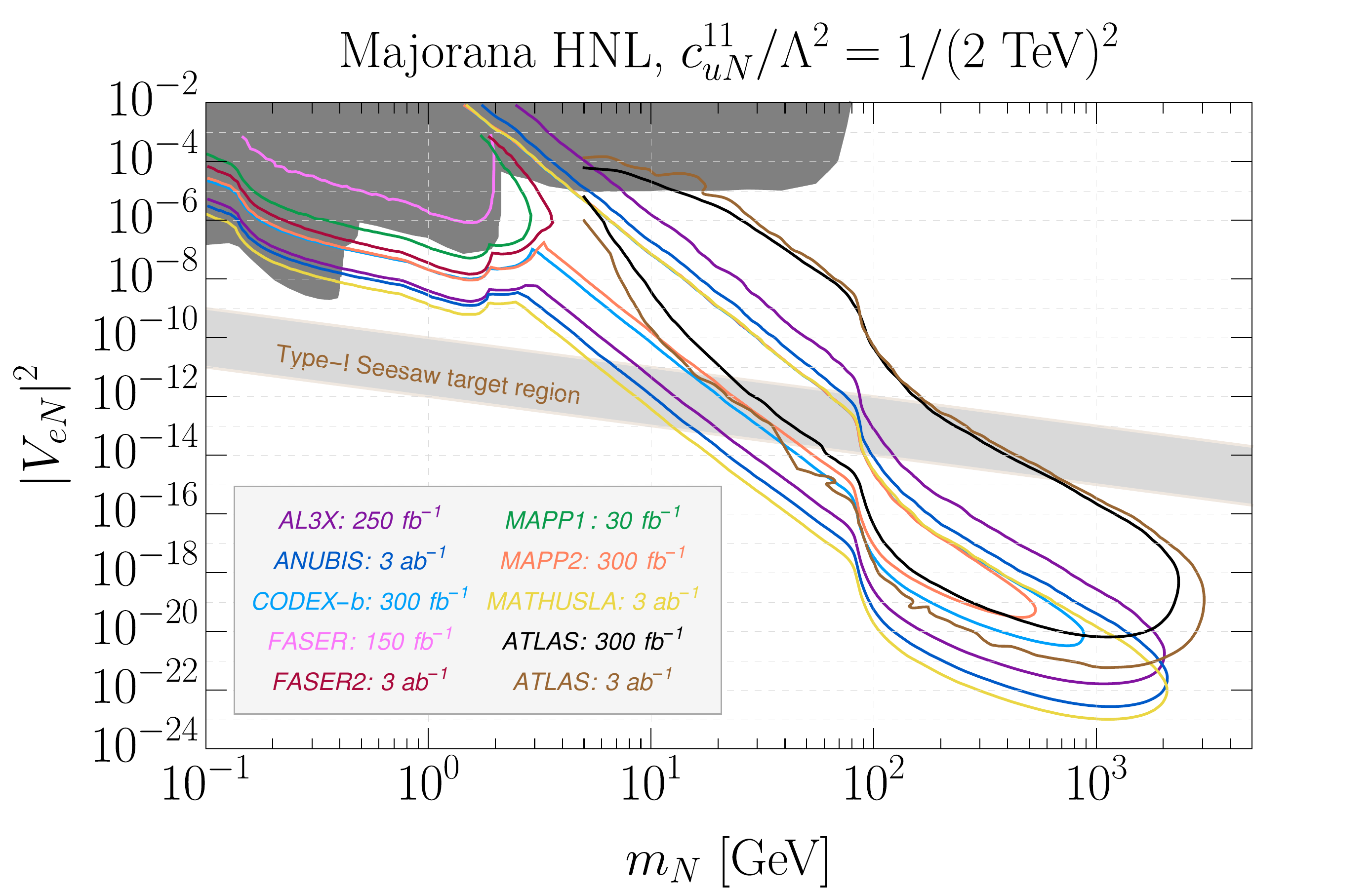}
	\\[0.4cm]
	\includegraphics[width=0.49\textwidth]{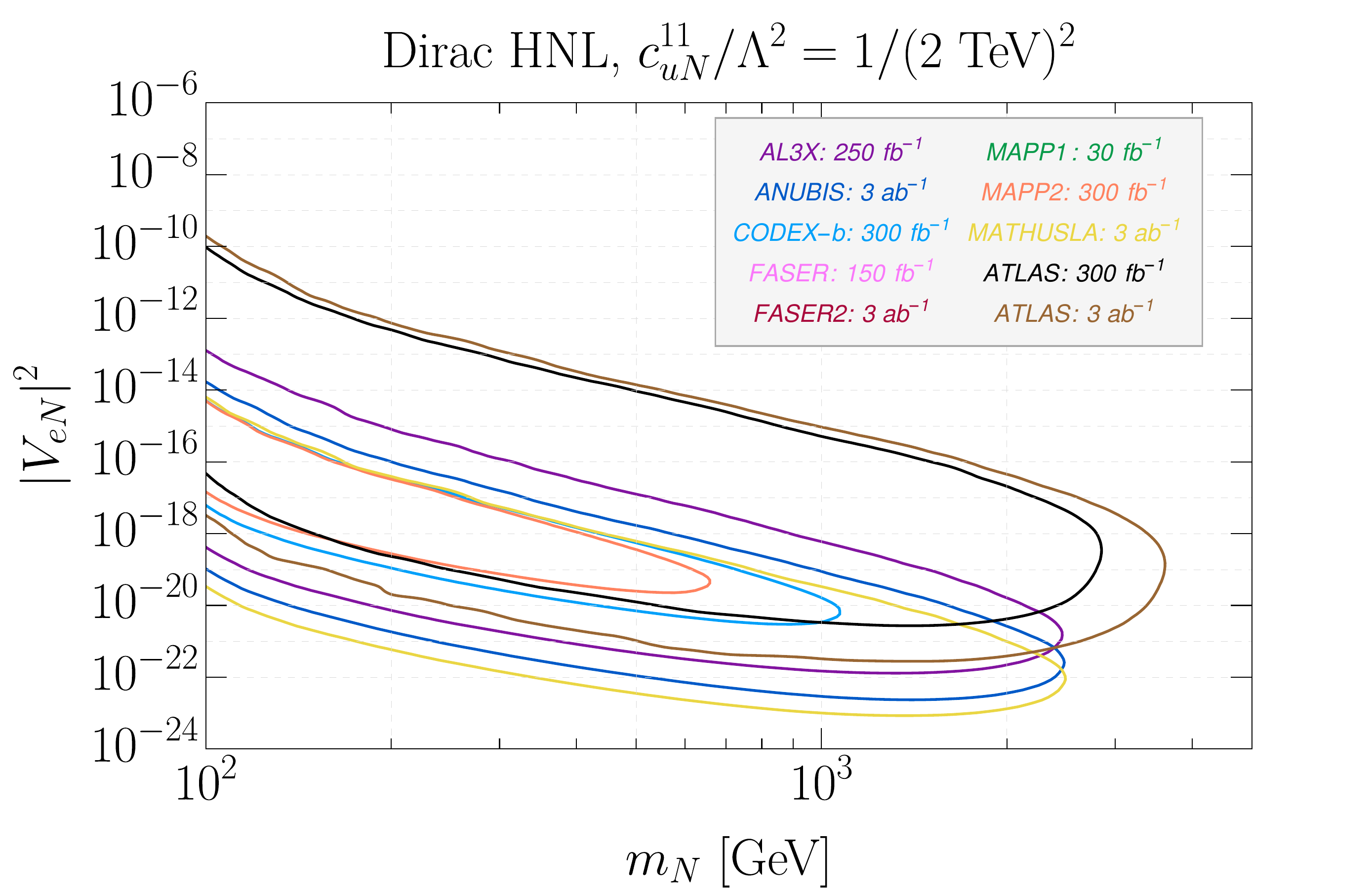}
	\includegraphics[width=0.49\textwidth]{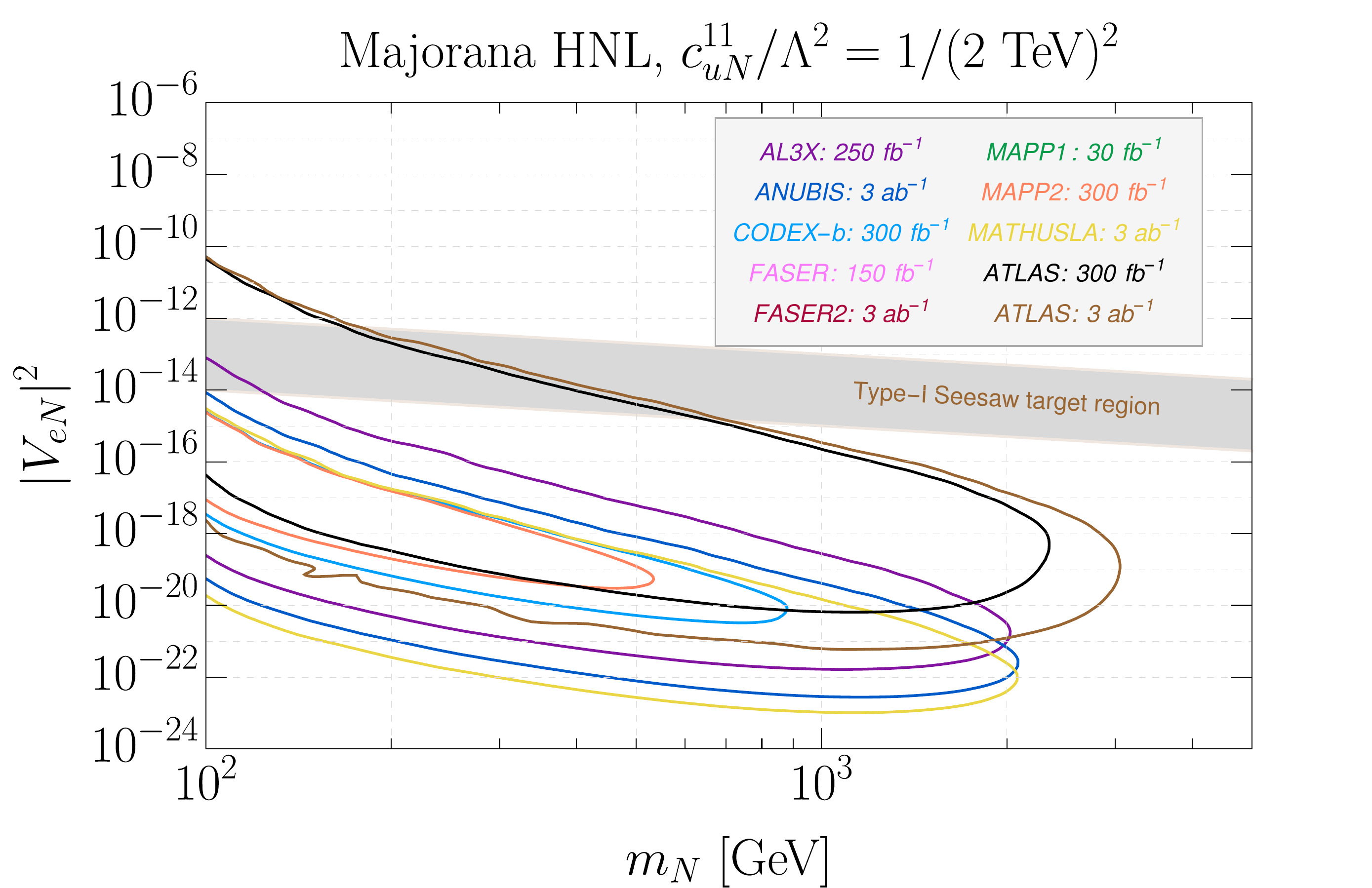}
	\caption{Experimental sensitivity reach for $|V_{eN}|^2$ as a
          function of $m_N$.  The plots consider one single operator
          ${\cal O}_{uN}$ only.  The values of the operator
          coefficient is fixed to $c_{uN}^{11} / \Lambda^2 = 1/(2
          \text{ TeV})^{2}$.  Limits have been obtained for Dirac
          (left) and Majorana (right) HNLs. The plots in the bottom
          row zoom in to the HNL masses larger than 100 GeV.
		\label{Fig:DiracMaj}}
\end{figure}
	
Figure~\ref{Fig:DiracMaj} shows the estimated experimental
sensitivities of both ATLAS and future far detectors in the
$|V_{eN}|^2$ versus $m_N$ plane, for one single effective operator
${\cal O}_{uN}$.  We switch on here only the coupling with the up
quark, $c_{uN}^{11}$.  The left (right) plots are for a Dirac
(Majorana) HNL~\footnote{For Dirac HNLs with mass below about 5 GeV,
  the limits are recast from the Majorana ones that exist in the
  literature~\cite{Kling:2018wct,Helo:2018qej,Dercks:2018wum,Hirsch:2020klk,deVries:2020qns}, by multiplying $|V_{eN}|^2$ with $\sqrt{2}$.}, for $c_{uN}^{11}/\Lambda^2 = 1/(2\text{ TeV})^{2}$.  As expected in this
case, the limits are slightly better than those for ${\cal O}_{dN}$
(the upper left plot of fig.~\ref{Fig:VmN}) and slightly worse than
the limits shown in the upper right plot of fig.~\ref{Fig:VmN} (where
all three operators for the first quark generation are switched on).  The
lower row of fig.~\ref{Fig:DiracMaj} contains plots which are the
zoomed-in version of those in the upper row, focusing on the
large-mass regime.  As we can see, the range of experimental
sensitivity on the HNL mass $m_N$ is larger for the Dirac case than
for the Majorana one.  For instance, the mass reach of MATHUSLA,
ANUBIS, and AL3X is at about 2.4 TeV for Dirac HNLs, compared to 2.0
TeV for Majorana HNLs.  ATLAS, on the other hand, for an integrated
luminosity of 3 ab$^{-1}$, can be sensitive up to masses of $m_N\sim
3.0~(3.6)$ TeV for a Majorana (Dirac) HNL.  This difference in the
mass reach between Dirac and Majorana HNL scenarios is due to the
different HNL pair production cross sections in the two cases. In
particular, this difference is more perceptible for larger $m_N$
(cf. eqs.~\eqref{eq:sigD} and \eqref{eq:sigM}, fig.~\ref{fig:XSec2}
and the related discussion in section~\ref{sect:models}).

Overall, the shape of the main results shown in figures~\ref{Fig:VmN} and~\ref{Fig:DiracMaj} can be qualitatively understood as the final efficiency being bounded by the cases when the HNLs are decaying either too promptly (the upper right corner) or too far away (the lower left corner), both outside the detectors' acceptance.
In addition, the upper mass reach is limited by the production cross sections.

We would like to note that Big Bang Nucleosynthesis can place further constraints on the active-heavy mixing angle in the low mass region $m_N \lesssim 2$~GeV (see, e.g., Refs.~\cite{Vincent:2014rja,Sabti:2020yrt,Boyarsky:2020dzc}). However, these constraints rely on the evolution of HNLs in the early Universe, discussion of which goes beyond the scope of our work. Thus, we remain agnostic about them and do not show the corresponding excluded regions in our figures.

\begin{figure}[t]
	\centering
	\includegraphics[width=0.49\textwidth]{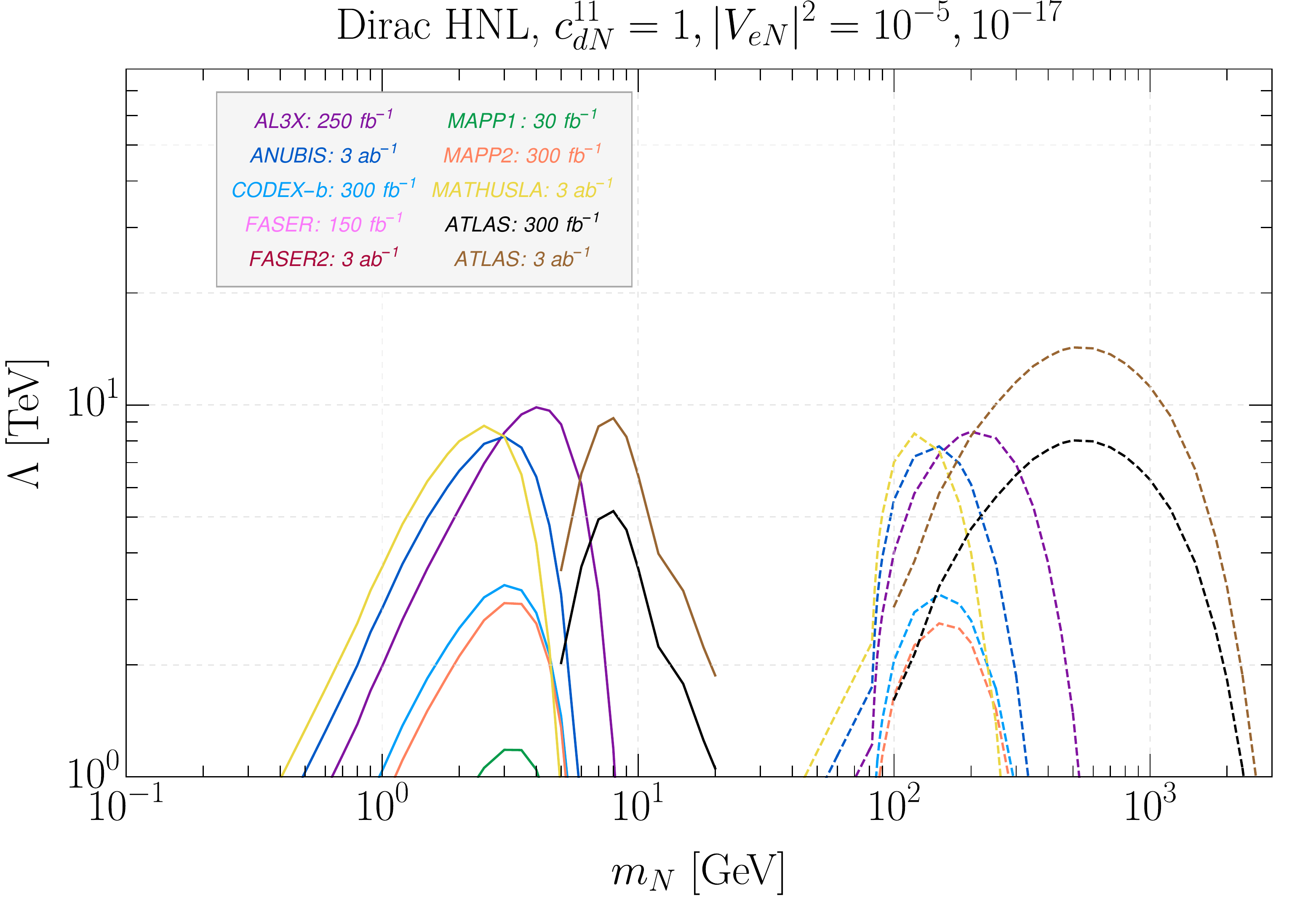}
	\includegraphics[width=0.49\textwidth]{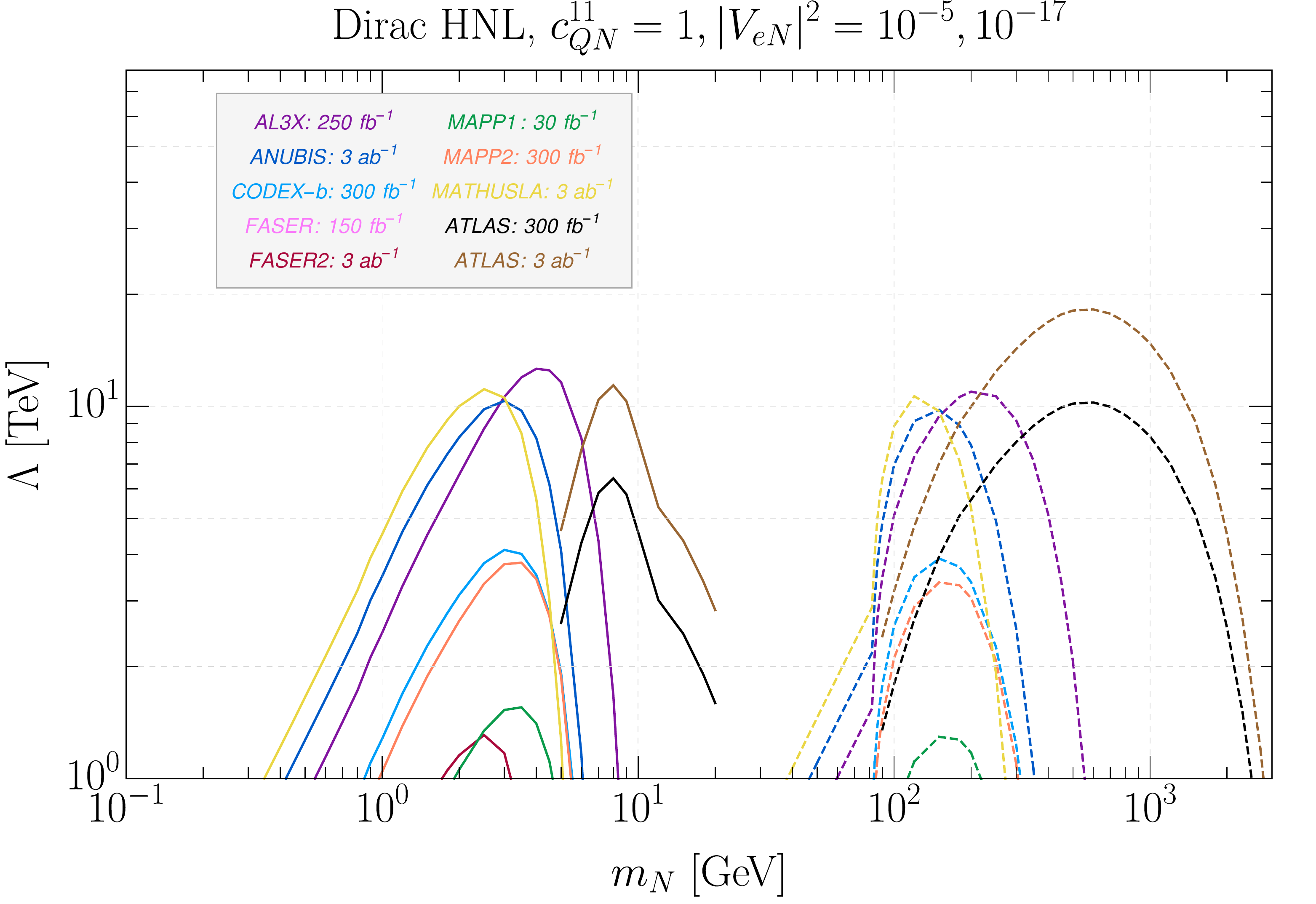}\\[0.4cm]	
	\includegraphics[width=0.49\textwidth]{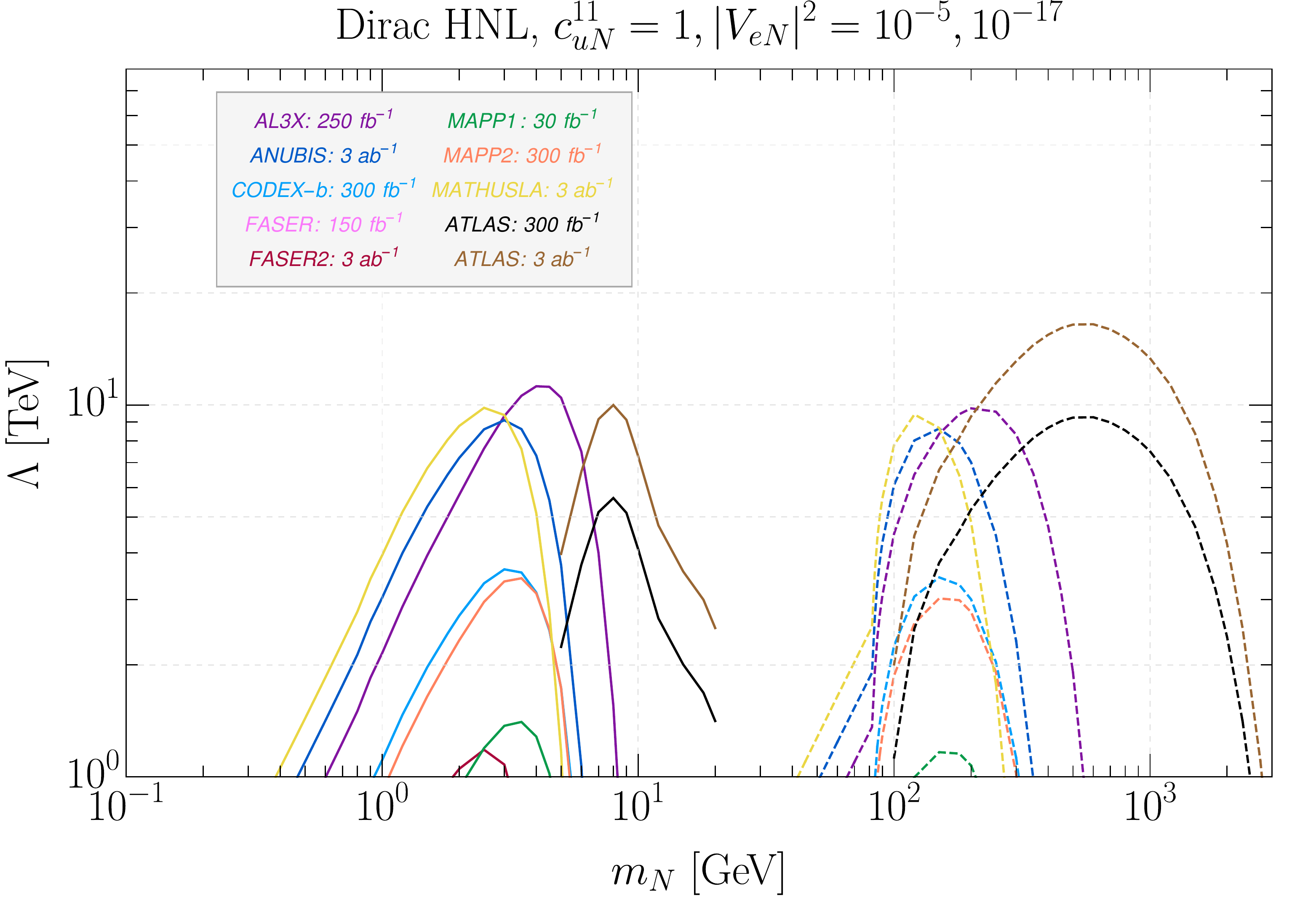}
	\includegraphics[width=0.49\textwidth]{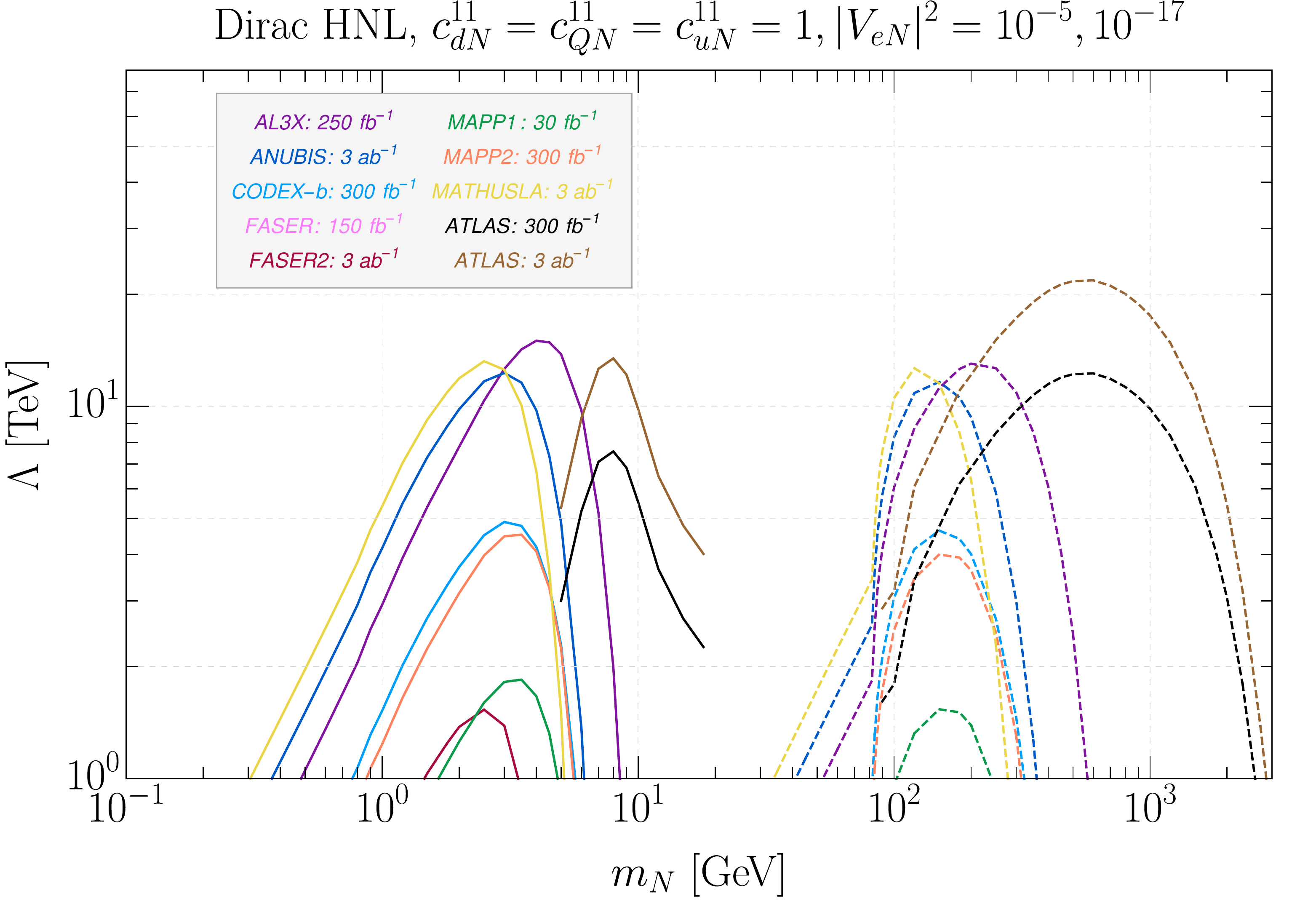}
	\caption{Experimental sensitivity reach on the new physics scale $\Lambda$ as a function of $m_N$ for Dirac HNLs.
		The values of the mixing with the active neutrinos have been fixed to $|V_{eN}|^2 = 10^{-5}$ (solid lines) and $10^{-17}$ (dashed lines).
		The four plots correspond to 4 different scenarios.
		\label{Fig:LambdamN}}
\end{figure}
Finally, fig.~\ref{Fig:LambdamN} depicts our limits in the plane
$\Lambda$ versus $m_N$ for a Dirac HNL for all four scenarios of types
of EFT operators we consider.  The values of the mixing with the
active neutrinos are fixed to be $|V_{eN}|^2 = 10^{-5}$ or $10^{-17}$.
(For values between these extremes, intermediate values of $m_N$ will
be tested.) In these four plots, we fix respectively $c_{dN}^{11}=1$,
$c_{QN}^{11}=1$, $c_{uN}^{11}=1$, or
$c_{dN}^{11}=c_{QN}^{11}=c_{uN}^{11}=1$.  In general, these plots show
that FD experiments MATHUSLA, ANUBIS, and AL3X, may probe
the new physics scale up to $\Lambda \lesssim 8-15$ TeV, depending on
the effective operator scenario we considered. ATLAS, on the other
hand, can probe new physics scale up to $\Lambda \lesssim 22$ TeV when
we simultaneously switch on the three types of effective operators,
and up to $\Lambda \lesssim 15$ TeV, when we consider the most
conservative scenario with one single effective operator ${\cal
  O}_{dN}$. 
  Assuming unit Yukawa couplings, our projected limits can be translated to bounds on the LQ masses of about $5-16$ TeV.
  The bounds on $\Lambda$ in the Majorana case are similar. 
  
  Currently the LHC could potentially reach values of $\Lambda\approx 3 \text{ TeV}$ when reinterpreting 13 TeV prompt searches.
  For a reinterpretation of the prompt same-sign dilepton
  	search performed by CMS in~Ref.~\cite{Sirunyan:2018xiv}, see
  	appendix~\ref{sect:appendix2}.

\section{Summary\label{sect:sum}}

In this work, we have studied the phenomenology of heavy neutral
leptons (HNLs) as long-lived particles (LLPs) in the framework of an
effective field theory (EFT).  Concretely, we have studied the effects
of $d=6$ four-fermion operators with a pair of HNLs and a pair of
  quarks in the EFT of a SM extension with HNLs as
``right-handed neutrinos'', $N_R$SMEFT.  We considered three different
types of these operators and have studied scenarios with only one or
all of them being present at the same time.  While quantitatively the
results depend on the assumptions made about type of operator and/or
Wilson coefficient present, qualitatively all HNL pair operators
behave similarly.  Operators with two HNLs and two quarks lead to
production cross sections at the LHC which are not suppressed by the
small mixing of the HNLs with the active neutrinos.  Instead, cross
sections are proportional to $\Lambda^{-4}$, where $\Lambda$ is the
energy scale of the non-renormalizable $d=6$ operators.

HNL pair operators will not cause (the lightest) HNL to decay.  Thus,
HNLs decay only via their mixing with the active neutrinos, $V^2$.
This scenario leads to one very important change in phenomenology with
respect to previous works, that considered HNLs at the LHC which were
produced via charged (and neutral) current diagrams induced by the
mixing of the HNLs only.  Namely, the total signal event number for
the different experiments in our setup scales at the smallest $V^2$
that one can probe only as $V^2$, instead of the usual $V^4$.

We have estimated the sensitivity range of various LHC experiments in
this setup: ATLAS and a series of proposed ``far detectors'', AL3X,
ANUBIS, CODEX-b, FASER, MATHUSLA, and MoEDAL-MAPP.  Our main result is
that for $\Lambda$ lower than roughly $\Lambda \simeq 10-15$ TeV,
depending on the operator, much larger HNL masses and much smaller
mixing angles, $V^2$, can be probed than in the ``standard'' case,
where both, production cross section and decay lengths, are determined
by the mixing angle (and the HNL mass) only.

We have also briefly discussed some ultraviolet completions for the
considered $d=6$ operators: leptoquarks (LQs) and a possible $Z'$.
While we presented our results only in terms of EFT parameters, it is
clear from the numbers found in our simulation that the reach in
LQ (or $Z'$) mass in these ``indirect'' searches will be much larger
than for direct on-shell production at the LHC.

We have also highlighted differences in the production cross sections
for Dirac and Majorana HNLs and shown both, analytically and
numerically, that cross sections at HNL masses larger than roughly
$100$ GeV are different in the two cases.  The relative suppression of
Majorana HNL cross section with respect to the Dirac one can not,
however, be easily used to distinguish the two cases: velocity
distributions for Majorana and Dirac HNLs are different at large HNL
masses, but measuring these would require a rather large event number.
However, this change in cross section results in a smaller sensitivity
reach of the different experiments for Majorana HNLs at the largest
HNL masses that can be probed at the LHC.
Rather, to distinguish between Majorana and Dirac HNLs one
  should search for events in ATLAS, in which both $N_R$'s decay inside the detector.  If such an event can be identified for Majorana HNLs the
  ratio of same-sign to opposite-sign dilepton events should equal 1.  This could be checked with just very few events, in
  principle.

It is possible to require reconstruction
  of both DV's in each signal events in ATLAS (not in the far
  detectors, however). This would also be more effective at
  suppressing background and hence allow for looser event selections
  cuts, enhancing the event selection efficiencies.  However, it would
  also reduce the signal event number, since the probability of a
  $N_R$ decaying inside the pixel detector is always smaller than one
  and in the extremes of parameter space that one can probe actually
  much smaller than one. As a result, while observing double DV events
  would be very interesting for establishing the Dirac/Majorana nature
  of the HNLs, searching for such events will not lead to the possibility
  to explore new parts of the parameter space of $N_R$SMEFT.

Finally, we close by mentioning that there are also $d=6$ operators
with only one HNL.  Parts of the parameter space where such operators
may be probed in the far detectors have already been studied in
Ref.~\cite{deVries:2020qns}.

\bigskip
\section*{Acknowledgements}

\medskip
We thank Oleg Brandt for useful discussions about ANUBIS.  This work
is supported by the Spanish grants FPA2017-85216-P (MINECO/AEI/FEDER,
UE) and PROMETEO/2018/165 grants (Generalitat
Valenciana). G.C. acknowledges support from ANID FONDECYT-Chile grant
No. 3190051. G.C. and J.C.H. also acknowledge support from grant ANID
FONDECYT-Chile No. 1201673. A.T. is supported by the “Generalitat
Valenciana” under grant PROMETEO/2019/087 and by the FEDER/MCIyU-AEI
grant FPA2017-84543-P.  Z.S.W. is supported by the Ministry of Science
and Technology (MoST) of Taiwan with grant number
MoST-109-2811-M-007-509. This work was funded by ANID – Millennium Science Initiative Program ICN2019\_044.

\bigskip


\appendix
\section{A timing cut at ANUBIS \label{sect:appendixA}}

In this appendix, we discuss briefly the effect of a potential timing
cut at the ANUBIS experiment on the expected exclusion limits for our
setup. ANUBIS \cite{Bauer:2019vqk} has been proposed to be installed
inside one of the service shafts above the ATLAS detector. The
relative proximity of ANUBIS to ATLAS would allow ANUBIS to trigger on
the readout of ATLAS.  Given the close distance from the IP, however,
it is not as straightforward to implement background vetos as in some
of the other proposed far-detector experiments; some discussion about 
backgrounds is given in the original proposal \cite{Bauer:2019vqk}.

One of the possible ways to reduce background events is to make use of
timing information.~%
\footnote{O. Brandt, private communication.} 
Since the hard collision can be timestamped, if a DV is formed significantly later than expected, the event can be rejected with this strategy, such that background events from e.g.\ beam-gas and beam-collimator interactions could be reduced to a large extent.
Such a timing cut is based on the idea that light LLPs (masses of a few GeV) are produced at the LHC with velocities $\beta\simeq 1$, where $c$ labels the speed of light.  In the EFT setup, discussed in this paper, however, HNLs with much larger masses can be probed and their velocity distributions will
depart from $\beta =1$, leading to a significant delay in arrival
times of the events in ANUBIS.

Here, for a rough estimate of the size of the effect, we consider two
possible acceptance time windows: 1 ns and 2 ns.  Note that these
effectively correspond to LLP speeds of $0.99\cdot c$ and
$0.98\cdot c$ for a distance of about 30 m which is roughly the distance of ANUBIS from the ATLAS IP.  As a benchmark, we take
Dirac HNLs with the single NRO $\mathcal{O}_{dN}^{11}$. Results for the other operators will be qualitatively very similar.

\begin{figure}[t]
	\centering 
\includegraphics[width=0.48\textwidth]{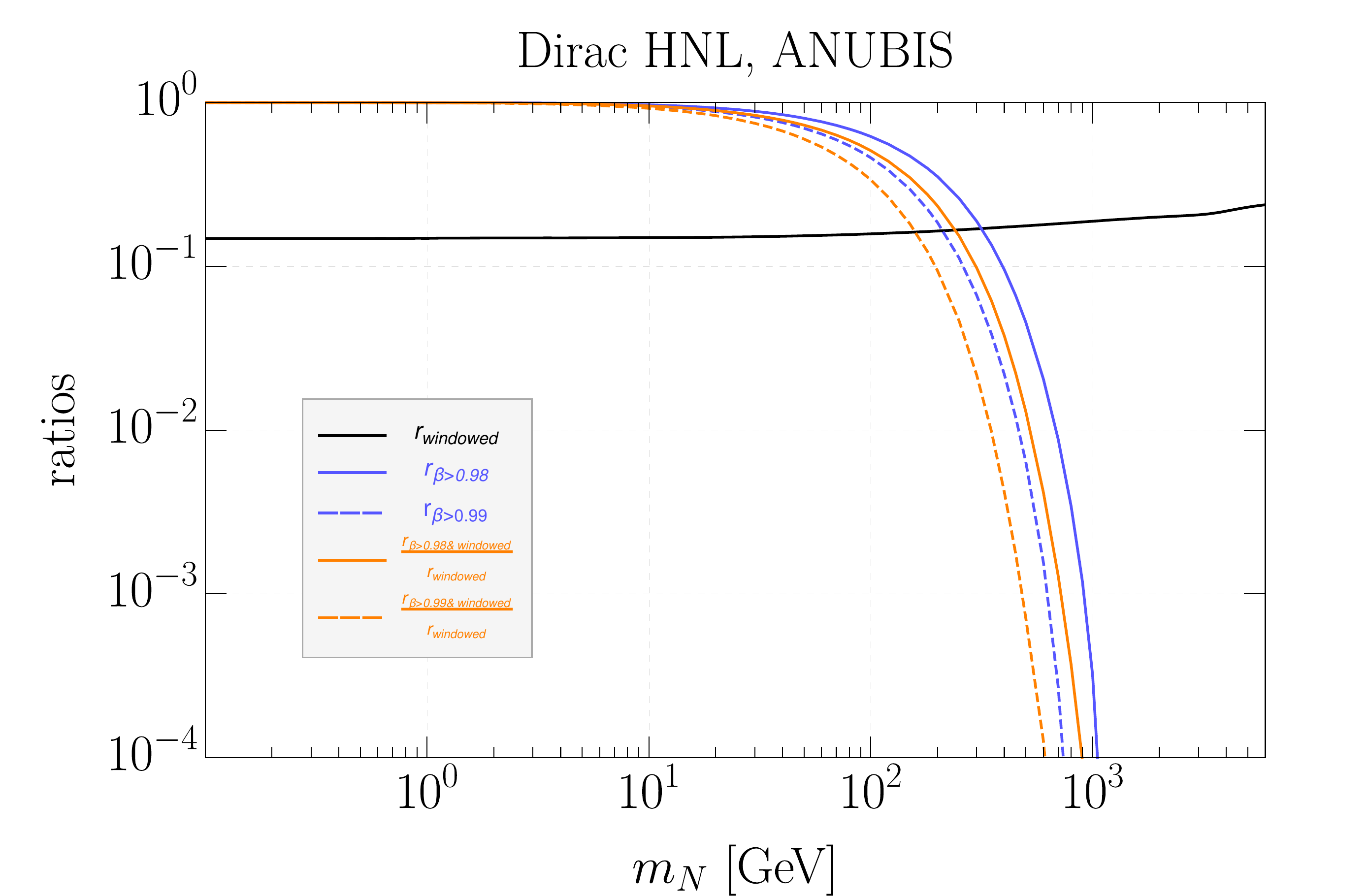} 
\includegraphics[width=0.48\textwidth]{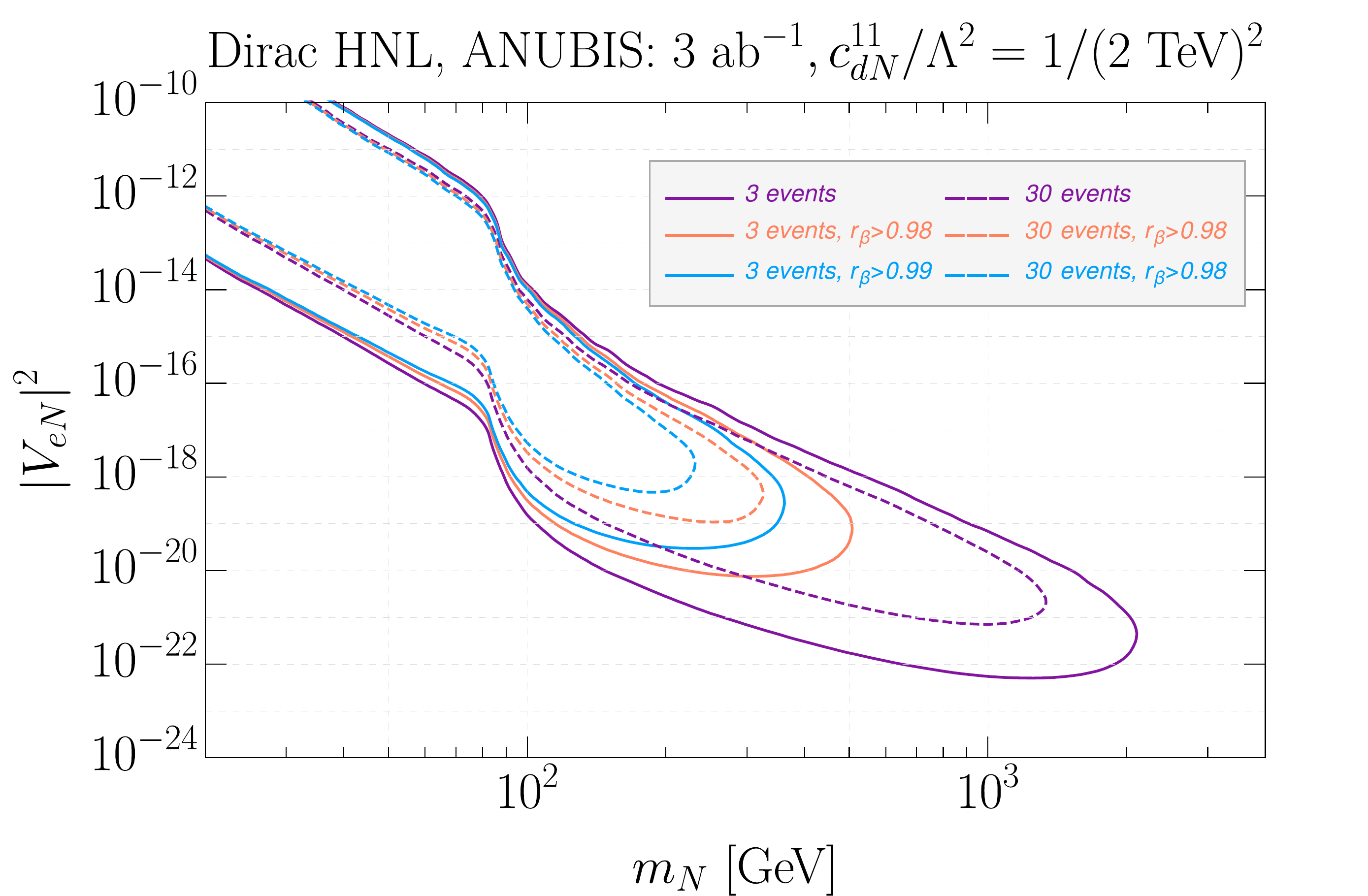} 
\caption{Left panel: ratios of the number of HNLs in the ANUBIS direction
to the total event number above a certain speed threshold, as a function
of the HNL mass. Here, ``windowed'' means the fraction of HNLs
traveling inside the ANUBIS solid-angle window without cuts on $\beta$,
where $\beta$ denotes the speed of the HNL. Two different $\beta$ cuts
are applied and compared in the plot. Right panel: sensitivity limits
to the Dirac HNLs with the operator $\mathcal{O}_{dN}^{11}$ with and
without timing cuts, for 3 or 30 signal
events.}  \label{fig:anubis_timing}
\end{figure}

Figure~\ref{fig:anubis_timing} shows two plots to exemplify the effect
of a timing cut.  The left plot shows various curves for the ratios of
the numbers of the HNLs traveling inside the solid angle coverage of
the bottom of ANUBIS with the speed above certain thresholds to the
total number of the produced HNLs, as a function of $m_N$.  We find
that for $m_N \gtrsim 200$ GeV the fraction of HNLs in the direction
of ANUBIS with a speed larger than $0.99\cdot c$ drops below $10\%$ of
the total number already, and for a mass close to 1 TeV the ratio even
drops down to the level of $10^{-4}$. The effect of this loss of
events on the sensitivity reach is demonstrated in the right panel.
For $m_N$ above about 200 GeV, a timing cut will reduce the exclusion
power by a significant amount.  For instance, if we require 3 signal
events, excluding HNLs with a speed smaller than $0.99\cdot c$, the
reach in $m_N$ shrinks from about 2.1 TeV to only 360 GeV and in
$|V_{eN}|^2$ from $5\times 10^{-23}$ to $3\times 10^{-20}$.

The plot on the right also shows curves for two different assumptions
of excluded events, 3 and 30 signal events, respectively. We conclude that for the largest HNL masses it could be advantageous to use a looser timing cut, since larger parts of parameter space could be probed this way, despite larger backgrounds.
We close this discussion by noting that a more detailed MC simulation of backgrounds in ANUBIS would be necessary in order to determine the optimal cuts, which is beyond the scope of our work.


\section{Reinterpretation of a prompt same-sign dilepton search
  \label{sect:appendix2}}

Our theoretical setup changes the production cross section for HNLs at
the LHC, relative to the simplest HNL models in which both, production
and decay, of the HNLs are determined by their mixing with the active
neutrinos. Any other LHC search for HNLs than the LLP searches we
discussed in this paper will be affected by the additional NROs in our
setup as well.  While only the experimental collaborations can make a
reliable search, specifically adapted to HNLs in $N_R$SMEFT, in this
appendix we will present a rough estimate of the sensitivity of a
recent CMS search \cite{Sirunyan:2018xiv} to $N_R$SMEFT
parameters. While other searches could be reinterpreted in a similar
way to the one discussed below, we have decided to concentrate on
this particular search, since it is currently the most sensitive HNL
search for masses $m_N \ge 100$ GeV.

CMS has presented a search \cite{Sirunyan:2018xiv} for same-sign
dileptons plus jets, based on ${\cal L}=35.9$ fb$^{-1}$ of statistics
taken at $\sqrt{s}=13$ TeV. The results are interpreted in a SM
extension with one HNL mixing with either the electron or muon
neutrino. No excesses were found in the data, thus upper limits on
$V_{eN}^2$ and $V_{\mu N}^2$ are presented as a function of HNL mass,
$m_N$.~%
\footnote{Also the ATLAS collaboration has published similar
  searches. However, \cite{Aad:2015xaa} is based on $\sqrt{s}=8$ TeV
  data, and thus no longer competitive. The analysis of
  \cite{Aaboud:2019wfg}, on the other hand, concentrates on a
  left-right extension of the SM in its analysis. The changes in cross
  section and kinematics for on-shell $W_R$ relative to the
  expectations for $N_R$SMEFT, however, would make a reinterpretation
  of this paper even more unreliable than our estimates for
  \cite{Sirunyan:2018xiv}. We therefore decided to concentrate on the
  CMS search here.}  Note that this search explicitly requires
same-sign leptons and thus is valid only for Majorana HNLs. One can
expect limits on Dirac HNLs to be significantly weaker in such a
search, since opposite-sign leptons have much larger SM backgrounds.

The search \cite{Sirunyan:2018xiv} assumes the HNLs decay promptly.  A
lower limit on $\Lambda$, calculated from this data, will therefore be
valid only if $V_{\alpha N}^2$ is large enough, such that a sufficient
number of events decay within the cuts used by CMS. We will quantify
the range of $V_{\alpha N}^2$ as function of $m_N$, for which this is
the case, in more details below. Let us first concentrate on how to 
estimate limits on $\Lambda$.

Since CMS decided to present their results only in the plane
($V_{\alpha N}^2, m_N$), the first step in our reinterpretation is to
calculate the corresponding number of excluded signal events as
function of $m_N$. In their modeling, CMS \cite{Sirunyan:2018xiv}
takes into account two types of Feynman diagrams for the production of
HNLs: (1) $s$-channel Drell–Yan (DY) process, mediated by a $W$-boson
and (2) ``photon-initiated'' process, which CMS also calls VBF
(vector-boson-fusion) channel. The importance of the latter at large
HNL masses was pointed out in Refs.~\cite{Dev:2013wba,Alva:2014gxa}.
This is due to the fact that VBF diagrams contain a $W$-boson in
$t$-channel, which has a softer $m_N$ dependence than the Drell-Yan
diagram at proton colliders.~%
\footnote{The final state of the VBF diagrams contains an additional
  jet, which will be mostly central.  The same final state can also be
  generated by ``gluon-initiated'' diagrams. These are numerically
  more important than the ``photon-initiated'' diagrams, except for
  $m_N$ in excess of roughly 1 TeV. (The exact number depends very
  strongly on the PDF.) However, CMS does not take into account these
  diagrams in their modeling, so we have to exclude them in our
  reinterpretation too.}
VBF diagrams do depend very sensitively on the photon content of the
proton. In a recent paper, Manohar et al. \cite{Manohar:2016nzj}
presented a set of PDFs that improve the calculation of the photon
density over previously existing PDFs. CMS uses in the calculation the
\texttt{LUXQED17\_{}PLUS\_{}PDF4LHC15\_{}NNLO\_{}100} PDF set
\cite{Manohar:2016nzj}, which we also use for our reinterpretation for
consistency.

CMS provides sample values of cross section for DY and VBF for
three choices of $m_N =40$, $100$ and $1000$ GeV. Using
\texttt{MadGraph5} \cite{Alwall:2007st,Alwall:2011uj,Alwall:2014hca},
we recalculated the cross sections for these points as a cross-check
of the reliability of our conversion procedure. While for $m_N=40$~GeV 
we reproduce very well the numbers quoted by CMS, for larger
$m_N$ our results depart from the numbers given by CMS. At $m_N=1$ TeV,
our VBF (DY) result is $\sim 75 \%$ ($80 \%$) larger than those
by CMS. We can only speculate that CMS uses some additional cuts in
the \texttt{run\_{}card.dat} of \texttt{MadGraph5}, that we were unable
to locate in the CMS paper. We decided therefore to use the difference
of our cross section results from those of CMS as an estimate of the
uncertainty for the total event number extraction in our reinterpretation
procedure.

The upper limits on $V_{eN}^2$ plus the calculated cross sections
allow us to estimate the number of excluded signal events (as a
function of $m_N$). This number, however, includes the detector
efficiencies indirectly. So, in order to extract limits on, for
example, $\Lambda$ in our EFT setup we have to assume that detection
efficiencies are the same for both models. While this is probably a
good approximation for small $m_N$, say below the $W-$threshold, our
operators produce pairs of HNLs, which at large $m_N$ affects the
kinematics, and thus efficiencies.  Given the uncertainty in total
cross sections discussed above, we believe this to be currently a less
important source of error for our estimate and will simply assume the
efficiencies are the same in both cases.

\begin{figure}[t]
\centering
\includegraphics[width=0.49\textwidth]{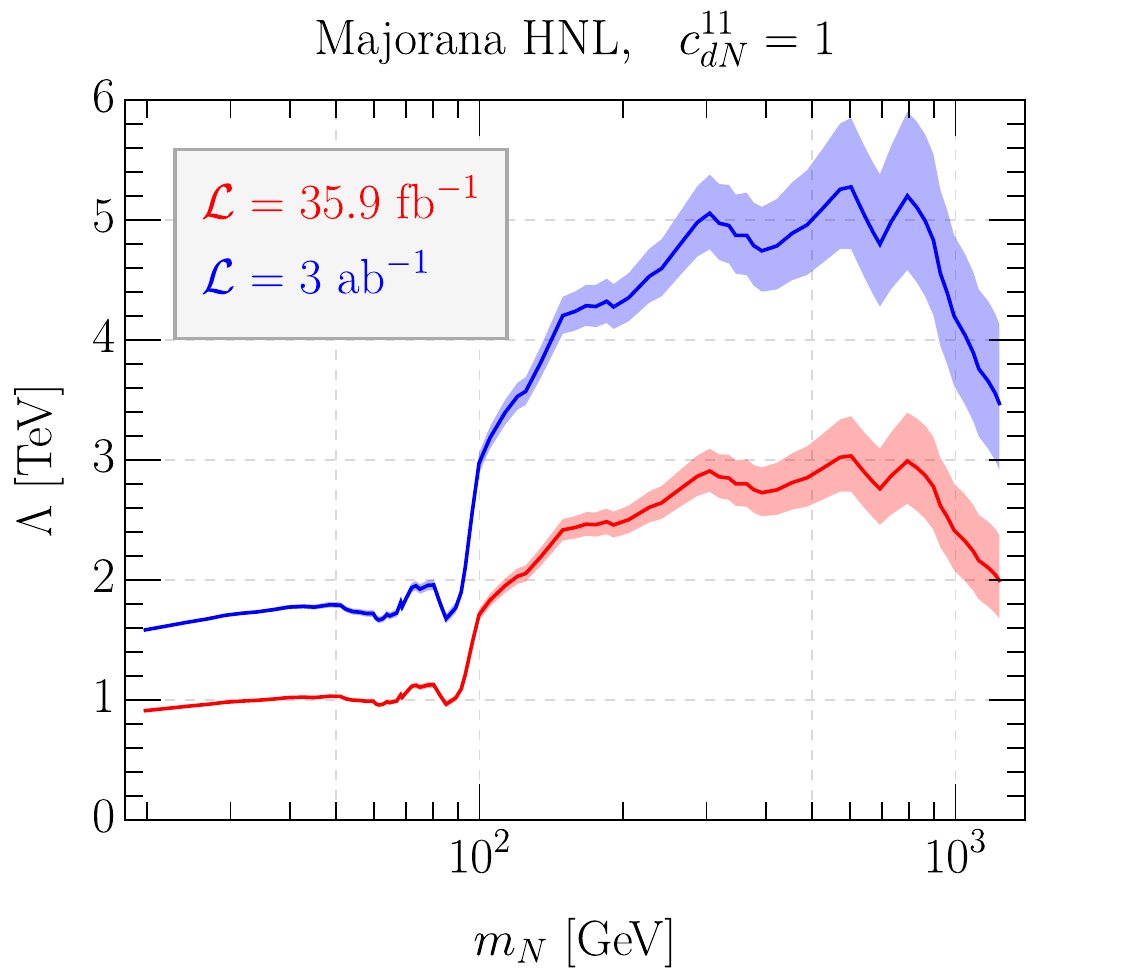}
\includegraphics[width=0.49\textwidth]{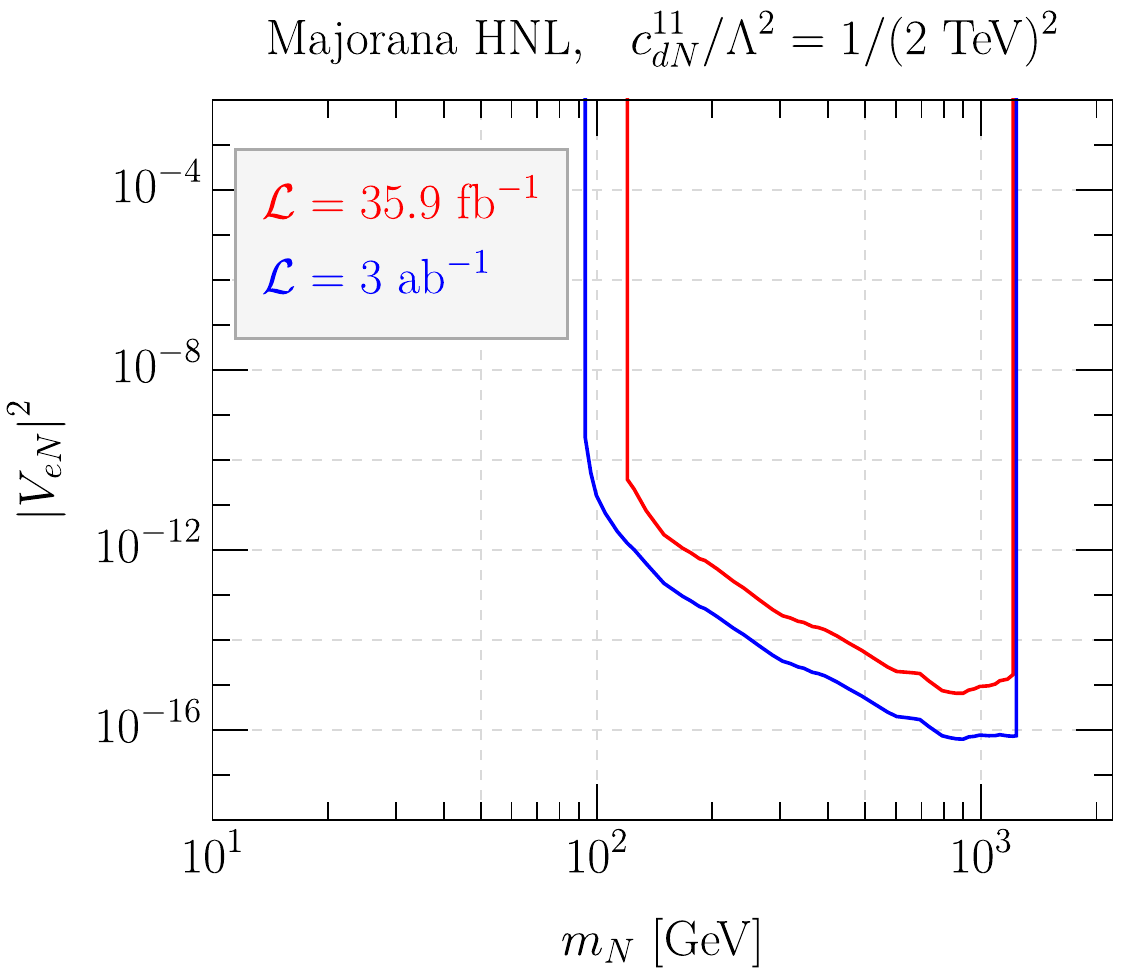}
\caption{To the left: Plot showing our estimate for the current limit
  on $\Lambda$ for the operator ${\cal O}_{dN}$ from the
  reinterpretation of the CMS same-sign dilepton search
  \cite{Sirunyan:2018xiv}, for electron final state. The red line is
  the current limit, the blue line is our estimate for a future
  luminosity of ${\cal L}=3$ ab$^{-1}$, based on this data. The coloured
  bands are a rough estimate of the error of our reinterpretation
  procedure.  To the right: Range of $m_N$ and $|V_{eN}|^2$ that could
  be probed for $\Lambda =2$ TeV by this search (see text for detail).
  \label{fig:ReInt}}
\end{figure}

We then show in fig.~\ref{fig:ReInt} to the left limits on $\Lambda$
for the operator ${\cal O}_{dN}$ as a function of $m_N$.
As in the main text, we consider only the first-generation quarks throughout this appendix.
The red line is our estimate for the sensitivity of Ref.~\cite{Sirunyan:2018xiv}.
Since this search is background dominated, we can estimate the future
sensitivity by scaling with the square root of the luminosity
increase. This results in the blue line, representing the final
high-luminosity LHC result.  The corresponding coloured bands are our
estimate of the uncertainty of our conversion procedure, based on the
cross section uncertainties. The bands are made symmetric, assuming
for simplicity that our cross section has a symmetric error of the
order of the difference of our calculation relative to the CMS
numbers.

As mentioned above, the search by CMS assumes that the HNL decays
promptly. The cuts on the impact parameter used by CMS correspond to
0.1 (0.4) mm in the transverse (longitudinal) coordinates. We
therefore simply require HNLs to decay with lengths smaller than
$L_\mathrm{exp} = 0.1$ mm. This allows us to estimate the range of
$V_{\alpha N}$ versus $m_N$ that could be probed, once we fix
$\Lambda$ (and the type of operator). An example for ${\cal O}_{dN}$
and $\Lambda=2$ TeV is shown in fig.~\ref{fig:ReInt} on the
right. Because of the fast decays for HNLs with masses larger than
about 100 GeV, very small values of $V_{\alpha N}$ could be probed 
this way, if $\Lambda$ is small enough to yield testable
  production cross sections. {\it We stress, however, that already for
$\Lambda$ in excess of 3 TeV, currently no new constraints on
$V_{\alpha N}$ could be set}, beyond those already found in the CMS
analysis. The results shown in fig.~\ref{fig:ReInt} are valid for
$\alpha=e$. Results for $\alpha=\mu$ should be quite similar. Also, in
the example calculation we used ${\cal O}_{dN}$ for production of
HNLs. Results for the other HNL pair operators are qualitatively
similar.

We close this discussion by stressing again that many HNL searches at
the LHC could be used to place limits on $N_R$SMEFT parameters. Our
reinterpreation of the CMS search \cite{Sirunyan:2018xiv}, however,
carries a large uncertainty and reinterpretation of other searches
will be similarly limited. We believe only the experimental
collaborations will be able to make reliable searches for our EFT
scenario. The main issues we found in this reinterpretation are: (i)
our conversion of the limits presented in Ref.~\cite{Sirunyan:2018xiv} into
limits on cross section times branching ratio has a surprisingly large
error. We therefore urge our experimental colleagues to present the
``model-independent'' cross section times branching ratio limits in
addition to the model-dependent final results in future
publications. And, (ii) for an improved estimate a better
understanding of the differences in efficiencies for HNL detection in
different models would be needed.  However, this will require a
complete Monte Carlo simulation, which is beyond the scope of our
present paper.

\bibliographystyle{utphys}
\bibliography{RefsEFT}

\providecommand{\href}[2]{#2}\begingroup\raggedright\begin{thebibliography}{100}

\bibitem{Alimena:2019zri}
J.~Alimena {\em et~al.}, ``{Searching for long-lived particles beyond the
  Standard Model at the Large Hadron Collider},''
  \href{http://dx.doi.org/10.1088/1361-6471/ab4574}{{\em J. Phys. G} {\bfseries
  47} no.~9, (2020) 090501}, \href{http://arxiv.org/abs/1903.04497}{{\ttfamily
  arXiv:1903.04497 [hep-ex]}}.

\bibitem{Lee:2018pag}
L.~Lee, C.~Ohm, A.~Soffer, and T.-T. Yu, ``{Collider Searches for Long-Lived
  Particles Beyond the Standard Model},''
  \href{http://dx.doi.org/10.1016/j.ppnp.2019.02.006}{{\em Prog. Part. Nucl.
  Phys.} {\bfseries 106} (2019) 210--255},
  \href{http://arxiv.org/abs/1810.12602}{{\ttfamily arXiv:1810.12602
  [hep-ph]}}.

\bibitem{Curtin:2018mvb}
D.~Curtin {\em et~al.}, ``{Long-Lived Particles at the Energy Frontier: The
  MATHUSLA Physics Case},''
  \href{http://dx.doi.org/10.1088/1361-6633/ab28d6}{{\em Rept. Prog. Phys.}
  {\bfseries 82} no.~11, (2019) 116201},
  \href{http://arxiv.org/abs/1806.07396}{{\ttfamily arXiv:1806.07396
  [hep-ph]}}.

\bibitem{Aaboud:2017mpt}
{\bfseries ATLAS} Collaboration, M.~Aaboud {\em et~al.}, ``{Search for
  long-lived charginos based on a disappearing-track signature in pp collisions
  at $ \sqrt{s}=13 $ TeV with the ATLAS detector},''
  \href{http://dx.doi.org/10.1007/JHEP06(2018)022}{{\em JHEP} {\bfseries 06}
  (2018) 022}, \href{http://arxiv.org/abs/1712.02118}{{\ttfamily
  arXiv:1712.02118 [hep-ex]}}.

\bibitem{Sirunyan:2018ldc}
{\bfseries CMS} Collaboration, A.~M. Sirunyan {\em et~al.}, ``{Search for
  disappearing tracks as a signature of new long-lived particles in
  proton-proton collisions at $\sqrt{s} =$ 13 TeV},''
  \href{http://dx.doi.org/10.1007/JHEP08(2018)016}{{\em JHEP} {\bfseries 08}
  (2018) 016}, \href{http://arxiv.org/abs/1804.07321}{{\ttfamily
  arXiv:1804.07321 [hep-ex]}}.

\bibitem{Sirunyan:2020pjd}
{\bfseries CMS} Collaboration, A.~M. Sirunyan {\em et~al.}, ``{Search for
  disappearing tracks in proton-proton collisions at $\sqrt{s} =$ 13 TeV},''
  \href{http://dx.doi.org/10.1016/j.physletb.2020.135502}{{\em Phys. Lett. B}
  {\bfseries 806} (2020) 135502},
  \href{http://arxiv.org/abs/2004.05153}{{\ttfamily arXiv:2004.05153
  [hep-ex]}}.

\bibitem{Dreiner:1997uz}
H.~K. Dreiner, ``{An Introduction to explicit R-parity violation},''
  \href{http://dx.doi.org/10.1142/9789814307505_0017}{{\em Adv. Ser. Direct.
  High Energy Phys.} {\bfseries 21} (2010) 565--583},
  \href{http://arxiv.org/abs/hep-ph/9707435}{{\ttfamily arXiv:hep-ph/9707435}}.

\bibitem{Barbier:2004ez}
R.~Barbier {\em et~al.}, ``{R-parity violating supersymmetry},''
  \href{http://dx.doi.org/10.1016/j.physrep.2005.08.006}{{\em Phys. Rept.}
  {\bfseries 420} (2005) 1--202},
  \href{http://arxiv.org/abs/hep-ph/0406039}{{\ttfamily arXiv:hep-ph/0406039}}.

\bibitem{Mohapatra:2015fua}
R.~N. Mohapatra, ``{Supersymmetry and R-parity: an Overview},''
  \href{http://dx.doi.org/10.1088/0031-8949/90/8/088004}{{\em Phys. Scripta}
  {\bfseries 90} (2015) 088004},
  \href{http://arxiv.org/abs/1503.06478}{{\ttfamily arXiv:1503.06478
  [hep-ph]}}.

\bibitem{Porod:2000hv}
W.~Porod, M.~Hirsch, J.~Romao, and J.~W.~F. Valle, ``{Testing neutrino mixing
  at future collider experiments},''
  \href{http://dx.doi.org/10.1103/PhysRevD.63.115004}{{\em Phys. Rev. D}
  {\bfseries 63} (2001) 115004},
  \href{http://arxiv.org/abs/hep-ph/0011248}{{\ttfamily arXiv:hep-ph/0011248}}.

\bibitem{Hirsch:2003fe}
M.~Hirsch and W.~Porod, ``{Neutrino properties and the decay of the lightest
  supersymmetric particle},''
  \href{http://dx.doi.org/10.1103/PhysRevD.68.115007}{{\em Phys. Rev. D}
  {\bfseries 68} (2003) 115007},
  \href{http://arxiv.org/abs/hep-ph/0307364}{{\ttfamily arXiv:hep-ph/0307364}}.

\bibitem{deSalas:2020pgw}
P.~F. de~Salas, D.~V. Forero, S.~Gariazzo, P.~Mart\'\i{}nez-Mirav\'e, O.~Mena,
  C.~A. Ternes, M.~T\'ortola, and J.~W.~F. Valle, ``{2020 global reassessment
  of the neutrino oscillation picture},''
  \href{http://dx.doi.org/10.1007/JHEP02(2021)071}{{\em JHEP} {\bfseries 02}
  (2021) 071}, \href{http://arxiv.org/abs/2006.11237}{{\ttfamily
  arXiv:2006.11237 [hep-ph]}}.

\bibitem{Esteban:2020cvm}
I.~Esteban, M.~C. Gonzalez-Garcia, M.~Maltoni, T.~Schwetz, and A.~Zhou, ``{The
  fate of hints: updated global analysis of three-flavor neutrino
  oscillations},'' \href{http://dx.doi.org/10.1007/JHEP09(2020)178}{{\em JHEP}
  {\bfseries 09} (2020) 178}, \href{http://arxiv.org/abs/2007.14792}{{\ttfamily
  arXiv:2007.14792 [hep-ph]}}.

\bibitem{Capozzi:2017ipn}
F.~Capozzi, E.~Di~Valentino, E.~Lisi, A.~Marrone, A.~Melchiorri, and
  A.~Palazzo, ``{Global constraints on absolute neutrino masses and their
  ordering},'' \href{http://dx.doi.org/10.1103/PhysRevD.95.096014}{{\em Phys.
  Rev. D} {\bfseries 95} no.~9, (2017) 096014},
  \href{http://arxiv.org/abs/2003.08511}{{\ttfamily arXiv:2003.08511
  [hep-ph]}}. [Addendum: Phys.Rev.D 101, 116013 (2020)].

\bibitem{Minkowski:1977sc}
P.~Minkowski, ``{$\mu \to e\gamma$ at a Rate of One Out of $10^{9}$ Muon
  Decays?},'' \href{http://dx.doi.org/10.1016/0370-2693(77)90435-X}{{\em Phys.
  Lett. B} {\bfseries 67} (1977) 421--428}.

\bibitem{Yanagida:1979as}
T.~Yanagida, ``{Horizontal gauge symmetry and masses of neutrinos},'' {\em
  Conf. Proc. C} {\bfseries 7902131} (1979) 95--99.

\bibitem{GellMann:1980vs}
M.~Gell-Mann, P.~Ramond, and R.~Slansky, ``{Complex Spinors and Unified
  Theories},'' {\em Conf. Proc. C} {\bfseries 790927} (1979) 315--321,
  \href{http://arxiv.org/abs/1306.4669}{{\ttfamily arXiv:1306.4669 [hep-th]}}.

\bibitem{Mohapatra:1979ia}
R.~N. Mohapatra and G.~Senjanovic, ``{Neutrino Mass and Spontaneous Parity
  Nonconservation},'' \href{http://dx.doi.org/10.1103/PhysRevLett.44.912}{{\em
  Phys. Rev. Lett.} {\bfseries 44} (1980) 912}.

\bibitem{Schechter:1980gr}
J.~Schechter and J.~W.~F. Valle, ``{Neutrino Masses in SU(2) x U(1)
  Theories},'' \href{http://dx.doi.org/10.1103/PhysRevD.22.2227}{{\em Phys.
  Rev. D} {\bfseries 22} (1980) 2227}.

\bibitem{Mohapatra:1986bd}
R.~Mohapatra and J.~Valle, ``{Neutrino Mass and Baryon Number Nonconservation
  in Superstring Models},''
\href{http://dx.doi.org/10.1103/PhysRevD.34.1642}{{\em Phys. Rev.} {\bfseries
  D34} (1986) 1642}.

\bibitem{Akhmedov:1995ip}
E.~K. Akhmedov, M.~Lindner, E.~Schnapka, and J.~Valle, ``{Left-right symmetry
  breaking in NJL approach},''
  \href{http://dx.doi.org/10.1016/0370-2693(95)01504-3}{{\em Phys.Lett.}
  {\bfseries B368} (1996) 270--280},
\href{http://arxiv.org/abs/hep-ph/9507275}{{\ttfamily arXiv:hep-ph/9507275
  [hep-ph]}}.

\bibitem{Akhmedov:1995vm}
E.~K. Akhmedov, M.~Lindner, E.~Schnapka, and J.~Valle, ``{Dynamical left-right
  symmetry breaking},'' \href{http://dx.doi.org/10.1103/PhysRevD.53.2752}{{\em
  Phys.Rev.} {\bfseries D53} (1996) 2752--2780},
\href{http://arxiv.org/abs/hep-ph/9509255}{{\ttfamily arXiv:hep-ph/9509255
  [hep-ph]}}.

\bibitem{Bergsma:1985is}
{\bfseries CHARM} Collaboration, F.~Bergsma {\em et~al.}, ``{A Search for
  Decays of Heavy Neutrinos in the Mass Range 0.5-{GeV} to 2.8-{GeV}},''
  \href{http://dx.doi.org/10.1016/0370-2693(86)91601-1}{{\em Phys. Lett. B}
  {\bfseries 166} (1986) 473--478}.

\bibitem{Bernardi:1987ek}
G.~Bernardi {\em et~al.}, ``{FURTHER LIMITS ON HEAVY NEUTRINO COUPLINGS},''
  \href{http://dx.doi.org/10.1016/0370-2693(88)90563-1}{{\em Phys. Lett. B}
  {\bfseries 203} (1988) 332--334}.

\bibitem{Baranov:1992vq}
S.~A. Baranov {\em et~al.}, ``{Search for heavy neutrinos at the IHEP-JINR
  neutrino detector},''
  \href{http://dx.doi.org/10.1016/0370-2693(93)90405-7}{{\em Phys. Lett. B}
  {\bfseries 302} (1993) 336--340}.

\bibitem{Abreu:1996pa}
{\bfseries DELPHI} Collaboration, P.~Abreu {\em et~al.}, ``{Search for neutral
  heavy leptons produced in Z decays},''
  \href{http://dx.doi.org/10.1007/s002880050370}{{\em Z. Phys. C} {\bfseries
  74} (1997) 57--71}. [Erratum: Z.Phys.C 75, 580 (1997)].

\bibitem{Sirunyan:2018xiv}
{\bfseries CMS} Collaboration, A.~M. Sirunyan {\em et~al.}, ``{Search for heavy
  Majorana neutrinos in same-sign dilepton channels in proton-proton collisions
  at $ \sqrt{s}=13 $ TeV},''
  \href{http://dx.doi.org/10.1007/JHEP01(2019)122}{{\em JHEP} {\bfseries 01}
  (2019) 122}, \href{http://arxiv.org/abs/1806.10905}{{\ttfamily
  arXiv:1806.10905 [hep-ex]}}.

\bibitem{Aad:2019kiz}
{\bfseries ATLAS} Collaboration, G.~Aad {\em et~al.}, ``{Search for heavy
  neutral leptons in decays of $W$ bosons produced in 13 TeV $pp$ collisions
  using prompt and displaced signatures with the ATLAS detector},''
  \href{http://dx.doi.org/10.1007/JHEP10(2019)265}{{\em JHEP} {\bfseries 10}
  (2019) 265}, \href{http://arxiv.org/abs/1905.09787}{{\ttfamily
  arXiv:1905.09787 [hep-ex]}}.

\bibitem{Abashian:2000cg}
{\bfseries Belle} Collaboration, A.~Abashian {\em et~al.}, ``{The Belle
  Detector},'' \href{http://dx.doi.org/10.1016/S0168-9002(01)02013-7}{{\em
  Nucl. Instrum. Meth. A} {\bfseries 479} (2002) 117--232}.

\bibitem{Abe:2010gxa}
{\bfseries Belle-II} Collaboration, T.~Abe {\em et~al.}, ``{Belle II Technical
  Design Report},'' \href{http://arxiv.org/abs/1011.0352}{{\ttfamily
  arXiv:1011.0352 [physics.ins-det]}}.

\bibitem{Kou:2018nap}
{\bfseries Belle-II} Collaboration, W.~Altmannshofer {\em et~al.}, ``{The Belle
  II Physics Book},'' \href{http://dx.doi.org/10.1093/ptep/ptz106}{{\em PTEP}
  {\bfseries 2019} no.~12, (2019) 123C01},
  \href{http://arxiv.org/abs/1808.10567}{{\ttfamily arXiv:1808.10567
  [hep-ex]}}. [Erratum: PTEP 2020, 029201 (2020)].

\bibitem{Liventsev:2013zz}
{\bfseries Belle} Collaboration, D.~Liventsev {\em et~al.}, ``{Search for heavy
  neutrinos at Belle},''
  \href{http://dx.doi.org/10.1103/PhysRevD.87.071102}{{\em Phys. Rev. D}
  {\bfseries 87} no.~7, (2013) 071102},
  \href{http://arxiv.org/abs/1301.1105}{{\ttfamily arXiv:1301.1105 [hep-ex]}}.
  [Erratum: Phys.Rev.D 95, 099903 (2017)].

\bibitem{Kobach:2014hea}
A.~Kobach and S.~Dobbs, ``{Heavy Neutrinos and the Kinematics of Tau Decays},''
  \href{http://dx.doi.org/10.1103/PhysRevD.91.053006}{{\em Phys. Rev. D}
  {\bfseries 91} no.~5, (2015) 053006},
  \href{http://arxiv.org/abs/1412.4785}{{\ttfamily arXiv:1412.4785 [hep-ph]}}.

\bibitem{Dib:2019tuj}
C.~O. Dib, J.~C. Helo, M.~Nayak, N.~A. Neill, A.~Soffer, and J.~Zamora-Saa,
  ``{Searching for a sterile neutrino that mixes predominantly with $\nu_\tau$
  at $B$ factories},''
  \href{http://dx.doi.org/10.1103/PhysRevD.101.093003}{{\em Phys. Rev. D}
  {\bfseries 101} no.~9, (2020) 093003},
  \href{http://arxiv.org/abs/1908.09719}{{\ttfamily arXiv:1908.09719
  [hep-ph]}}.

\bibitem{Cottin:2018kmq}
G.~Cottin, J.~C. Helo, and M.~Hirsch, ``{Searches for light sterile neutrinos
  with multitrack displaced vertices},''
  \href{http://dx.doi.org/10.1103/PhysRevD.97.055025}{{\em Phys. Rev. D}
  {\bfseries 97} no.~5, (2018) 055025},
  \href{http://arxiv.org/abs/1801.02734}{{\ttfamily arXiv:1801.02734
  [hep-ph]}}.

\bibitem{Cottin:2018nms}
G.~Cottin, J.~C. Helo, and M.~Hirsch, ``{Displaced vertices as probes of
  sterile neutrino mixing at the LHC},''
  \href{http://dx.doi.org/10.1103/PhysRevD.98.035012}{{\em Phys. Rev. D}
  {\bfseries 98} no.~3, (2018) 035012},
  \href{http://arxiv.org/abs/1806.05191}{{\ttfamily arXiv:1806.05191
  [hep-ph]}}.

\bibitem{Drewes:2019fou}
M.~Drewes and J.~Hajer, ``{Heavy Neutrinos in displaced vertex searches at the
  LHC and HL-LHC},'' \href{http://dx.doi.org/10.1007/JHEP02(2020)070}{{\em
  JHEP} {\bfseries 02} (2020) 070},
  \href{http://arxiv.org/abs/1903.06100}{{\ttfamily arXiv:1903.06100
  [hep-ph]}}.

\bibitem{Bondarenko:2019tss}
K.~Bondarenko, A.~Boyarsky, M.~Ovchynnikov, O.~Ruchayskiy, and L.~Shchutska,
  ``{Probing new physics with displaced vertices: muon tracker at CMS},''
  \href{http://dx.doi.org/10.1103/PhysRevD.100.075015}{{\em Phys. Rev. D}
  {\bfseries 100} no.~7, (2019) 075015},
  \href{http://arxiv.org/abs/1903.11918}{{\ttfamily arXiv:1903.11918
  [hep-ph]}}.

\bibitem{Liu:2019ayx}
J.~Liu, Z.~Liu, L.-T. Wang, and X.-P. Wang, ``{Seeking for sterile neutrinos
  with displaced leptons at the LHC},''
  \href{http://dx.doi.org/10.1007/JHEP07(2019)159}{{\em JHEP} {\bfseries 07}
  (2019) 159}, \href{http://arxiv.org/abs/1904.01020}{{\ttfamily
  arXiv:1904.01020 [hep-ph]}}.

\bibitem{Bauer:2019vqk}
M.~Bauer, O.~Brandt, L.~Lee, and C.~Ohm, ``{ANUBIS: Proposal to search for
  long-lived neutral particles in CERN service shafts},''
  \href{http://arxiv.org/abs/1909.13022}{{\ttfamily arXiv:1909.13022
  [physics.ins-det]}}.

\bibitem{Gligorov:2018vkc}
V.~V. Gligorov, S.~Knapen, B.~Nachman, M.~Papucci, and D.~J. Robinson,
  ``{Leveraging the ALICE/L3 cavern for long-lived particle searches},''
  \href{http://dx.doi.org/10.1103/PhysRevD.99.015023}{{\em Phys. Rev. D}
  {\bfseries 99} no.~1, (2019) 015023},
  \href{http://arxiv.org/abs/1810.03636}{{\ttfamily arXiv:1810.03636
  [hep-ph]}}.

\bibitem{Gligorov:2017nwh}
V.~V. Gligorov, S.~Knapen, M.~Papucci, and D.~J. Robinson, ``{Searching for
  Long-lived Particles: A Compact Detector for Exotics at LHCb},''
  \href{http://dx.doi.org/10.1103/PhysRevD.97.015023}{{\em Phys. Rev. D}
  {\bfseries 97} no.~1, (2018) 015023},
  \href{http://arxiv.org/abs/1708.09395}{{\ttfamily arXiv:1708.09395
  [hep-ph]}}.

\bibitem{Feng:2017uoz}
J.~L. Feng, I.~Galon, F.~Kling, and S.~Trojanowski, ``{ForwArd Search
  ExpeRiment at the LHC},''
  \href{http://dx.doi.org/10.1103/PhysRevD.97.035001}{{\em Phys. Rev. D}
  {\bfseries 97} no.~3, (2018) 035001},
  \href{http://arxiv.org/abs/1708.09389}{{\ttfamily arXiv:1708.09389
  [hep-ph]}}.

\bibitem{Ariga:2018uku}
{\bfseries FASER} Collaboration, A.~Ariga {\em et~al.},
  ``{FASER\textquoteright{}s physics reach for long-lived particles},''
  \href{http://dx.doi.org/10.1103/PhysRevD.99.095011}{{\em Phys. Rev. D}
  {\bfseries 99} no.~9, (2019) 095011},
  \href{http://arxiv.org/abs/1811.12522}{{\ttfamily arXiv:1811.12522
  [hep-ph]}}.

\bibitem{Pinfold:2019nqj}
J.~L. Pinfold, ``{The MoEDAL Experiment at the LHC\textemdash{}A Progress
  Report},'' \href{http://dx.doi.org/10.3390/universe5020047}{{\em Universe}
  {\bfseries 5} no.~2, (2019) 47}.

\bibitem{Pinfold:2019zwp}
J.~L. Pinfold, ``{The MoEDAL experiment: a new light on the high-energy
  frontier},'' \href{http://dx.doi.org/10.1098/rsta.2019.0382}{{\em Phil.
  Trans. Roy. Soc. Lond. A} {\bfseries 377} no.~2161, (2019) 20190382}.

\bibitem{Chou:2016lxi}
J.~P. Chou, D.~Curtin, and H.~J. Lubatti, ``{New Detectors to Explore the
  Lifetime Frontier},''
  \href{http://dx.doi.org/10.1016/j.physletb.2017.01.043}{{\em Phys. Lett. B}
  {\bfseries 767} (2017) 29--36},
  \href{http://arxiv.org/abs/1606.06298}{{\ttfamily arXiv:1606.06298
  [hep-ph]}}.

\bibitem{Alpigiani:2020tva}
{\bfseries MATHUSLA} Collaboration, C.~Alpigiani {\em et~al.}, ``{An Update to
  the Letter of Intent for MATHUSLA: Search for Long-Lived Particles at the
  HL-LHC},'' \href{http://arxiv.org/abs/2009.01693}{{\ttfamily arXiv:2009.01693
  [physics.ins-det]}}.

\bibitem{Kling:2018wct}
F.~Kling and S.~Trojanowski, ``{Heavy Neutral Leptons at FASER},''
  \href{http://dx.doi.org/10.1103/PhysRevD.97.095016}{{\em Phys. Rev. D}
  {\bfseries 97} no.~9, (2018) 095016},
  \href{http://arxiv.org/abs/1801.08947}{{\ttfamily arXiv:1801.08947
  [hep-ph]}}.

\bibitem{Helo:2018qej}
J.~C. Helo, M.~Hirsch, and Z.~S. Wang, ``{Heavy neutral fermions at the
  high-luminosity LHC},'' \href{http://dx.doi.org/10.1007/JHEP07(2018)056}{{\em
  JHEP} {\bfseries 07} (2018) 056},
  \href{http://arxiv.org/abs/1803.02212}{{\ttfamily arXiv:1803.02212
  [hep-ph]}}.

\bibitem{Dercks:2018wum}
D.~Dercks, H.~K. Dreiner, M.~Hirsch, and Z.~S. Wang, ``{Long-Lived Fermions at
  AL3X},'' \href{http://dx.doi.org/10.1103/PhysRevD.99.055020}{{\em Phys. Rev.
  D} {\bfseries 99} no.~5, (2019) 055020},
  \href{http://arxiv.org/abs/1811.01995}{{\ttfamily arXiv:1811.01995
  [hep-ph]}}.

\bibitem{Hirsch:2020klk}
M.~Hirsch and Z.~S. Wang, ``{Heavy neutral leptons at ANUBIS},''
  \href{http://dx.doi.org/10.1103/PhysRevD.101.055034}{{\em Phys. Rev. D}
  {\bfseries 101} no.~5, (2020) 055034},
  \href{http://arxiv.org/abs/2001.04750}{{\ttfamily arXiv:2001.04750
  [hep-ph]}}.

\bibitem{deVries:2020qns}
J.~De~Vries, H.~K. Dreiner, J.~Y. G\"unther, Z.~S. Wang, and G.~Zhou,
  ``{Long-lived Sterile Neutrinos at the LHC in Effective Field Theory},''
  \href{http://dx.doi.org/10.1007/JHEP03(2021)148}{{\em JHEP} {\bfseries 03}
  (2021) 148}, \href{http://arxiv.org/abs/2010.07305}{{\ttfamily
  arXiv:2010.07305 [hep-ph]}}.

\bibitem{Deppisch:2018eth}
F.~F. Deppisch, W.~Liu, and M.~Mitra, ``{Long-lived Heavy Neutrinos from Higgs
  Decays},'' \href{http://dx.doi.org/10.1007/JHEP08(2018)181}{{\em JHEP}
  {\bfseries 08} (2018) 181}, \href{http://arxiv.org/abs/1804.04075}{{\ttfamily
  arXiv:1804.04075 [hep-ph]}}.

\bibitem{Amrith:2018yfb}
S.~Amrith, J.~M. Butterworth, F.~F. Deppisch, W.~Liu, A.~Varma, and D.~Yallup,
  ``{LHC Constraints on a $B-L$ Gauge Model using Contur},''
  \href{http://dx.doi.org/10.1007/JHEP05(2019)154}{{\em JHEP} {\bfseries 05}
  (2019) 154}, \href{http://arxiv.org/abs/1811.11452}{{\ttfamily
  arXiv:1811.11452 [hep-ph]}}.

\bibitem{Deppisch:2019kvs}
F.~Deppisch, S.~Kulkarni, and W.~Liu, ``{Heavy neutrino production via $Z'$ at
  the lifetime frontier},''
  \href{http://dx.doi.org/10.1103/PhysRevD.100.035005}{{\em Phys. Rev. D}
  {\bfseries 100} no.~3, (2019) 035005},
  \href{http://arxiv.org/abs/1905.11889}{{\ttfamily arXiv:1905.11889
  [hep-ph]}}.

\bibitem{Chiang:2019ajm}
C.-W. Chiang, G.~Cottin, A.~Das, and S.~Mandal, ``{Displaced heavy neutrinos
  from $Z'$ decays at the LHC},''
  \href{http://dx.doi.org/10.1007/JHEP12(2019)070}{{\em JHEP} {\bfseries 12}
  (2019) 070}, \href{http://arxiv.org/abs/1908.09838}{{\ttfamily
  arXiv:1908.09838 [hep-ph]}}.

\bibitem{Dorsner:2016wpm}
I.~Dor\v{s}ner, S.~Fajfer, A.~Greljo, J.~F. Kamenik, and N.~Ko\v{s}nik,
  ``{Physics of leptoquarks in precision experiments and at particle
  colliders},'' \href{http://dx.doi.org/10.1016/j.physrep.2016.06.001}{{\em
  Phys. Rept.} {\bfseries 641} (2016) 1--68},
  \href{http://arxiv.org/abs/1603.04993}{{\ttfamily arXiv:1603.04993
  [hep-ph]}}.

\bibitem{Cottin:2021tfo}
G.~Cottin, O.~Fischer, S.~Mandal, M.~Mitra, and R.~Padhan, ``{Displaced
  Neutrino Jets at the LHeC},''
  \href{http://arxiv.org/abs/2104.13578}{{\ttfamily arXiv:2104.13578
  [hep-ph]}}.

\bibitem{delAguila:2008ir}
F.~del Aguila, S.~Bar-Shalom, A.~Soni, and J.~Wudka, ``{Heavy Majorana
  Neutrinos in the Effective Lagrangian Description: Application to Hadron
  Colliders},'' \href{http://dx.doi.org/10.1016/j.physletb.2008.11.031}{{\em
  Phys. Lett. B} {\bfseries 670} (2009) 399--402},
  \href{http://arxiv.org/abs/0806.0876}{{\ttfamily arXiv:0806.0876 [hep-ph]}}.

\bibitem{Aparici:2009fh}
A.~Aparici, K.~Kim, A.~Santamaria, and J.~Wudka, ``{Right-handed neutrino
  magnetic moments},'' \href{http://dx.doi.org/10.1103/PhysRevD.80.013010}{{\em
  Phys. Rev. D} {\bfseries 80} (2009) 013010},
  \href{http://arxiv.org/abs/0904.3244}{{\ttfamily arXiv:0904.3244 [hep-ph]}}.

\bibitem{Liao:2016qyd}
Y.~Liao and X.-D. Ma, ``{Operators up to Dimension Seven in Standard Model
  Effective Field Theory Extended with Sterile Neutrinos},''
  \href{http://dx.doi.org/10.1103/PhysRevD.96.015012}{{\em Phys. Rev. D}
  {\bfseries 96} no.~1, (2017) 015012},
  \href{http://arxiv.org/abs/1612.04527}{{\ttfamily arXiv:1612.04527
  [hep-ph]}}.

\bibitem{Bell:2005kz}
N.~F. Bell, V.~Cirigliano, M.~J. Ramsey-Musolf, P.~Vogel, and M.~B. Wise,
  ``{How magnetic is the Dirac neutrino?},''
  \href{http://dx.doi.org/10.1103/PhysRevLett.95.151802}{{\em Phys. Rev. Lett.}
  {\bfseries 95} (2005) 151802},
  \href{http://arxiv.org/abs/hep-ph/0504134}{{\ttfamily arXiv:hep-ph/0504134}}.

\bibitem{Graesser:2007yj}
M.~L. Graesser, ``{Broadening the Higgs boson with right-handed neutrinos and a
  higher dimension operator at the electroweak scale},''
  \href{http://dx.doi.org/10.1103/PhysRevD.76.075006}{{\em Phys. Rev. D}
  {\bfseries 76} (2007) 075006},
  \href{http://arxiv.org/abs/0704.0438}{{\ttfamily arXiv:0704.0438 [hep-ph]}}.

\bibitem{Graesser:2007pc}
M.~L. Graesser, ``{Experimental Constraints on Higgs Boson Decays to TeV-scale
  Right-Handed Neutrinos},'' \href{http://arxiv.org/abs/0705.2190}{{\ttfamily
  arXiv:0705.2190 [hep-ph]}}.

\bibitem{Chala:2020vqp}
M.~Chala and A.~Titov, ``{One-loop matching in the SMEFT extended with a
  sterile neutrino},'' \href{http://dx.doi.org/10.1007/JHEP05(2020)139}{{\em
  JHEP} {\bfseries 05} (2020) 139},
  \href{http://arxiv.org/abs/2001.07732}{{\ttfamily arXiv:2001.07732
  [hep-ph]}}.

\bibitem{Caputo:2017pit}
A.~Caputo, P.~Hernandez, J.~Lopez-Pavon, and J.~Salvado, ``{The seesaw portal
  in testable models of neutrino masses},''
  \href{http://dx.doi.org/10.1007/JHEP06(2017)112}{{\em JHEP} {\bfseries 06}
  (2017) 112}, \href{http://arxiv.org/abs/1704.08721}{{\ttfamily
  arXiv:1704.08721 [hep-ph]}}.

\bibitem{Cirigliano:2012ab}
V.~Cirigliano, M.~Gonzalez-Alonso, and M.~L. Graesser, ``{Non-standard Charged
  Current Interactions: beta decays versus the LHC},''
  \href{http://dx.doi.org/10.1007/JHEP02(2013)046}{{\em JHEP} {\bfseries 02}
  (2013) 046}, \href{http://arxiv.org/abs/1210.4553}{{\ttfamily arXiv:1210.4553
  [hep-ph]}}.

\bibitem{Alcaide:2019pnf}
J.~Alcaide, S.~Banerjee, M.~Chala, and A.~Titov, ``{Probes of the Standard
  Model effective field theory extended with a right-handed neutrino},''
  \href{http://dx.doi.org/10.1007/JHEP08(2019)031}{{\em JHEP} {\bfseries 08}
  (2019) 031}, \href{http://arxiv.org/abs/1905.11375}{{\ttfamily
  arXiv:1905.11375 [hep-ph]}}.

\bibitem{Duarte:2015iba}
L.~Duarte, J.~Peressutti, and O.~A. Sampayo, ``{Majorana neutrino decay in an
  Effective Approach},''
  \href{http://dx.doi.org/10.1103/PhysRevD.92.093002}{{\em Phys. Rev. D}
  {\bfseries 92} no.~9, (2015) 093002},
  \href{http://arxiv.org/abs/1508.01588}{{\ttfamily arXiv:1508.01588
  [hep-ph]}}.

\bibitem{Duarte:2016miz}
L.~Duarte, I.~Romero, J.~Peressutti, and O.~A. Sampayo, ``{Effective Majorana
  neutrino decay},''
  \href{http://dx.doi.org/10.1140/epjc/s10052-016-4301-8}{{\em Eur. Phys. J. C}
  {\bfseries 76} no.~8, (2016) 453},
  \href{http://arxiv.org/abs/1603.08052}{{\ttfamily arXiv:1603.08052
  [hep-ph]}}.

\bibitem{Butterworth:2019iff}
J.~M. Butterworth, M.~Chala, C.~Englert, M.~Spannowsky, and A.~Titov, ``{Higgs
  phenomenology as a probe of sterile neutrinos},''
  \href{http://dx.doi.org/10.1103/PhysRevD.100.115019}{{\em Phys. Rev. D}
  {\bfseries 100} no.~11, (2019) 115019},
  \href{http://arxiv.org/abs/1909.04665}{{\ttfamily arXiv:1909.04665
  [hep-ph]}}.

\bibitem{Biekotter:2020tbd}
A.~Biek\"otter, M.~Chala, and M.~Spannowsky, ``{The effective field theory of
  low scale see-saw at colliders},''
  \href{http://dx.doi.org/10.1140/s10052-020-8339-2}{{\em Eur. Phys. J. C}
  {\bfseries 80} no.~8, (2020) 743},
  \href{http://arxiv.org/abs/2007.00673}{{\ttfamily arXiv:2007.00673
  [hep-ph]}}.

\bibitem{Duarte:2014zea}
L.~Duarte, G.~A. Gonz\'alez-Sprinberg, and O.~A. Sampayo, ``{Majorana neutrinos
  production at LHeC in an effective approach},''
  \href{http://dx.doi.org/10.1103/PhysRevD.91.053007}{{\em Phys. Rev. D}
  {\bfseries 91} no.~5, (2015) 053007},
  \href{http://arxiv.org/abs/1412.1433}{{\ttfamily arXiv:1412.1433 [hep-ph]}}.

\bibitem{Duarte:2016caz}
L.~Duarte, J.~Peressutti, and O.~A. Sampayo, ``{Not-that-heavy Majorana
  neutrino signals at the LHC},''
  \href{http://dx.doi.org/10.1088/1361-6471/aa99f5}{{\em J. Phys. G} {\bfseries
  45} no.~2, (2018) 025001}, \href{http://arxiv.org/abs/1610.03894}{{\ttfamily
  arXiv:1610.03894 [hep-ph]}}.

\bibitem{Han:2020pff}
T.~Han, J.~Liao, H.~Liu, and D.~Marfatia, ``{Scalar and tensor neutrino
  interactions},'' \href{http://dx.doi.org/10.1007/JHEP07(2020)207}{{\em JHEP}
  {\bfseries 07} (2020) 207}, \href{http://arxiv.org/abs/2004.13869}{{\ttfamily
  arXiv:2004.13869 [hep-ph]}}.

\bibitem{Barducci:2020ncz}
D.~Barducci, E.~Bertuzzo, A.~Caputo, and P.~Hernandez, ``{Minimal flavor
  violation in the see-saw portal},''
  \href{http://dx.doi.org/10.1007/JHEP06(2020)185}{{\em JHEP} {\bfseries 06}
  (2020) 185}, \href{http://arxiv.org/abs/2003.08391}{{\ttfamily
  arXiv:2003.08391 [hep-ph]}}.

\bibitem{Barducci:2020icf}
D.~Barducci, E.~Bertuzzo, A.~Caputo, P.~Hernandez, and B.~Mele, ``{The see-saw
  portal at future Higgs Factories},''
  \href{http://dx.doi.org/10.1007/JHEP03(2021)117}{{\em JHEP} {\bfseries 03}
  (2021) 117}, \href{http://arxiv.org/abs/2011.04725}{{\ttfamily
  arXiv:2011.04725 [hep-ph]}}.

\bibitem{Bischer:2019ttk}
I.~Bischer and W.~Rodejohann, ``{General neutrino interactions from an
  effective field theory perspective},''
  \href{http://dx.doi.org/10.1016/j.nuclphysb.2019.114746}{{\em Nucl. Phys. B}
  {\bfseries 947} (2019) 114746},
  \href{http://arxiv.org/abs/1905.08699}{{\ttfamily arXiv:1905.08699
  [hep-ph]}}.

\bibitem{Li:2020lba}
T.~Li, X.-D. Ma, and M.~A. Schmidt, ``{General neutrino interactions with
  sterile neutrinos in light of coherent neutrino-nucleus scattering and meson
  invisible decays},'' \href{http://dx.doi.org/10.1007/JHEP07(2020)152}{{\em
  JHEP} {\bfseries 07} (2020) 152},
  \href{http://arxiv.org/abs/2005.01543}{{\ttfamily arXiv:2005.01543
  [hep-ph]}}.

\bibitem{Li:2020wxi}
T.~Li, X.-D. Ma, and M.~A. Schmidt, ``{Constraints on the charged currents in
  general neutrino interactions with sterile neutrinos},''
  \href{http://dx.doi.org/10.1007/JHEP10(2020)115}{{\em JHEP} {\bfseries 10}
  (2020) 115}, \href{http://arxiv.org/abs/2007.15408}{{\ttfamily
  arXiv:2007.15408 [hep-ph]}}.

\bibitem{Bolton:2020xsm}
P.~D. Bolton, F.~F. Deppisch, and C.~Hati, ``{Probing new physics with
  long-range neutrino interactions: an effective field theory approach},''
  \href{http://dx.doi.org/10.1007/JHEP07(2020)013}{{\em JHEP} {\bfseries 07}
  (2020) 013}, \href{http://arxiv.org/abs/2004.08328}{{\ttfamily
  arXiv:2004.08328 [hep-ph]}}.

\bibitem{Dekens:2020ttz}
W.~Dekens, J.~de~Vries, K.~Fuyuto, E.~Mereghetti, and G.~Zhou, ``{Sterile
  neutrinos and neutrinoless double beta decay in effective field theory},''
  \href{http://dx.doi.org/10.1007/JHEP06(2020)097}{{\em JHEP} {\bfseries 06}
  (2020) 097}, \href{http://arxiv.org/abs/2002.07182}{{\ttfamily
  arXiv:2002.07182 [hep-ph]}}.

\bibitem{Cirigliano:2021peb}
V.~Cirigliano, W.~Dekens, J.~de~Vries, K.~Fuyuto, E.~Mereghetti, and R.~Ruiz,
  ``{Leptonic anomalous magnetic moments in $\nu$SMEFT},''
  \href{http://arxiv.org/abs/2105.11462}{{\ttfamily arXiv:2105.11462
  [hep-ph]}}.

\bibitem{Chala:2020pbn}
M.~Chala and A.~Titov, ``{One-loop running of dimension-six Higgs-neutrino
  operators and implications of a large neutrino dipole moment},''
  \href{http://dx.doi.org/10.1007/JHEP09(2020)188}{{\em JHEP} {\bfseries 09}
  (2020) 188}, \href{http://arxiv.org/abs/2006.14596}{{\ttfamily
  arXiv:2006.14596 [hep-ph]}}.

\bibitem{Datta:2020ocb}
A.~Datta, J.~Kumar, H.~Liu, and D.~Marfatia, ``{Anomalous dimensions from gauge
  couplings in SMEFT with right-handed neutrinos},''
  \href{http://dx.doi.org/10.1007/JHEP02(2021)015}{{\em JHEP} {\bfseries 02}
  (2021) 015}, \href{http://arxiv.org/abs/2010.12109}{{\ttfamily
  arXiv:2010.12109 [hep-ph]}}.

\bibitem{Datta:2021akg}
A.~Datta, J.~Kumar, H.~Liu, and D.~Marfatia, ``{Anomalous dimensions from
  Yukawa couplings in SMNEFT: four-fermion operators},''
  \href{http://arxiv.org/abs/2103.04441}{{\ttfamily arXiv:2103.04441
  [hep-ph]}}.

\bibitem{Anelli:2015pba}
{\bfseries SHiP} Collaboration, M.~Anelli {\em et~al.}, ``{A facility to Search
  for Hidden Particles (SHiP) at the CERN SPS},''
  \href{http://arxiv.org/abs/1504.04956}{{\ttfamily arXiv:1504.04956
  [physics.ins-det]}}.

\bibitem{SHiP:2018yqc}
{\bfseries SHiP} Collaboration, C.~Ahdida {\em et~al.}, ``{The experimental
  facility for the Search for Hidden Particles at the CERN SPS},''
  \href{http://dx.doi.org/10.1088/1748-0221/14/03/P03025}{{\em JINST}
  {\bfseries 14} no.~03, (2019) P03025},
  \href{http://arxiv.org/abs/1810.06880}{{\ttfamily arXiv:1810.06880
  [physics.ins-det]}}.

\bibitem{Aad:2008zzm}
{\bfseries ATLAS} Collaboration, G.~Aad {\em et~al.}, ``{The ATLAS Experiment
  at the CERN Large Hadron Collider},''
  \href{http://dx.doi.org/10.1088/1748-0221/3/08/S08003}{{\em JINST} {\bfseries
  3} (2008) S08003}.

\bibitem{Alwall:2007st}
J.~Alwall, P.~Demin, S.~de~Visscher, R.~Frederix, M.~Herquet, F.~Maltoni,
  T.~Plehn, D.~L. Rainwater, and T.~Stelzer, ``{MadGraph/MadEvent v4: The New
  Web Generation},''
  \href{http://dx.doi.org/10.1088/1126-6708/2007/09/028}{{\em JHEP} {\bfseries
  09} (2007) 028}, \href{http://arxiv.org/abs/0706.2334}{{\ttfamily
  arXiv:0706.2334 [hep-ph]}}.

\bibitem{Alwall:2011uj}
J.~Alwall, M.~Herquet, F.~Maltoni, O.~Mattelaer, and T.~Stelzer, ``{MadGraph 5
  : Going Beyond},'' \href{http://dx.doi.org/10.1007/JHEP06(2011)128}{{\em
  JHEP} {\bfseries 06} (2011) 128},
  \href{http://arxiv.org/abs/1106.0522}{{\ttfamily arXiv:1106.0522 [hep-ph]}}.

\bibitem{Alwall:2014hca}
J.~Alwall, R.~Frederix, S.~Frixione, V.~Hirschi, F.~Maltoni, O.~Mattelaer,
  H.~S. Shao, T.~Stelzer, P.~Torrielli, and M.~Zaro, ``{The automated
  computation of tree-level and next-to-leading order differential cross
  sections, and their matching to parton shower simulations},''
  \href{http://dx.doi.org/10.1007/JHEP07(2014)079}{{\em JHEP} {\bfseries 07}
  (2014) 079}, \href{http://arxiv.org/abs/1405.0301}{{\ttfamily arXiv:1405.0301
  [hep-ph]}}.

\bibitem{Wolfenstein:1981kw}
L.~Wolfenstein, ``{Different Varieties of Massive Dirac Neutrinos},''
  \href{http://dx.doi.org/10.1016/0550-3213(81)90096-1}{{\em Nucl. Phys. B}
  {\bfseries 186} (1981) 147--152}.

\bibitem{Petcov:1982ya}
S.~T. Petcov, ``{On Pseudodirac Neutrinos, Neutrino Oscillations and
  Neutrinoless Double beta Decay},''
  \href{http://dx.doi.org/10.1016/0370-2693(82)91246-1}{{\em Phys. Lett. B}
  {\bfseries 110} (1982) 245--249}.

\bibitem{Valle:1982yw}
J.~W.~F. Valle, ``{Neutrinoless Double Beta Decay With Quasi Dirac
  Neutrinos},'' \href{http://dx.doi.org/10.1103/PhysRevD.27.1672}{{\em Phys.
  Rev. D} {\bfseries 27} (1983) 1672--1674}.

\bibitem{Mohapatra:1986aw}
R.~N. Mohapatra, ``{Mechanism for Understanding Small Neutrino Mass in
  Superstring Theories},''
  \href{http://dx.doi.org/10.1103/PhysRevLett.56.561}{{\em Phys. Rev. Lett.}
  {\bfseries 56} (1986) 561--563}.

\bibitem{Barr:2003nn}
S.~M. Barr, ``{A Different seesaw formula for neutrino masses},''
  \href{http://dx.doi.org/10.1103/PhysRevLett.92.101601}{{\em Phys. Rev. Lett.}
  {\bfseries 92} (2004) 101601},
  \href{http://arxiv.org/abs/hep-ph/0309152}{{\ttfamily arXiv:hep-ph/0309152}}.

\bibitem{Grimus:2013tva}
W.~Grimus and L.~Lavoura, ``{Double seesaw mechanism and lepton mixing},''
  \href{http://dx.doi.org/10.1007/JHEP03(2014)004}{{\em JHEP} {\bfseries 03}
  (2014) 004}, \href{http://arxiv.org/abs/1309.3186}{{\ttfamily arXiv:1309.3186
  [hep-ph]}}.

\bibitem{Anamiati:2016uxp}
G.~Anamiati, M.~Hirsch, and E.~Nardi, ``{Quasi-Dirac neutrinos at the LHC},''
  \href{http://dx.doi.org/10.1007/JHEP10(2016)010}{{\em JHEP} {\bfseries 10}
  (2016) 010}, \href{http://arxiv.org/abs/1607.05641}{{\ttfamily
  arXiv:1607.05641 [hep-ph]}}.

\bibitem{Casas:2001sr}
J.~A. Casas and A.~Ibarra, ``{Oscillating neutrinos and $\mu \to e, \gamma$},''
  \href{http://dx.doi.org/10.1016/S0550-3213(01)00475-8}{{\em Nucl. Phys. B}
  {\bfseries 618} (2001) 171--204},
  \href{http://arxiv.org/abs/hep-ph/0103065}{{\ttfamily arXiv:hep-ph/0103065}}.

\bibitem{Cordero-Carrion:2018xre}
I.~Cordero-Carri\'on, M.~Hirsch, and A.~Vicente, ``{Master Majorana neutrino
  mass parametrization},''
  \href{http://dx.doi.org/10.1103/PhysRevD.99.075019}{{\em Phys. Rev. D}
  {\bfseries 99} no.~7, (2019) 075019},
  \href{http://arxiv.org/abs/1812.03896}{{\ttfamily arXiv:1812.03896
  [hep-ph]}}.

\bibitem{Cordero-Carrion:2019qtu}
I.~Cordero-Carri\'on, M.~Hirsch, and A.~Vicente, ``{General parametrization of
  Majorana neutrino mass models},''
  \href{http://dx.doi.org/10.1103/PhysRevD.101.075032}{{\em Phys. Rev. D}
  {\bfseries 101} no.~7, (2020) 075032},
  \href{http://arxiv.org/abs/1912.08858}{{\ttfamily arXiv:1912.08858
  [hep-ph]}}.

\bibitem{Atre:2009rg}
A.~Atre, T.~Han, S.~Pascoli, and B.~Zhang, ``{The Search for Heavy Majorana
  Neutrinos},'' \href{http://dx.doi.org/10.1088/1126-6708/2009/05/030}{{\em
  JHEP} {\bfseries 05} (2009) 030},
  \href{http://arxiv.org/abs/0901.3589}{{\ttfamily arXiv:0901.3589 [hep-ph]}}.

\bibitem{Bondarenko:2018ptm}
K.~Bondarenko, A.~Boyarsky, D.~Gorbunov, and O.~Ruchayskiy, ``{Phenomenology of
  GeV-scale Heavy Neutral Leptons},''
  \href{http://dx.doi.org/10.1007/JHEP11(2018)032}{{\em JHEP} {\bfseries 11}
  (2018) 032}, \href{http://arxiv.org/abs/1805.08567}{{\ttfamily
  arXiv:1805.08567 [hep-ph]}}.

\bibitem{Bhattacharya:2015vja}
S.~Bhattacharya and J.~Wudka, ``{Dimension-seven operators in the standard
  model with right handed neutrinos},''
  \href{http://dx.doi.org/10.1103/PhysRevD.94.055022}{{\em Phys. Rev. D}
  {\bfseries 94} no.~5, (2016) 055022},
  \href{http://arxiv.org/abs/1505.05264}{{\ttfamily arXiv:1505.05264
  [hep-ph]}}. [Erratum: Phys.Rev.D 95, 039904 (2017)].

\bibitem{Li:2021tsq}
H.-L. Li, Z.~Ren, M.-L. Xiao, J.-H. Yu, and Y.-H. Zheng, ``{Operator Bases in
  Effective Field Theories with Sterile Neutrinos: $d \leq 9$},''
  \href{http://arxiv.org/abs/2105.09329}{{\ttfamily arXiv:2105.09329
  [hep-ph]}}.

\bibitem{Weinberg:1979sa}
S.~Weinberg, ``{Baryon and Lepton Nonconserving Processes},''
\href{http://dx.doi.org/10.1103/PhysRevLett.43.1566}{{\em Phys. Rev. Lett.}
  {\bfseries 43} (1979) 1566--1570}.

\bibitem{Fonseca:2017lem}
R.~M. Fonseca, ``{The Sym2Int program: going from symmetries to
  interactions},'' \href{http://dx.doi.org/10.1088/1742-6596/873/1/012045}{{\em
  J. Phys. Conf. Ser.} {\bfseries 873} no.~1, (2017) 012045},
  \href{http://arxiv.org/abs/1703.05221}{{\ttfamily arXiv:1703.05221
  [hep-ph]}}.

\bibitem{Fonseca:2019yya}
R.~M. Fonseca, ``{Enumerating the operators of an effective field theory},''
  \href{http://dx.doi.org/10.1103/PhysRevD.101.035040}{{\em Phys. Rev. D}
  {\bfseries 101} no.~3, (2020) 035040},
  \href{http://arxiv.org/abs/1907.12584}{{\ttfamily arXiv:1907.12584
  [hep-ph]}}.

\bibitem{Christensen:2008py}
N.~D. Christensen and C.~Duhr, ``{FeynRules - Feynman rules made easy},''
  \href{http://dx.doi.org/10.1016/j.cpc.2009.02.018}{{\em Comput. Phys.
  Commun.} {\bfseries 180} (2009) 1614--1641},
  \href{http://arxiv.org/abs/0806.4194}{{\ttfamily arXiv:0806.4194 [hep-ph]}}.

\bibitem{Alloul:2013bka}
A.~Alloul, N.~D. Christensen, C.~Degrande, C.~Duhr, and B.~Fuks, ``{FeynRules
  2.0 - A complete toolbox for tree-level phenomenology},''
  \href{http://dx.doi.org/10.1016/j.cpc.2014.04.012}{{\em Comput. Phys.
  Commun.} {\bfseries 185} (2014) 2250--2300},
  \href{http://arxiv.org/abs/1310.1921}{{\ttfamily arXiv:1310.1921 [hep-ph]}}.

\bibitem{Degrande:2011ua}
C.~Degrande, C.~Duhr, B.~Fuks, D.~Grellscheid, O.~Mattelaer, and T.~Reiter,
  ``{UFO - The Universal FeynRules Output},''
  \href{http://dx.doi.org/10.1016/j.cpc.2012.01.022}{{\em Comput. Phys.
  Commun.} {\bfseries 183} (2012) 1201--1214},
  \href{http://arxiv.org/abs/1108.2040}{{\ttfamily arXiv:1108.2040 [hep-ph]}}.

\bibitem{Degrande:2016aje}
C.~Degrande, O.~Mattelaer, R.~Ruiz, and J.~Turner, ``{Fully-Automated Precision
  Predictions for Heavy Neutrino Production Mechanisms at Hadron Colliders},''
  \href{http://dx.doi.org/10.1103/PhysRevD.94.053002}{{\em Phys. Rev. D}
  {\bfseries 94} no.~5, (2016) 053002},
  \href{http://arxiv.org/abs/1602.06957}{{\ttfamily arXiv:1602.06957
  [hep-ph]}}.

\bibitem{Coloma:2020lgy}
P.~Coloma, E.~Fern\'andez-Mart\'\i{}nez, M.~Gonz\'alez-L\'opez,
  J.~Hern\'andez-Garc\'\i{}a, and Z.~Pavlovic, ``{GeV-scale neutrinos:
  interactions with mesons and DUNE sensitivity},''
  \href{http://dx.doi.org/10.1140/epjc/s10052-021-08861-y}{{\em Eur. Phys. J.
  C} {\bfseries 81} no.~1, (2021) 78},
  \href{http://arxiv.org/abs/2007.03701}{{\ttfamily arXiv:2007.03701
  [hep-ph]}}.

\bibitem{Staub:2012pb}
F.~Staub, ``{SARAH 3.2: Dirac Gauginos, UFO output, and more},''
  \href{http://dx.doi.org/10.1016/j.cpc.2013.02.019}{{\em Comput. Phys.
  Commun.} {\bfseries 184} (2013) 1792--1809},
  \href{http://arxiv.org/abs/1207.0906}{{\ttfamily arXiv:1207.0906 [hep-ph]}}.

\bibitem{Staub:2013tta}
F.~Staub, ``{SARAH 4 : A tool for (not only SUSY) model builders},''
  \href{http://dx.doi.org/10.1016/j.cpc.2014.02.018}{{\em Comput. Phys.
  Commun.} {\bfseries 185} (2014) 1773--1790},
  \href{http://arxiv.org/abs/1309.7223}{{\ttfamily arXiv:1309.7223 [hep-ph]}}.

\bibitem{Artoisenet:2012st}
P.~Artoisenet, R.~Frederix, O.~Mattelaer, and R.~Rietkerk, ``{Automatic
  spin-entangled decays of heavy resonances in Monte Carlo simulations},''
  \href{http://dx.doi.org/10.1007/JHEP03(2013)015}{{\em JHEP} {\bfseries 03}
  (2013) 015}, \href{http://arxiv.org/abs/1212.3460}{{\ttfamily arXiv:1212.3460
  [hep-ph]}}.

\bibitem{Sjostrand:2014zea}
T.~Sj\"ostrand, S.~Ask, J.~R. Christiansen, R.~Corke, N.~Desai, P.~Ilten,
  S.~Mrenna, S.~Prestel, C.~O. Rasmussen, and P.~Z. Skands, ``{An introduction
  to PYTHIA 8.2},'' \href{http://dx.doi.org/10.1016/j.cpc.2015.01.024}{{\em
  Comput. Phys. Commun.} {\bfseries 191} (2015) 159--177},
  \href{http://arxiv.org/abs/1410.3012}{{\ttfamily arXiv:1410.3012 [hep-ph]}}.

\bibitem{Aad:2015rba}
{\bfseries ATLAS} Collaboration, G.~Aad {\em et~al.}, ``{Search for massive,
  long-lived particles using multitrack displaced vertices or displaced lepton
  pairs in pp collisions at $\sqrt{s}$ = 8 TeV with the ATLAS detector},''
  \href{http://dx.doi.org/10.1103/PhysRevD.92.072004}{{\em Phys. Rev. D}
  {\bfseries 92} no.~7, (2015) 072004},
  \href{http://arxiv.org/abs/1504.05162}{{\ttfamily arXiv:1504.05162
  [hep-ex]}}.

\bibitem{Aaboud:2017iio}
{\bfseries ATLAS} Collaboration, M.~Aaboud {\em et~al.}, ``{Search for
  long-lived, massive particles in events with displaced vertices and missing
  transverse momentum in $\sqrt{s}$ = 13 TeV $pp$ collisions with the ATLAS
  detector},'' \href{http://dx.doi.org/10.1103/PhysRevD.97.052012}{{\em Phys.
  Rev. D} {\bfseries 97} no.~5, (2018) 052012},
  \href{http://arxiv.org/abs/1710.04901}{{\ttfamily arXiv:1710.04901
  [hep-ex]}}.

\bibitem{Aad:2020srt}
{\bfseries ATLAS} Collaboration, G.~Aad {\em et~al.}, ``{Search for long-lived,
  massive particles in events with a displaced vertex and a muon with large
  impact parameter in $pp$ collisions at $\sqrt{s} = 13$ TeV with the ATLAS
  detector},'' \href{http://dx.doi.org/10.1103/PhysRevD.102.032006}{{\em Phys.
  Rev. D} {\bfseries 102} no.~3, (2020) 032006},
  \href{http://arxiv.org/abs/2003.11956}{{\ttfamily arXiv:2003.11956
  [hep-ex]}}.

\bibitem{Vincent:2014rja}
A.~C. Vincent, E.~F. Martinez, P.~Hern\'andez, M.~Lattanzi, and O.~Mena,
  ``{Revisiting cosmological bounds on sterile neutrinos},''
  \href{http://dx.doi.org/10.1088/1475-7516/2015/04/006}{{\em JCAP} {\bfseries
  04} (2015) 006}, \href{http://arxiv.org/abs/1408.1956}{{\ttfamily
  arXiv:1408.1956 [astro-ph.CO]}}.

\bibitem{Sabti:2020yrt}
N.~Sabti, A.~Magalich, and A.~Filimonova, ``{An Extended Analysis of Heavy
  Neutral Leptons during Big Bang Nucleosynthesis},''
  \href{http://dx.doi.org/10.1088/1475-7516/2020/11/056}{{\em JCAP} {\bfseries
  11} (2020) 056}, \href{http://arxiv.org/abs/2006.07387}{{\ttfamily
  arXiv:2006.07387 [hep-ph]}}.

\bibitem{Boyarsky:2020dzc}
A.~Boyarsky, M.~Ovchynnikov, O.~Ruchayskiy, and V.~Syvolap, ``{Improved big
  bang nucleosynthesis constraints on heavy neutral leptons},''
  \href{http://dx.doi.org/10.1103/PhysRevD.104.023517}{{\em Phys. Rev. D}
  {\bfseries 104} no.~2, (2021) 023517},
  \href{http://arxiv.org/abs/2008.00749}{{\ttfamily arXiv:2008.00749
  [hep-ph]}}.

\bibitem{Aad:2015xaa}
{\bfseries ATLAS} Collaboration, G.~Aad {\em et~al.}, ``{Search for heavy
  Majorana neutrinos with the ATLAS detector in pp collisions at $ \sqrt{s}=8 $
  TeV},'' \href{http://dx.doi.org/10.1007/JHEP07(2015)162}{{\em JHEP}
  {\bfseries 07} (2015) 162}, \href{http://arxiv.org/abs/1506.06020}{{\ttfamily
  arXiv:1506.06020 [hep-ex]}}.

\bibitem{Aaboud:2019wfg}
{\bfseries ATLAS} Collaboration, M.~Aaboud {\em et~al.}, ``{Search for a
  right-handed gauge boson decaying into a high-momentum heavy neutrino and a
  charged lepton in $pp$ collisions with the ATLAS detector at $\sqrt{s}=13$
  TeV},'' \href{http://dx.doi.org/10.1016/j.physletb.2019.134942}{{\em Phys.
  Lett. B} {\bfseries 798} (2019) 134942},
  \href{http://arxiv.org/abs/1904.12679}{{\ttfamily arXiv:1904.12679
  [hep-ex]}}.

\bibitem{Dev:2013wba}
P.~S.~B. Dev, A.~Pilaftsis, and U.-k. Yang, ``{New Production Mechanism for
  Heavy Neutrinos at the LHC},''
  \href{http://dx.doi.org/10.1103/PhysRevLett.112.081801}{{\em Phys. Rev.
  Lett.} {\bfseries 112} no.~8, (2014) 081801},
  \href{http://arxiv.org/abs/1308.2209}{{\ttfamily arXiv:1308.2209 [hep-ph]}}.

\bibitem{Alva:2014gxa}
D.~Alva, T.~Han, and R.~Ruiz, ``{Heavy Majorana neutrinos from $W\gamma$ fusion
  at hadron colliders},'' \href{http://dx.doi.org/10.1007/JHEP02(2015)072}{{\em
  JHEP} {\bfseries 02} (2015) 072},
  \href{http://arxiv.org/abs/1411.7305}{{\ttfamily arXiv:1411.7305 [hep-ph]}}.

\bibitem{Manohar:2016nzj}
A.~Manohar, P.~Nason, G.~P. Salam, and G.~Zanderighi, ``{How bright is the
  proton? A precise determination of the photon parton distribution
  function},'' \href{http://dx.doi.org/10.1103/PhysRevLett.117.242002}{{\em
  Phys. Rev. Lett.} {\bfseries 117} no.~24, (2016) 242002},
  \href{http://arxiv.org/abs/1607.04266}{{\ttfamily arXiv:1607.04266
  [hep-ph]}}.

\end{thebibliography}\endgroup

\end{document}